\documentclass[fleqn,usenatbib]{mnras}

\usepackage{newtxtext,newtxmath}
\usepackage[T1]{fontenc}

\DeclareRobustCommand{\VAN}[3]{#2}
\let\VANthebibliography\thebibliography
\def\thebibliography{\DeclareRobustCommand{\VAN}[3]{##3}\VANthebibliography}

\usepackage{graphicx}	% Including figure files
\usepackage{amsmath}	% Advanced maths commands
\usepackage{amssymb}	% Extra maths symbols
\usepackage{subfigure}

\usepackage[dvipsnames]{xcolor}     % colorful comments 
\usepackage{ulem}
\usepackage{hyperref}
\usepackage{multirow}
\newcommand{\mpch}{{{\rm Mpc}~h^{-1}}}

\newcommand{\denweight}{}

\title[DECaLS photo-$z$ self-calibration]{Self-calibration of photometric redshift scatter from DECaLS DR8 power spectrum and validation with simulated catalogues}

\author[H. Peng et al.]{
Hui Peng,$^{1,2}$
Haojie Xu,$^{1,2}$
Le Zhang,$^{3,4,5}$
Zhao Chen,$^{1,2}$
Yu Yu$^{1,2}$\thanks{E-mail:\href{mailto:yuyu22@sjtu.edu.cn}{yuyu22@sjtu.edu.cn}}
\\
$^{1}$Department of Astronomy, School of Physics and Astronomy, Shanghai Jiao Tong University, 800 Dongchuan Road, Shanghai 200240, China\\
$^{2}$Key Laboratory for Particle Astrophysics and Cosmology (MOE)/Shanghai Key Laboratory for Particle Physics and Cosmology, Shanghai 200240, China\\
$^{3}$School of Physics and Astronomy, Sun Yat-Sen University, 2 Daxue Road, Zhuhai 519082, China\\
$^{4}$CSST Science Center for the Guangdong-Hong Kong-Macau Greater Bay Area, Zhuhai 519082, China\\
$^{5}$Peng Cheng Laboratory, No.2, Xingke 1st Street, Shenzhen 518000, China
}

\date{Accepted 2022 September 20. Received 2022 September 16; in original form  2022 March 28}

\pubyear{2022}

\begin{document}
\label{firstpage}
\pagerange{\pageref{firstpage}--\pageref{lastpage}}
\maketitle

\begin{abstract}
The uncertainty in the photometric redshift estimation is one of the major systematics in weak lensing cosmology.
The self-calibration method is able to reduce this systematics without assuming strong priors.
We improve the recently proposed self-calibration algorithm to enhance the stability and robustness with the noisy measurement.
The improved algorithm is tested on the power spectra measured from the simulated catalogues constructed according to DECaLS DR8 photometric catalogue.
For the fiducial analysis with 5 equal-width redshift bins over $0<z<1$ and 6 bands over scales $100 \leq \ell<1000$, we find that the improved algorithm successfully reconstructs the scatter rates and the auto power spectrum in true redshift bins at the level of $\sim0.015$ and $\sim4.4$ per cent, respectively.
The bias of the mean redshift is reduced by more than 50 per cent compared to the photo-$z$ without self-calibration, especially for the cases with catastrophic photo-$z$ errors.
The reconstructed results of DECaLS DR8 galaxy sample are in line with the expectations from the simulation validation.
The self-calibration code is publicly available at \href{https://github.com/PengHui-hub/FP-NMF}{https://github.com/PengHui-hub/FP-NMF}.
\end{abstract}


%%%%%%%%%%%%%%%%%%%%%%%%%%%%%%%%%%%%%%%%%%%%%%%%%%

%%%%%%%%%%%%%%%%% BODY OF PAPER %%%%%%%%%%%%%%%%%%

\section{Introduction}
\label{sec:intro}

Weak gravitational lensing is a powerful cosmological probe to study the distribution of dark matter and the properties of dark energy \citep{Albrecht:2006tg}.
More stringent constraints on cosmological model and galaxy evolution is being made thanks to the recent and ongoing surveys such as the Dark Energy Survey \citep[DES;][]{The-Dark-Energy-Survey-Collaboration:2005vv}, the Hyper Suprime-Cam survey \citep{Aihara:2018tn}, the Kilo-Degree Survey \citep{Kuijken:2019vi}, the Dark Energy Spectroscopic Instrument \citep[DESI;][]{DESI-Collaboration:2016vy,DESI-Collaboration:2016vs}, and the \textit{Chinese Space Station Telescope} \citep[\textit{CSST};][]{Gong:2019tb}.

However, weak-lensing cosmology suffers from various sources of systematic errors and the major one among them is the photometric redshift (photo-$z$) errors \citep{Huterer:2006th,Ma:2006wj,Bernstein:2010wc}.
For a typical weak lensing survey with large number of sources, it is very time-consuming and infeasible to obtain a complete spectroscopic redshift measurement.
The majority of objects rely on the inaccurate photometric redshift measurement, and therefore it is crucial to calibrate the redshift as precisely as possible in the cosmological analysis \citep{Newman:2022vy}.

Several methods have been developed, such as the direct calibration technique by attempting to construct a representative spectroscopic sub-sample \citep{Lima:2008vl,Bonnett:2016wb,Hildebrandt:2017ua}, the cross-correlation method by cross correlating with spectroscopic samples in overlapping survey areas \citep[e.g.][]{Newman:2008vv,Matthews:2010un,Schmidt:2013va,McQuinn:2013uh,Menard:2013uh,Kovetz:2017vm,McLeod:2017vz,Gatti:2018tg,van-den-Busch:2020ts,Gatti:2022va,Rau:2022ud} and the self-organising maps method \citep[SOMs;][]{Kohonen1982} which is an unsupervised neural network that preserves topology \citep[e.g.][]{Masters:2015wp,Buchs:2019tx,Davidzon:2019vh,Wright:2020vs,Hildebrandt:2021vt,Hartley:2022wj}.
These techniques either require a representative and complete spectroscopic subsample or adopt some cosmological priors, which may lead to some potential bias in the derived relationship \citep{Newman:2022vy}.

On the contrary, the self-calibration techniques do not rely on external spectroscopic samples \citep{Schneider:2006ta,Benjamin:2010ue,Zhang:2010wr,Zhang:2017um,Schaan:2020up,Stolzner:2021va}.
In the absence of photo-$z$ errors, the galaxy–galaxy correlation between two distinguish redshift bins should vanish under the limber approximation,
and there is no correlation between the low-redshift cosmic shear and high-redshift galaxy distribution.
Therefore, we can use these non-vanishing correlations to infer the photo-$z$ outlier rate of redshift bins.
The Fisher analysis presented in \citet{Zhang:2010wr} shows that the scatters between the photo-$z$ and true-$z$ bins can be accurately reconstructed.
However, finding a stable and efficient algorithm to solve the non-convex nonlinear optimization problem with numerous unknown parameters and multiple constraints is a non-trivial task.
\citet{Zhang:2017um} developed a non-negative matrix factorization (NMF) based algorithm and the validation succeeded on the theoretical data.
Further validation and optimization are required before applying to surveys.

In this work, we construct lightcone simulations to systematically investigate the performance of the self-calibration algorithm based on \citet{Zhang:2017um}.
To deal with the complexity in the simulated galaxy catalogues and observations, the algorithm is improved by modifying the randomness of the initial guess, adjusting the convergence criterion, and assigning weightings on different scales.
After the successful implementations in the simulated data for various configurations, the algorithm is applied to calibrate the photo-$z$ scatter of DECaLS DATA RELEASE 8 (DR8) galaxy sample.

This paper is organized as follows. In Section \ref{sec:method}, we give a brief overview of the self-calibration method and the improved algorithm.
Section \ref{sec:data} describes the simulation and survey data we use to test the algorithm.
The implementations of the algorithm and the main results are presented in Section \ref{sec:results}.
Finally, we conclude and discuss the possibilities to improve the accuracy in Section \ref{sec:conclusion}.
% In our companion paper, Xu et al. (in preparation) also constrained the photometric redshift scatter from the same DR8 galaxy sample, but using the angular two-point correlation function instead of power spectrum.
% The main algorithm of the two are same, but differ in details due to the different statistical properties in configuration space and Fourier space.
% The results are consistent with each other, which reflects the rationality of the self-calibration method.

%%%%%%%%%%%%%
%%%%%%%%%%%%%
\section{The Self-calibration Method and the improved Algorithm}
\label{sec:method}

\subsection{The self-calibration method}

Due to the large uncertainties, typically $\sigma_z=0.05(1+z)$, in the photometric redshift measurement, a non-negligible amount of galaxies in the observed photo-$z$ bin come from other redshifts.
Assume that we split galaxies into $n$ photo-$z$/true-$z$ bins.
Following the notations in \citet{Zhang:2010wr}, we denote the probability of a galaxy in true redshift bin $i$ but observed in photometric redshift bin $j$ as $P_{ij}\,{\equiv}\,N_{i{\rightarrow}j}/N_j^P$.
Here, $N_{i{\rightarrow}j}$ is the number of galaxies scattered from the $i$th true-$z$ bin to the $j$th photo-$z$ bin, and $N_j^P$ is the total number of galaxies in the $j$th photo-$z$ bin.
We have the normalization $\sum_iP_{ij}=1$.
The cross power spectrum of two photo-$z$ bins, $C_{ij}^{gg,P}(\ell)$, and the galaxy power spectrum of true-$z$ bins, $C_{ij}^{gg,R}(\ell)$, is related by
\begin{equation}
    C_{ij}^{gg,P}(\ell)=\sum_kP_{ki}P_{kj}C_{kk}^{gg,R}(\ell)+\delta{N_{ij}^{gg,P}(\ell)}\ .
    \label{eqn:CggPsum}
\end{equation}
This equation has approximated $C_{k{\neq}m}^{gg,R}(\ell)=0$, as the galaxy cross-correlation between non-overlapping redshift bins would vanish under the Limber approximation and in the absence of the lensing magnification.
The last term $\delta{N_{ij}^{gg,P}(\ell)}$ is the associated shot noise ﬂuctuation after the subtraction of its ensemble average from the power spectrum measurements.
The shot noise level is related to the bin size $\Delta\ell$, the sky fraction $f_\mathrm{sky}$ and the number density $\bar{n}_i$ and $\bar{n}_j$ in $i$th and $j$th photo-$z$ bin,
\begin{equation}
    \left(\sigma_{i j}^{g g, P}\right)^{2}=\frac{1}{(2 \ell+1) \Delta \ell f_{\text{sky}}} \frac{1}{\bar{n}_{i} \bar{n}_{j}}\left(1+\delta_{i j}\right)\ . 
    \label{sigma_gg}
\end{equation}
Here, $1+\delta_{i j}$ is the Kronecker Delta.

Similarly, the cross power spectrum between the lensing convergence in the $i$th photo-$z$ bin and the galaxy number density in the $j$th photo-$z$ bin reads
\begin{equation}
    C_{ij}^{Gg,P}(\ell)=\sum_{k{\geq}m}P_{ki}P_{mj}C_{km}^{Gg,R}(\ell)+\delta{N_{ij}^{Gg,P}(\ell)}\ ,
    \label{eqn:CGgPsum}
\end{equation}
where we expect $C_{k<m}^{Gg,R}(\ell)=0$ in the absence of lensing magniﬁcation bias.
The shot noise in this measurement also depends on the galaxy shape noise in the cosmic shear measurement,
\begin{equation}
    \left(\sigma_{i j}^{G g, P}\right)^{2}=\frac{1}{(2 \ell+1) \Delta \ell f_{\text{sky}}} \frac{\gamma_{\text {rms }}^{2}}{\bar{n}_{i} \bar{n}_{j}}\ .
\end{equation}

With the measurements on different scales, the number of observables ($C_{ij}^{Gg,P}$, $C_{ij}^{gg,P}$) is larger than the number of unknowns ($P_{ij}$, $C_{kk}^{gg,R}$, $C_{km}^{Gg,R}$ with $k{\geq}m$).
In principle, the above problem can be solved as long as the system is not degenerate.
Through the Fisher matrix analysis, \citet{Zhang:2010wr} estimated that the statistical errors of photo-$z$ outlier rates can be determined to $\text{0.01 per cent} - \text{1 per cent}$.
However, it is not easy to find an efficient and accurate algorithm to solve this constrained nonlinear optimization problem.

\subsection{The non-negative matrix factorization approach}

For a given $\ell$, we can rewrite the above equations in matrix form,
\begin{align}
    C_{\ell}^{gg,P}&=P^{T}C_{\ell}^{gg,R}P+\delta{N_{\ell}^{gg,P}}\ , \label{eqn:Cggp} \\
    C_{\ell}^{Gg,P}&=P^{T}C_{\ell}^{Gg,R}P+\delta{N_{\ell}^{Gg,P}}\ . \label{eqn:CGgp}
\end{align}
We call the matrix $P$ as scattering matrix, and it is non-negative.
\citet{Zhang:2017um} developed an algorithm based on the non-negative matrix factorization.
The technique consists of two parts, Algorithm 1 and Algorithm 2, which both attempt to find the global optimal result by minimizing
\begin{equation}
    \mathcal{J}\left(P ; C_{\ell=1, \ldots, N_{\ell}}^{g g, R}\right) \equiv \frac{1}{2} \sum_{\ell}\left\|C_{\ell}^{g g, P}-P^{T} C_{\ell}^{g g, R} P\right\|_{F}^{2}\ ,
    \label{define:J}
\end{equation}
where $\|.\|_F$ is the Frobenius form.
$\mathcal{J}$ measures the accumulation of decomposition error across all data matrices between the observations and reconstructions, where $C_{\ell}^{gg,P}$ is the observational data, $P$ and $C_{\ell}^{gg,R}$ are the derived results after completing the matrix factorization.
Algorithm 1 is a fixed-point-based algorithm, which obtains an estimation of the matrix $P$ iteratively from an initial guess.
Algorithm 2 is based on NMF, which can further iterate over the matrix reconstructed by Algorithm 1 to further reduce the $\mathcal{J}$ and ﬁnd the optimal solution.
Note that the iterative rule of Algorithm 2 is designed according to the definition of $\mathcal{J}$.
We refer the reader interested on the iteration design to the Appendix of \citet{Zhang:2017um}.

The definition of $\mathcal{J}$ here is similar to a $\chi^2$ minimization problem.
However, this simple version neglects the variance of each data point and the correlations between the data.
We do not use the $\chi^2$ definition since it is non-trivial to combine the minimum variance weighting into the matrix factorization algorithm.
The $\mathcal{J}$ defined here has the property that for a given input power spectrum data, the result with smaller $\mathcal{J}$ represents the better factorization when we try to optimize the algorithm.
%we can evaluate the performance of different reconstructions by comparing the corresponding $\mathcal{J}$ values.
However, note that the definition of $\mathcal{J}$ also neglects the complexity such as the number of $\ell$ bins, the number of redshift bins and the amplitude of the data points in different cases.
Thus, we caution that the direct comparison of $\mathcal{J}$ values for different binning or weighting schemes is misleading.

In \citet{Zhang:2017um}, the NMF based algorithm successfully obtained the correct and accurate solution from a mock data with theoretical signal and Gaussian measurement noise.
The configuration considered there is quiet optimal, with the sky coverage $f_{\text{sky}}=0.5$, the galaxy number density $\bar{n}_{g}=40\,\text{arcmin}^{-2}$ and the galaxy shape noise $\gamma_{\text {rms}}=0.2$.

\subsection{The improved algorithm}

As the DECaLS DR8 sample is far from the optimal case in the previous work, based on the repeating tests, we improve the algorithm in this work to handle the complexities in the realistic mocks and real observation.

Note that in \citet{Zhang:2017um}, the lensing-galaxy cross-correlation measurement, $C_{\ell}^{Gg,P}$ in equation~(\ref{eqn:CGgp}), is only used to swap the rows of the resulting $P$, which is iterated from a random positive column-sum-to-one matrix in Algorithm 1.
We find that a more realistic initial guess, i.e. a diagonal dominating initial $P$, can automatically break the row order degeneracy in Algorithm 1.
Thus, we use the more realistic initial values for $P$, with larger diagonal elements, and only use the galaxy-galaxy correlation measurement $C_{\ell}^{gg,P}$.

Due to the much large noise in DECaLS DR8 sample data we use ($n_g\sim 2\ \mathrm{arcmin}^{-2}$ and $f_{\text{sky}}=0.125$), the algorithm is less stable with the convergence criterion proposed in \citet{Zhang:2017um}, such as the change of each element in $P$ is less than $10^{-8}$.
We upgrade the convergence criterion in our new flow, and the new condition is mainly determined by $\mathcal{J}$.

Because of the large statistical error in the current observation condition, 
the matrix decomposition problem may not be solved stably.
Starting from some initial guess, the fixed-point iteration algorithm may oscillate, or even worse diverge in the end.
To avoid these situations, the maximum number of iterations for Algorithm 1 is set to 10000, and we break the iteration once the $\mathcal{J}$ begins to increase significantly.
Now the stability of the algorithm is guaranteed and it does not crash or results in singular matrix in any cases.
Algorithm 1 runs 1000 times with different initial matrices.
We use all the obtained $P$ from Algorithm 1 to initialize Algorithm 2.
Again, the maximum number of steps is limited to 10000, and the iteration is ended if $\mathcal{J}$ starts to bounce.
%In order to enhance the stability of the algorithm, we do not initialize $C_{\ell}^{g g, R}={\rm Abs} \left\{P^{T-1} C_{\ell}^{g g, P} P^{-1}\right\}$ in Algorithm 1.

%In fact, $\mathcal{J}$ will decrease rapidly and then increase in most simulation cases, reaching the minimum value after thousands or just several iterations, and the required number of iterations can be roughly determined through a single test.

%Although the Algorithm 2 is based on NMF, the negative values in the input power spectrum have no impact on the stability.

We note that there is no guarantee that the solution with minimum $\mathcal{J}$ is the optimal result in the presence of large noise.
Thus, we approximate the optimal solution as the average over the solutions with low $\mathcal{J}$ values.
This choice is reasonable as long as the results found by Algorithm 2 are randomly distributed around the true answer.
The details of the selection will be shown in Section \ref{result:normal}.

Using 50 cores, the whole algorithm runs less than an hour for each application.
The cost is totally acceptable.
The improved algorithm is first tested on the noiseless theoretical power spectrum.
The 1000 solutions are very clustered within a small compact region, and the solution with minimum $\mathcal{J}$ can reach an accuracy of $\sim 10^{-4}$ for all elements in scattering matrix $P$.

%%%%%%%%%%%%%
%%%%%%%%%%%%%
\section{Data}
\label{sec:data}

\subsection{DECaLS DR8}
\label{sec:obs}

\begin{figure}
    \centering
    \includegraphics[width=0.90\columnwidth]{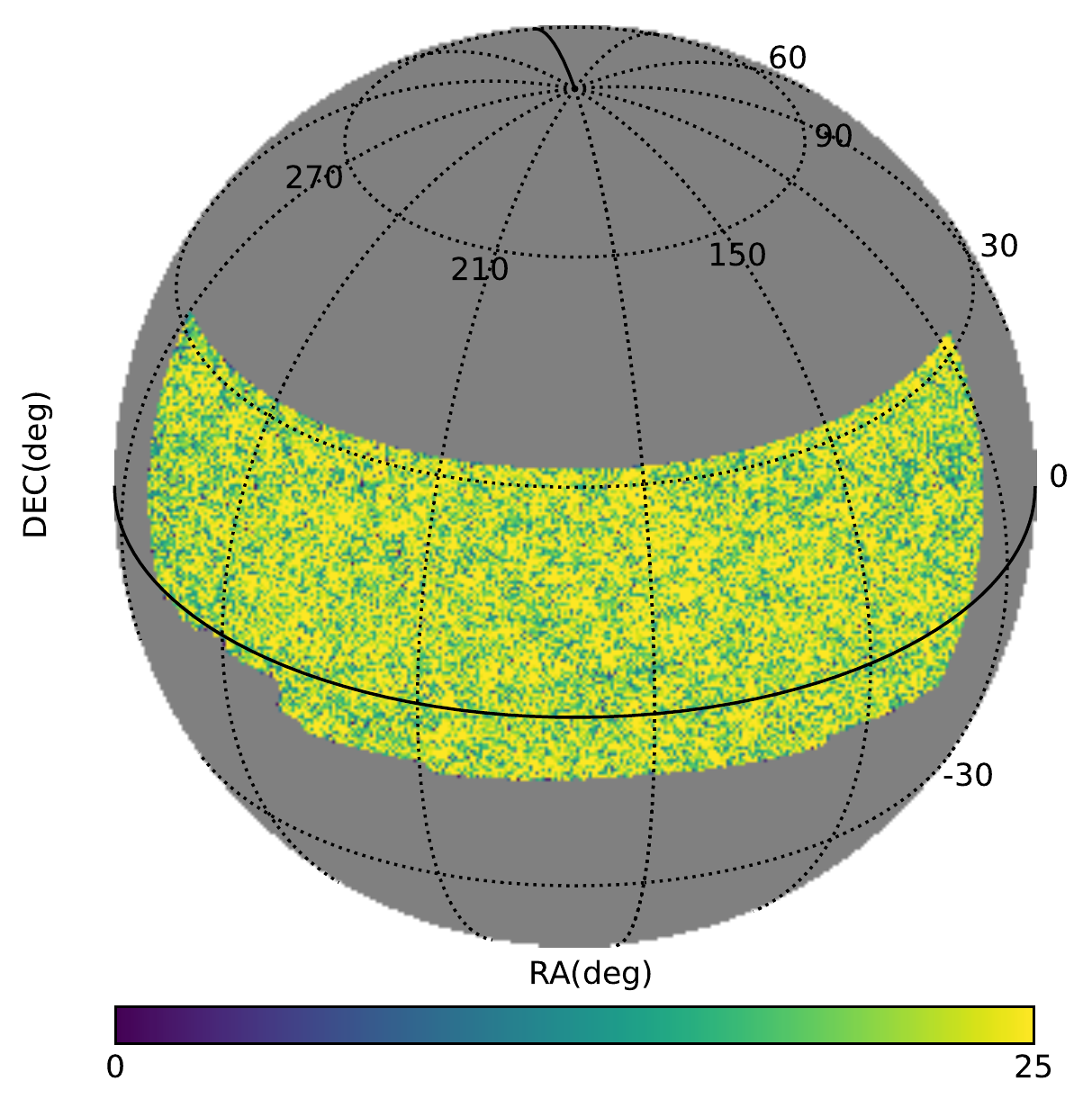}
    
	\includegraphics[width=0.95\columnwidth]{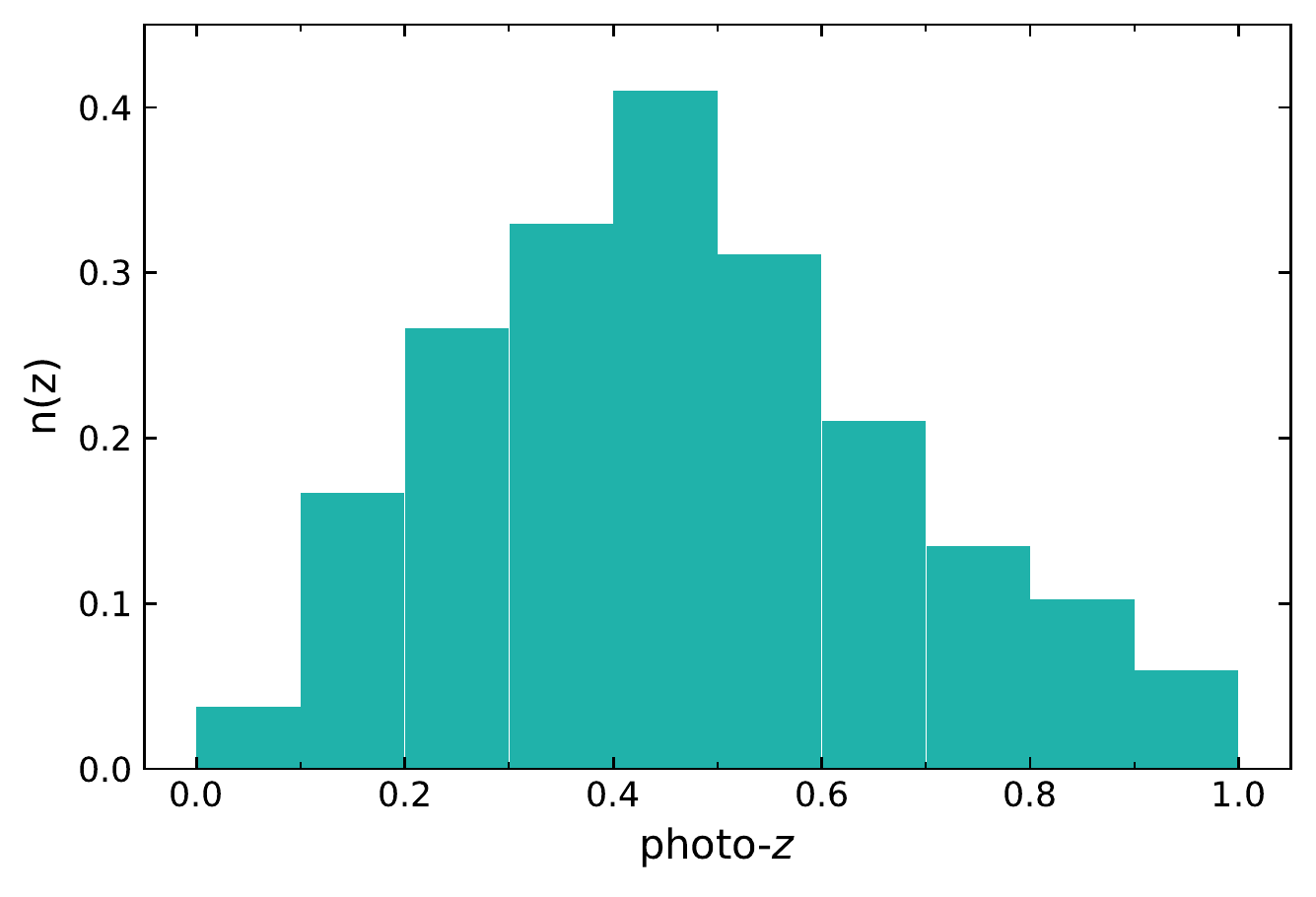}
    \caption{Top panel: sky coverage of the DECaLS NGC DR8 sample. The map is constructed in \textsc{healpix} with resolution parameter \textsc{nside} = 1024 ($\sim 3.4$ arcmin).
    Lower panel: the redshift distribution (surface density in photometric redshift bins) of the DECaLS DR8 galaxy sample. The y-axis is the number of objects per $\text{arcmin}^2$ in the redshift bin of width 0.1.}
    \label{fig:n_z}
\end{figure}

DESI \citep{DESI-Collaboration:2016vy} is a stage-IV spectroscopic galaxy survey which is devoted to producing an unprecedented three-dimensional sky map and reveal the expansion history of the Universe as well as the nature of dark energy\footnote{\href{https://www.desi.lbl.gov/}{https://www.desi.lbl.gov/}}.
Based on DESI Legacy imaging surveys DR8 \citep{Dey:2019uk}, the publicly available photometric redshift catalogue in \citet{Zhou:2021uw} provides photometric redshifts for objects over more than $16000\,\text{deg}^2$ of sky.
DR8 includes data from two regions, the Dark Energy Camera Legacy Survey (DECaLS) survey in the southern portion of the planned DESI footprint ($\text{Dec}. \leq 34^{\circ}$ ) and the MzLS and BASS surveys in the northern sky ($\text{Dec.} \geq 32^{\circ}$).

For $z_{\text{mag}}<21$ objects in the DECaLS region, the scatter between photo-$z$ and spectroscopic redshift is $\thicksim0.013$ and the outlier rate is 1.5 per cent \citep{Zhou:2021uw}.
In this work we use the galaxy photo-$z$ catalogue from DECaLS North Galactic Cap (NGC) region over the redshift range $0<z<1$ and $z$-band magnitude limit $z_{\text{mag}}<21$.
The galaxy samples are constructed according to the selection criterion in \cite{Yang:2021vo} and the morphological types from the \textsc{tractor} software \citep{Lang:2016wy} are used to select out the galaxies.
The selected objects have at least one exposure in each optical band and their fluxes are not affected by the globular clusters, large galaxies or bright stars.
The total galaxy number is 37\,588\,320 over $5146.31\,\text{deg}^2$ of sky and the mean galaxy surface density of the total redshift range $\bar{n}_g=2.03\,\text{arcmin}^{-2}$.
Note that this surface density is much lower than the validation tests in \citet{Zhang:2017um}.
We use 5 redshift bins as the fiducial case in the analysis, with bin width $\Delta z=0.2$.
The \denweight mean photometric redshifts are $[0.139, 0.308, 0.491, 0.686, 0.884]$.
The sky coverage and redshift distribution are presented in Fig.~\ref{fig:n_z}.
To show the robustness of the self-calibration method, we also test the algorithm with 10 redshift bins.

We use the \textsc{healpix} \citep{Gorski:2005te} to construct galaxy overdensity map, with $\textsc{nside}=1024$, corresponding to a spatial resolution of $\sim 3.4$ arcmin.
The angular power spectrum of galaxies is measured with the function \textsc{compute\_coupled\_cell} in \textsc{namaster} \citep{Alonso:2019aa}.

\subsection{Simulation}
\label{sec:simu}

We use one of the high-resolution $N$-body simulations in \citet{Jing:2019uw} to construct mock galaxy catalogues.
The cosmological model adopted is a flat $\Lambda$CDM cosmology consistent with the \textit{WMAP} observations \citep{Komatsu:2011un, Hinshaw:2013vg}, with $\Omega_m=0.268$, $\Omega_\Lambda=0.732$, $h=0.71$, $n_s=0.968$ and $\sigma_8=0.83$.
Inside a $(600~\mpch)^3$ comoving volume with periodical condition, the simulation evovles $3072^3$ particles by a particle-particle-particle-mesh gravity solver from $z=144$.
We cut out curved slices with $300~\mpch$ thickness from the snapshots at various redshifts, and stack them to construct light-cones upto $z\sim 2.48$.
To prevent the repeating structures along line-of-sight, each box is randomly shifted and rotated before slicing.
Totally, we made 300 pseudo-independent light-cones, including dark matter particle distributions, friends-of-friends haloes, and lensing convergence maps.

We use 77 simulated maps, with $67.13\,\text{deg}^2$ each, to cover a similar sky portion of DECaLS DR8 sample.
To construct mock catalogues with the same photo-$z$ distribution and similar noise level as the observation, we use the following steps.
Firstly, we set the number of galaxies in each photo-$z$ bin as the same as the observed DECaLS DR8 sample.
Then, according to the scattering matrix $P^{\text{true}}$ we preset, the number of galaxies in each true-$z$ bin can be fixed.
We regard the halos generated in simulation as the galaxies we need.
In each true-$z$ bin, we pick the galaxies in descending order of halo mass until the number is satisfied.
Finally, to match the elements of matrix, we randomly assign corresponding fraction of galaxies in true-$z$ bins to each photo-$z$ bin.
We use the same setting described in section \ref{sec:obs} to calculate the power spectrum for the simulated catalogues.

%%%%%%%%%%%%%
%%%%%%%%%%%%%
\section{Results}
\label{sec:results}

At sufﬁciently large angular scales, i.e. low-$\ell$ region, the Limber approximation fails and the intrinsic cross-correlation $C_{k{\neq}m}^{gg,R}(\ell)\neq0$.
Meanwhile, the power spectrum measurements at high-$\ell$ are dominated by the shot noise for the data we use.
%It is also necessary to avoid the small scale data which may suffer from intrinsic alignment contamination (if the lensing-galaxy correlation is used), complicated baryonic eﬀects and nonlinear bias between galaxy number density and the underlying dark matter density.
Therefore, we conservatively exclude the modes with $\ell<100$ or $\ell\geq1000$.
We divide the total $\ell$ range $100 \leq \ell<1000$ into $N_{\ell}=6$ broad bands, $[100, 418)$, $[418, 583)$, $[583, 711)$, $[711, 819)$, $[819, 914)$, $[914, 1000)$.
This binning scheme approximates the same shot noise fluctuations in each broad band.

To obtain the solution, we require the number of $\ell$-bins larger than 2.  The constraint on the scattering matrix comes from multiple scales.
Actually, we have the freedom to assign different weighting for different scales.
The iterative rule of the algorithms includes the summation over
all the scales. 
This summation implies that the relative power spectrum amplitude on different scales would influence the overall performance of the factorization. 
Generally speaking, the scales with larger amplitude provide more constraining power, and thus have better reconstruction performance.
However, it is non-trivial to find the best weighting scheme, as for each $C_{ij}(\ell)$, the power spectrum has different $\ell$-dependence. 
Generally, the power spectrum $C^{gg,P}$ is a decreasing function toward small scale.
In the fiducial analysis, we simply multiply the power spectrum measurement by $\ell$ to balance the contribution from large scale and small scale.
Note that this weighting amplifies the noise in the small scale for the above $\ell$-binning scheme.
We present the test on the weighting scheme later.

In Section \ref{result:normal}, we first apply the improved algorithm on the case without catastrophic photo-$z$ errors.
We make decision on the solution selection criterion according to the result of $\mathcal{J}$ in this case.
Then we apply the same analysis to the simulations with catastrophic photo-$z$ errors in Section \ref{result:cata}.
In Section \ref{result:redshift bias}, we discuss the reduction of the mean redshift bias after the self-calibration.
At last, the result from the DECaLS DR8 galaxy sample is presented in Section \ref{result:DECaLS}.
In Appendix we investigate the performance using another binning scheme, and compare the result with the consideration of magnification bias.  Both of them has insignificant impact on the reconstruction performance.

\subsection{Validation on the case without catastrophic photo-$z$ errors}
%outline
\label{result:normal}

\begin{figure}
	\includegraphics[width=\columnwidth]{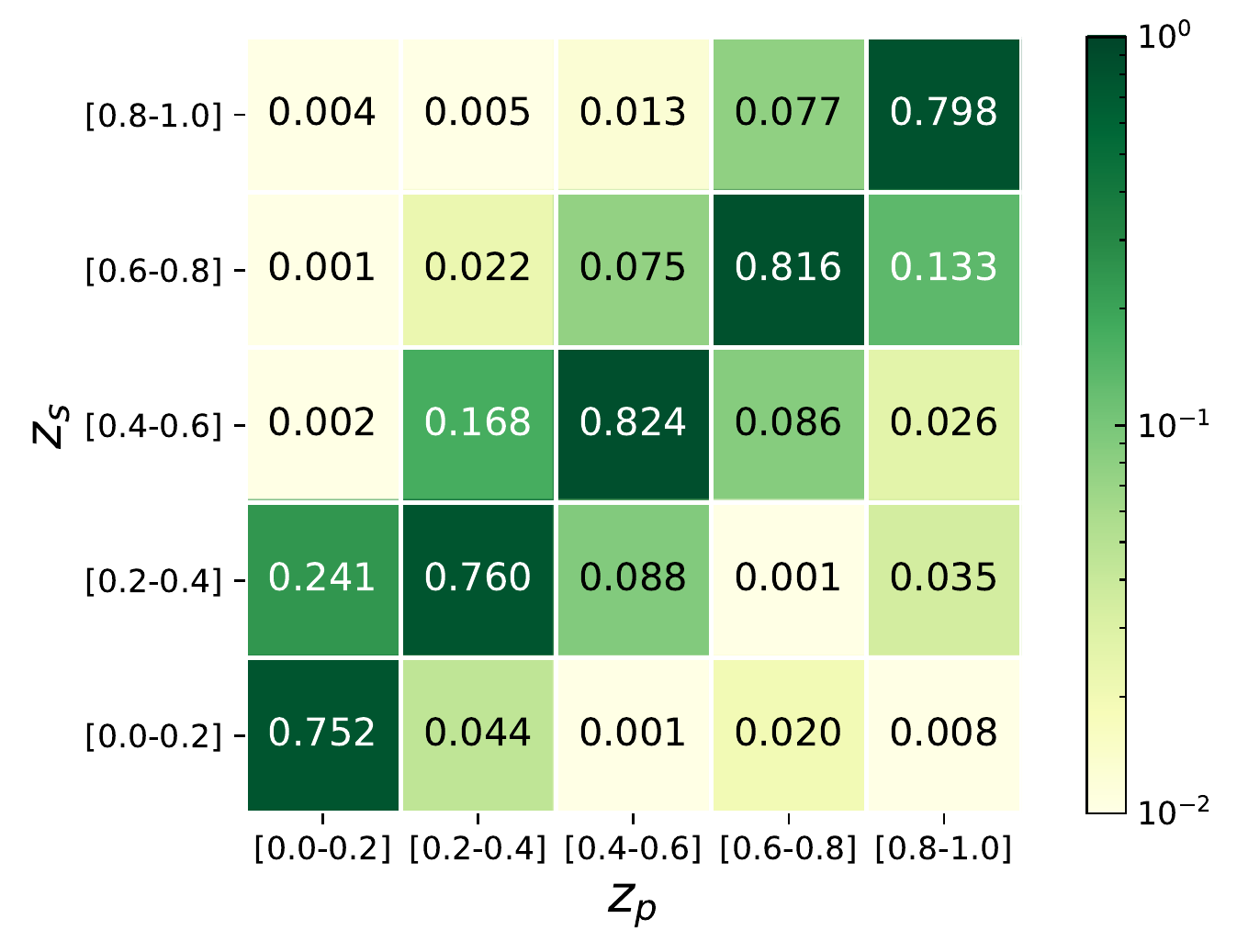}
    \caption{The scattering matrix $P^{\rm true}$ we preset for the validation.
    The horizontal axis $z_p$ denotes the photometric redshift bins and the vertical axis $z_s$ labels the true redshift bins. 
    The values present the scatter rates from the true redshift bin to the photometric redshift bin, under the column-sum-to-one constrain.}
    \label{fig:sim_normal_pture}
\end{figure}

\begin{figure}
	\includegraphics[width=\columnwidth]{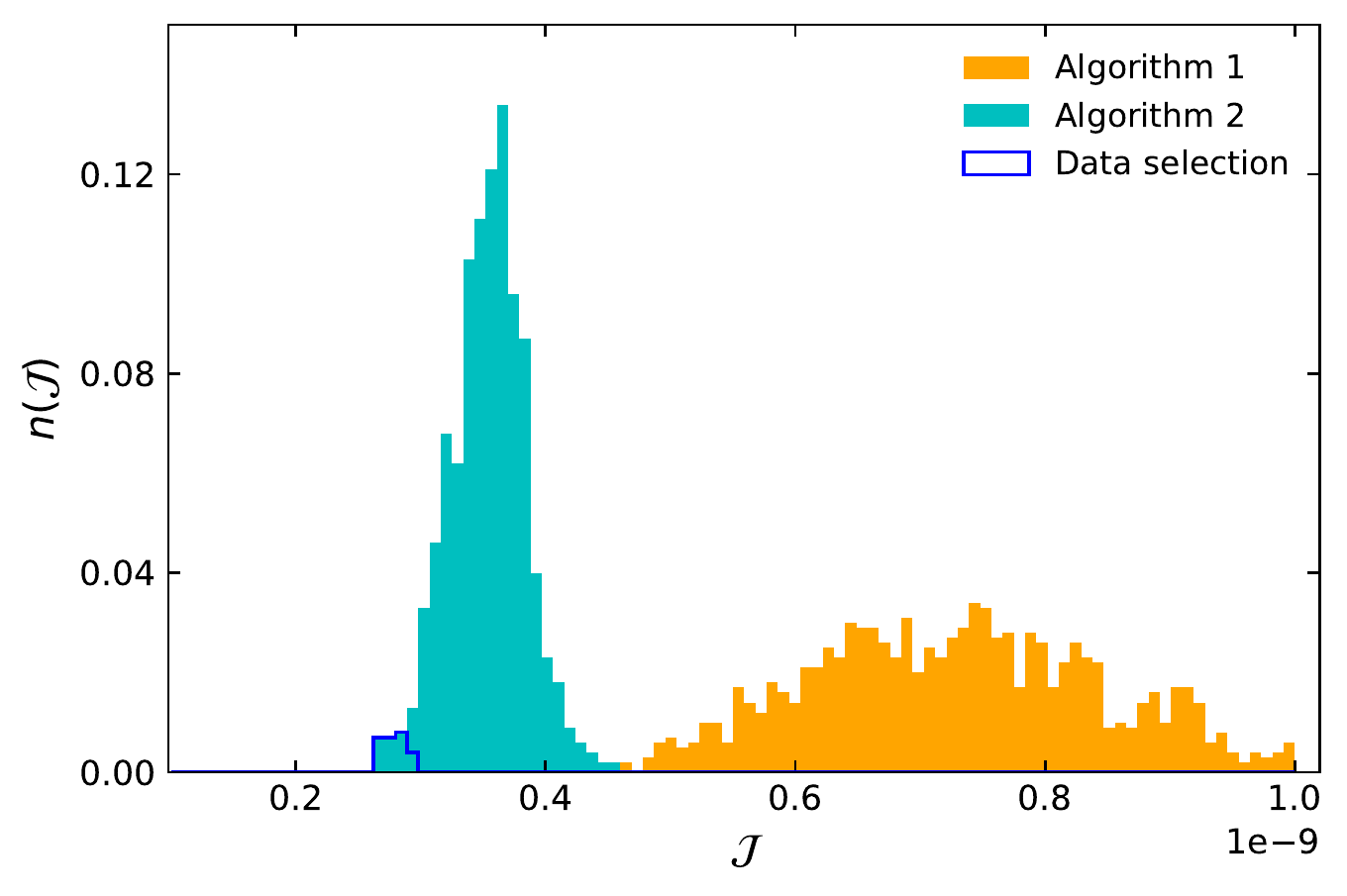}
    \caption{The $\mathcal{J}$ distribution of Algorithm 1 (orange) and Algorithm 2 (cyan) for the case without catastrophic photo-$z$ errors.
    The input power spectrum is $\ell C_{\ell}^{gg,P}$.
    Algorithm 2 further reduces $\mathcal{J}$.
    The blue box area means the solutions selected via equation~(\ref{define:J_selection}) and the final result is the average over these selected outputs.
    }
    \label{fig:sim_normal_l1_J}
\end{figure}

\begin{figure}
	\includegraphics[width=\columnwidth]{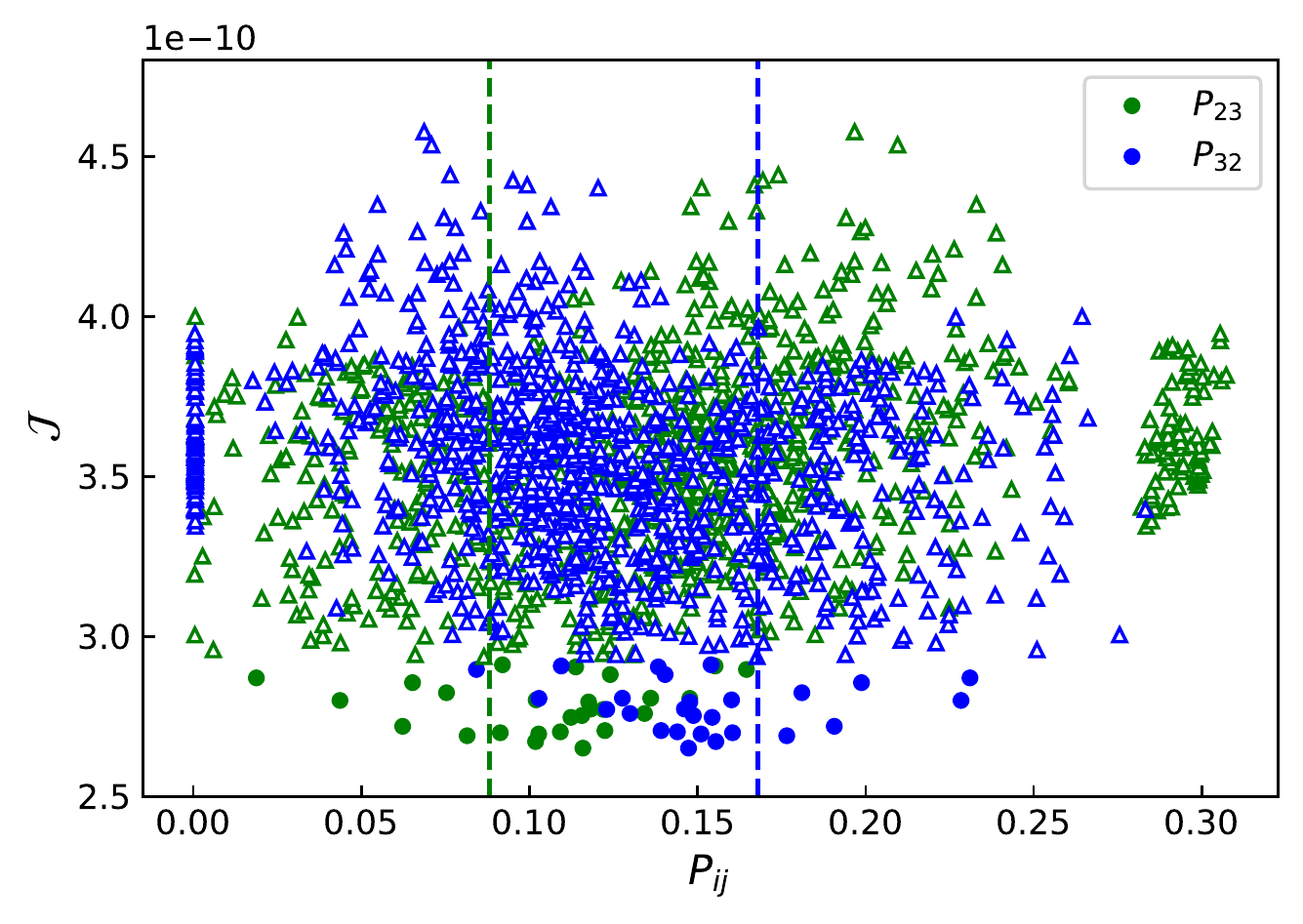}
    \caption{The 1000 output results of $P_{23}$ (green) and $P_{32}$ (blue) from Algorithm 2. The circles are the selected results via equation~(\ref{define:J_selection}) and the triangles are the discarded data. The dashed lines indicate the truth values we preset.
    }
    \label{fig:sim_normal_l1_J_pij}
\end{figure}

\begin{figure*}
\centering
\setcounter {subfigure} {0}
\subfigure[average value: $\langle P \rangle$]{
    \includegraphics[width=\columnwidth]{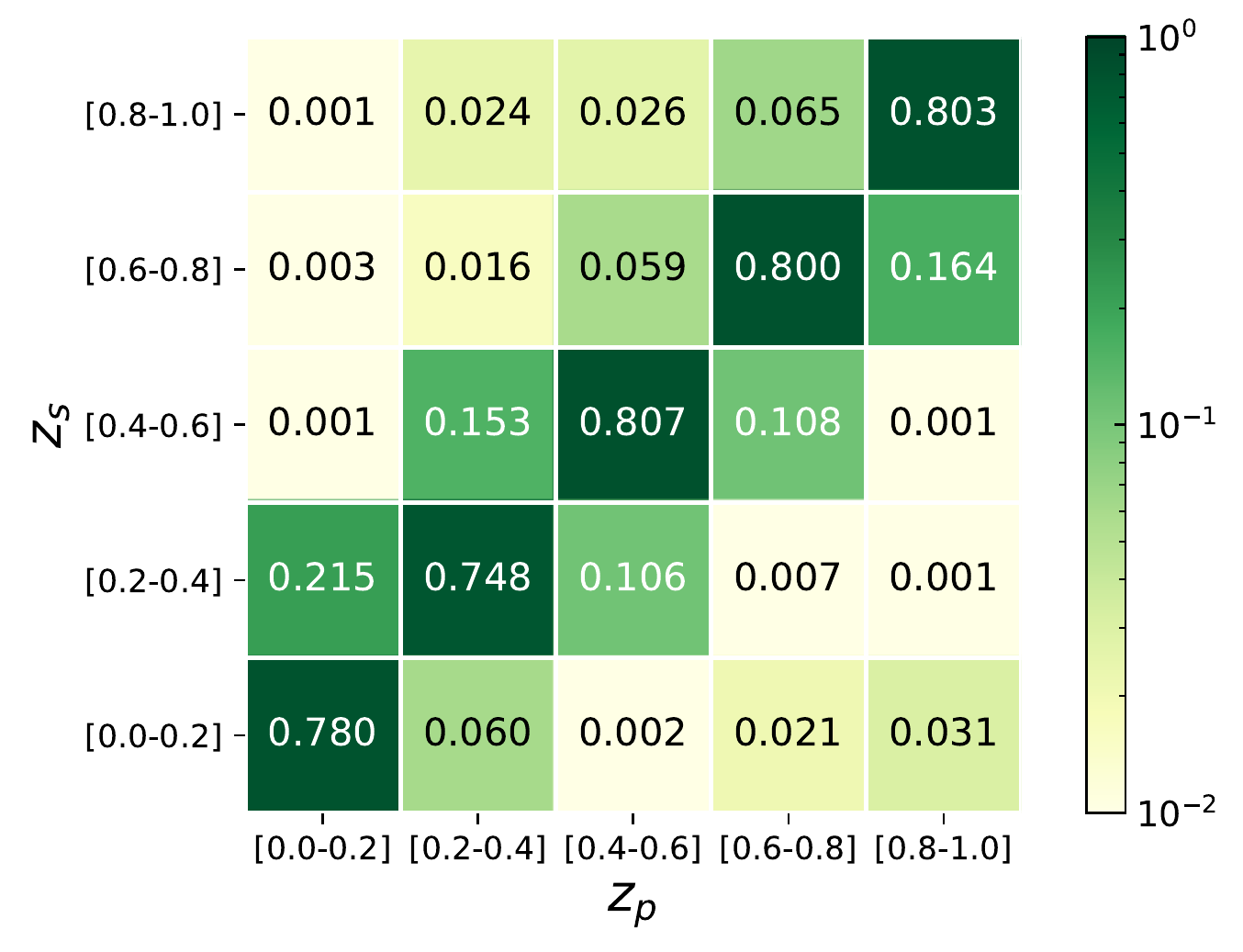}}
%\quad
\centering
\setcounter {subfigure} {3}
\subfigure[reconstructed diagonal $C_{\ell}^{gg,P}$ and $C_{\ell,{\rm obs}}^{gg,P}$]{\raisebox{0.3cm}{
    \includegraphics[width=\columnwidth]{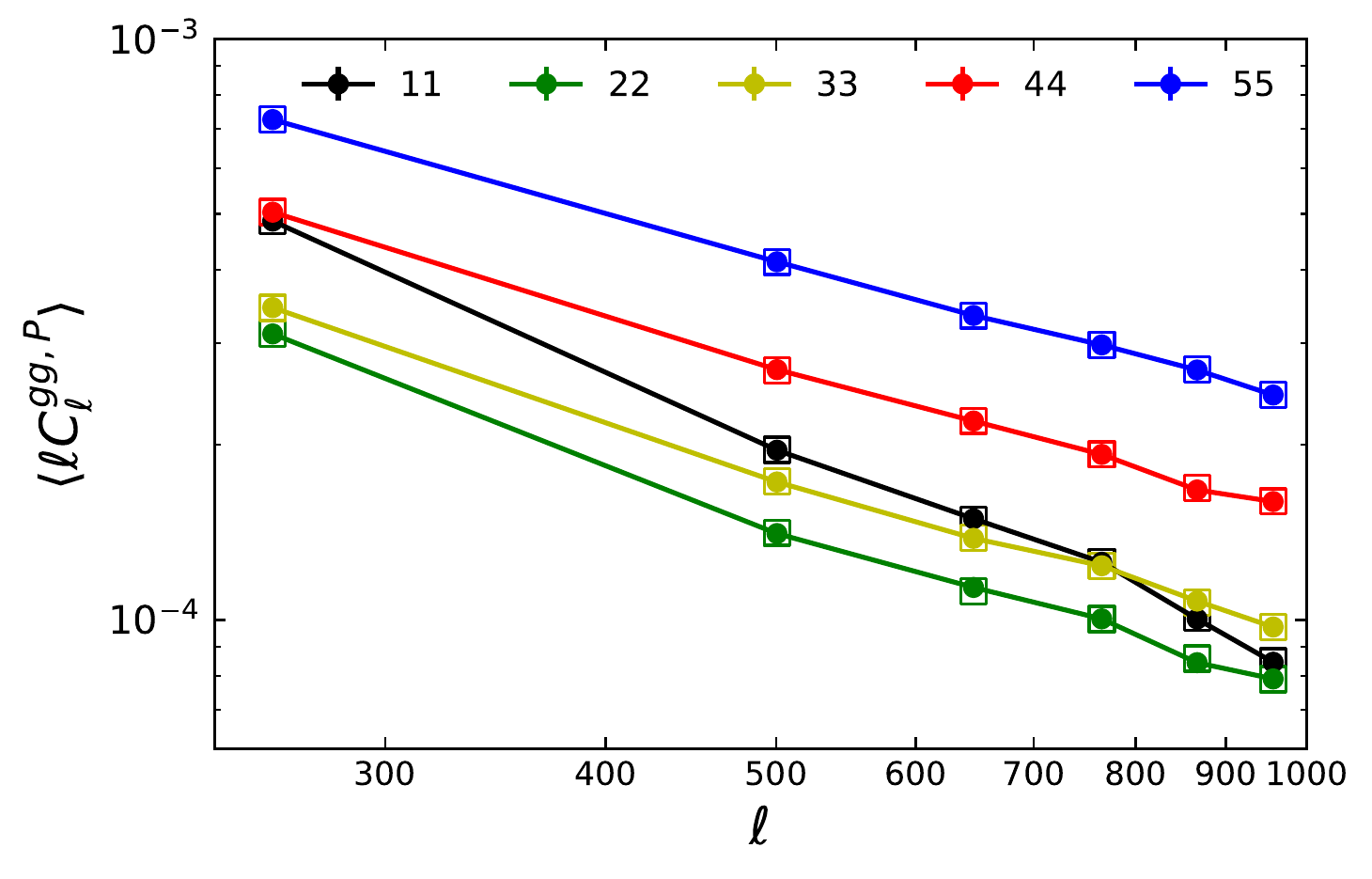}}
}
\centering
\setcounter {subfigure} {1}
\subfigure[bias: $\langle P \rangle-P^{\rm true}$]{
    \includegraphics[width=\columnwidth]{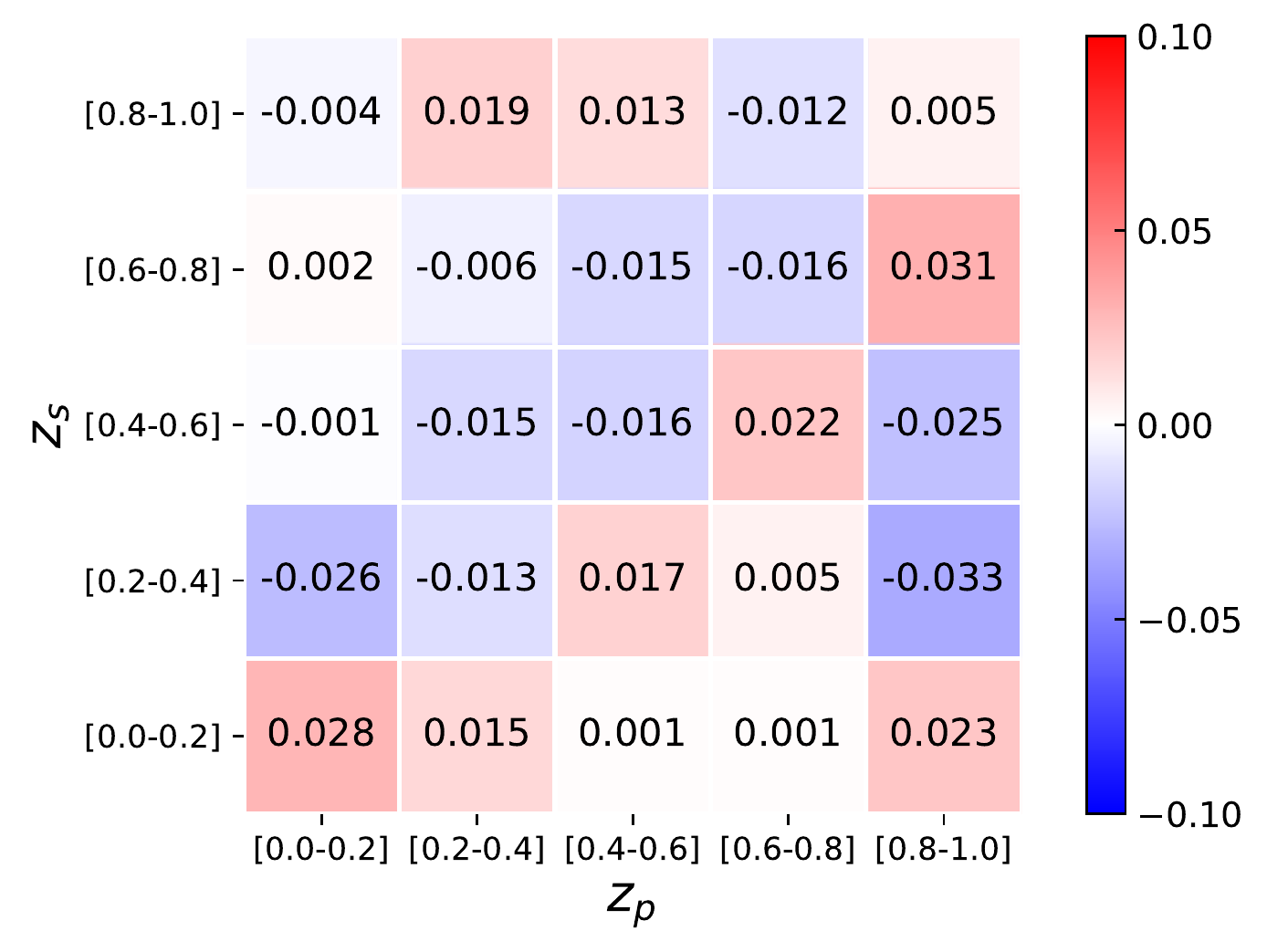}
}
%\quad
\centering
\setcounter {subfigure} {4}
\subfigure[reconstructed sub-diagonal $C_{\ell}^{gg,P}$ and $C_{\ell,{\rm obs}}^{gg,P}$]{\raisebox{0.3cm}{
    \includegraphics[width=\columnwidth]{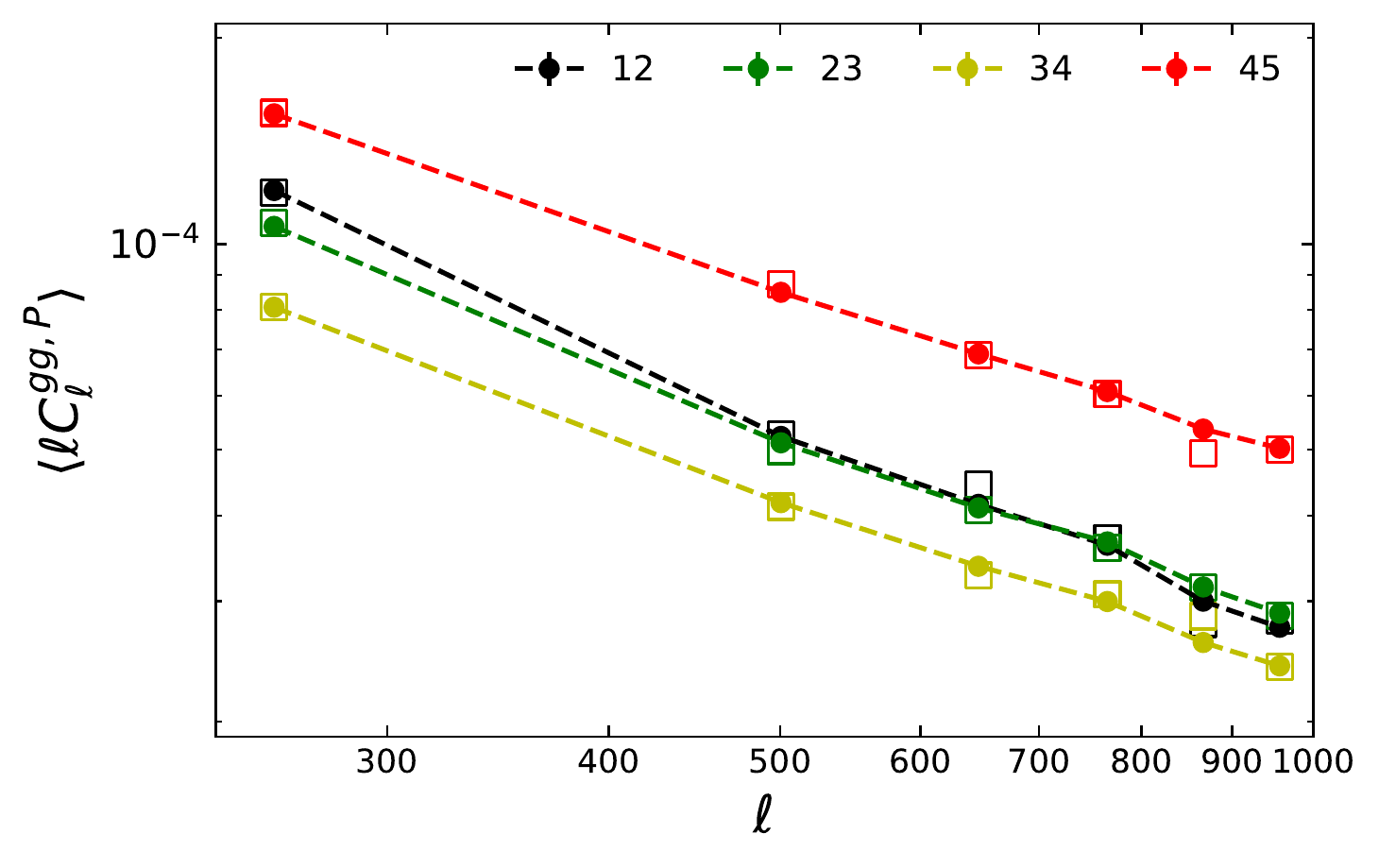}}
}
%\quad
\centering
\setcounter {subfigure} {2}
\subfigure[standard deviation: $\sigma_P$]{
    \includegraphics[width=\columnwidth]{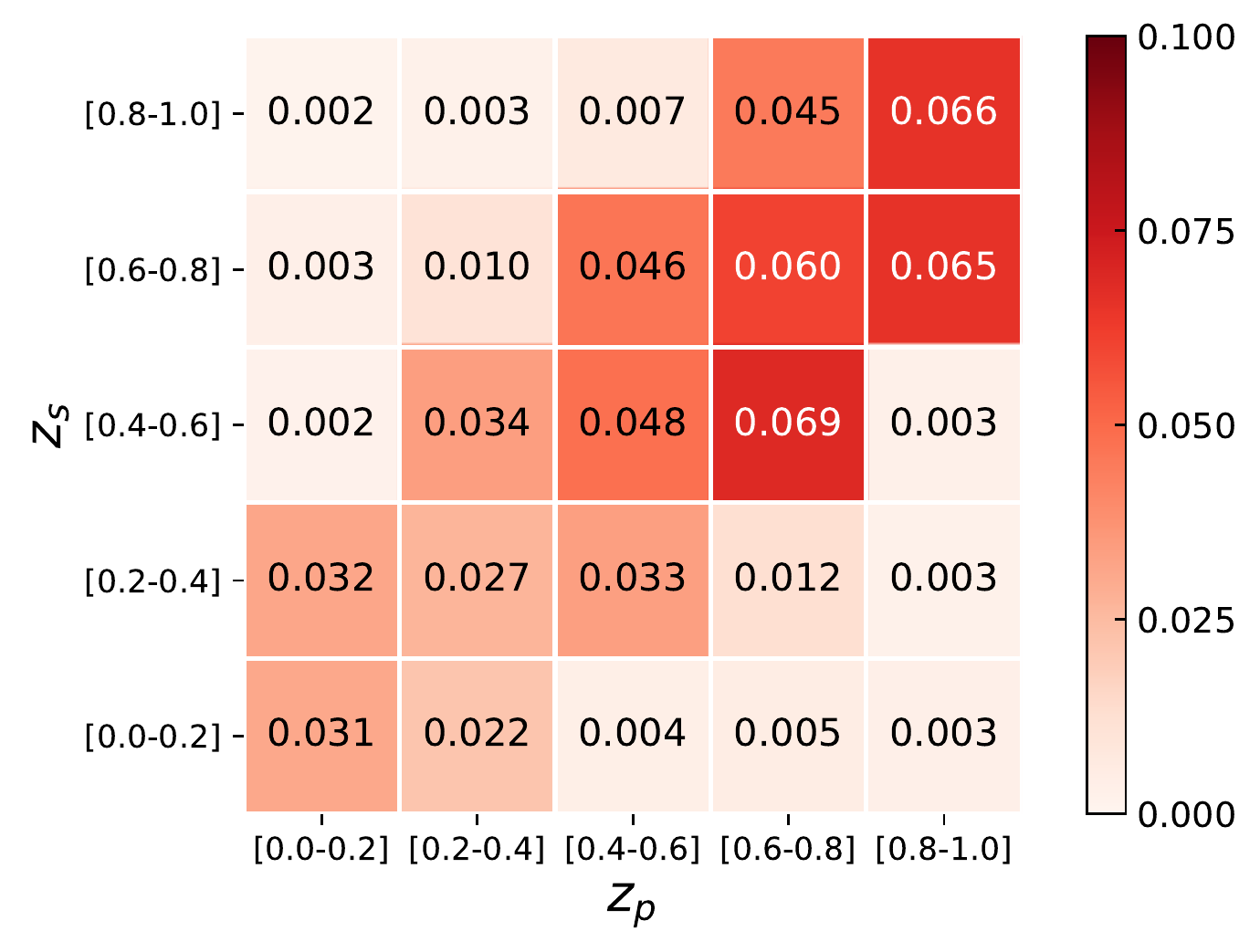}
}
%\quad
\centering
\setcounter {subfigure} {5}
\subfigure[reconstructed $C_{\ell}^{gg,R}$ and $C_{\ell,{\rm true}}^{gg,R}$]{\raisebox{0.3cm}{
    \includegraphics[width=\columnwidth]{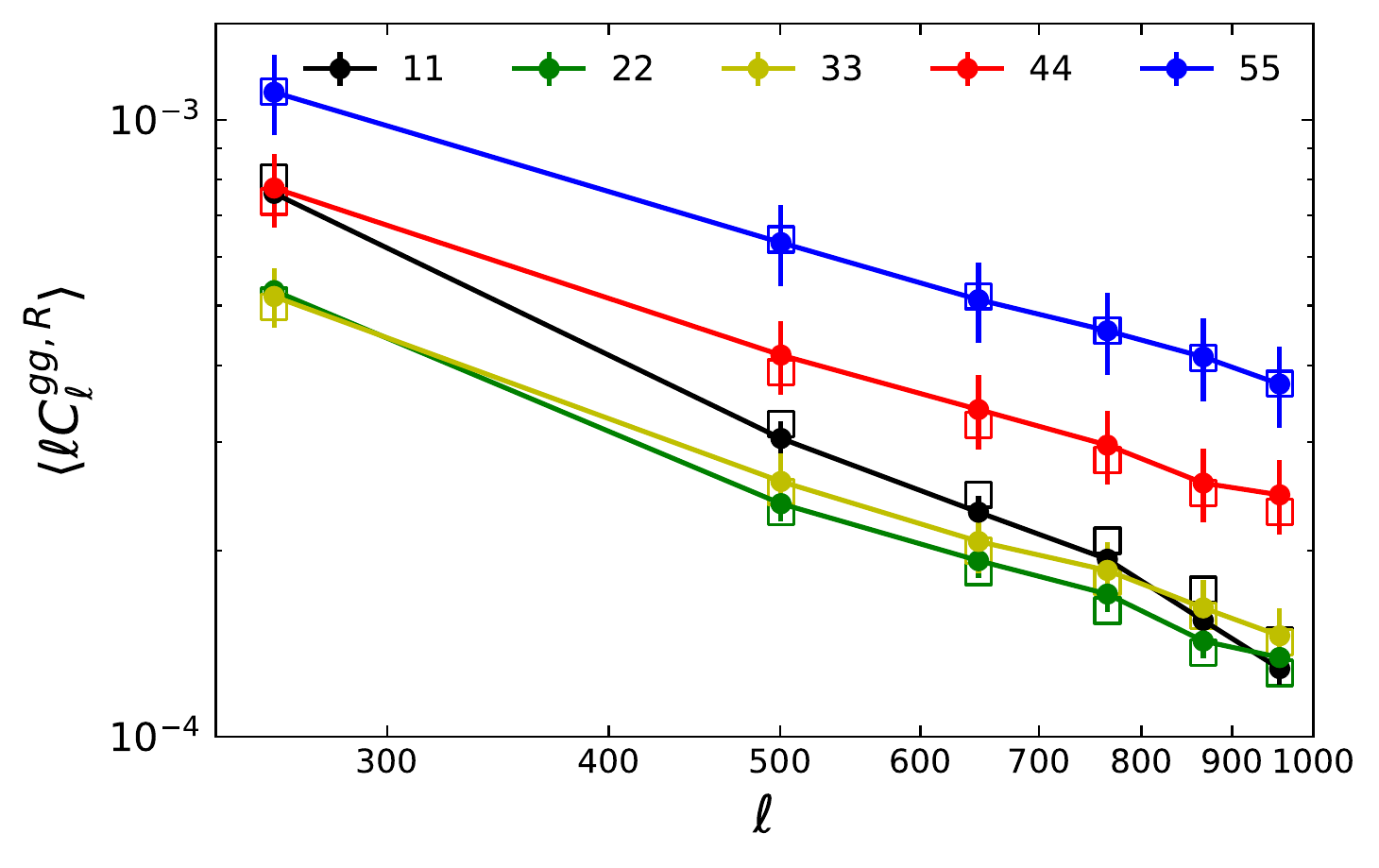}}
}
\caption{The results from the self-calibration method are presented by the reconstructed scattering matrix $P$ [panel (a)], the bias from $P^{\mathrm{true}}$ [panel (b)], the uncertainty in the recovered $P$ [panel (c)], the reconstructed power spectrum for photo-$z$ bins [panels (d) and (e)] and for true-$z$ bins [panel (f)].
The final results are the average over the 26 selected solutions with criterion equation~(\ref{define:J_selection}).
The lines with error bars in panels (d), (e) and (f) are the results from the self-calibration algorithm, which are compared to the direct measurements from the simulation ($\mathbf {squares}$).
For the power spectrum between photo-$z$ bins, we only show the diagonal [solid lines in panel (d)] and sub-diagonal [dashed lines in panel (e)] elements.
The number pair $ij$ in the legend of panels (d) and (e) means the power spectrum between the $i$th and $j$th photo-$z$ bin, and the number $ii$ in the legend of panel (f) means the galaxy auto power spectrum in the $i$th true-$z$ bin.
To show the biases and uncertainties of the reconstructed $P$ and $C_{\ell}^{gg,R}$ in a more clear way, the same information is plotted in another form in the panel (b) of Fig.~\ref{fig:sim_normal_differ_l_result}.}
\label{fig:sim_normal_result}
\end{figure*}

\begin{figure}
\centering
    \includegraphics[width=0.9\columnwidth]{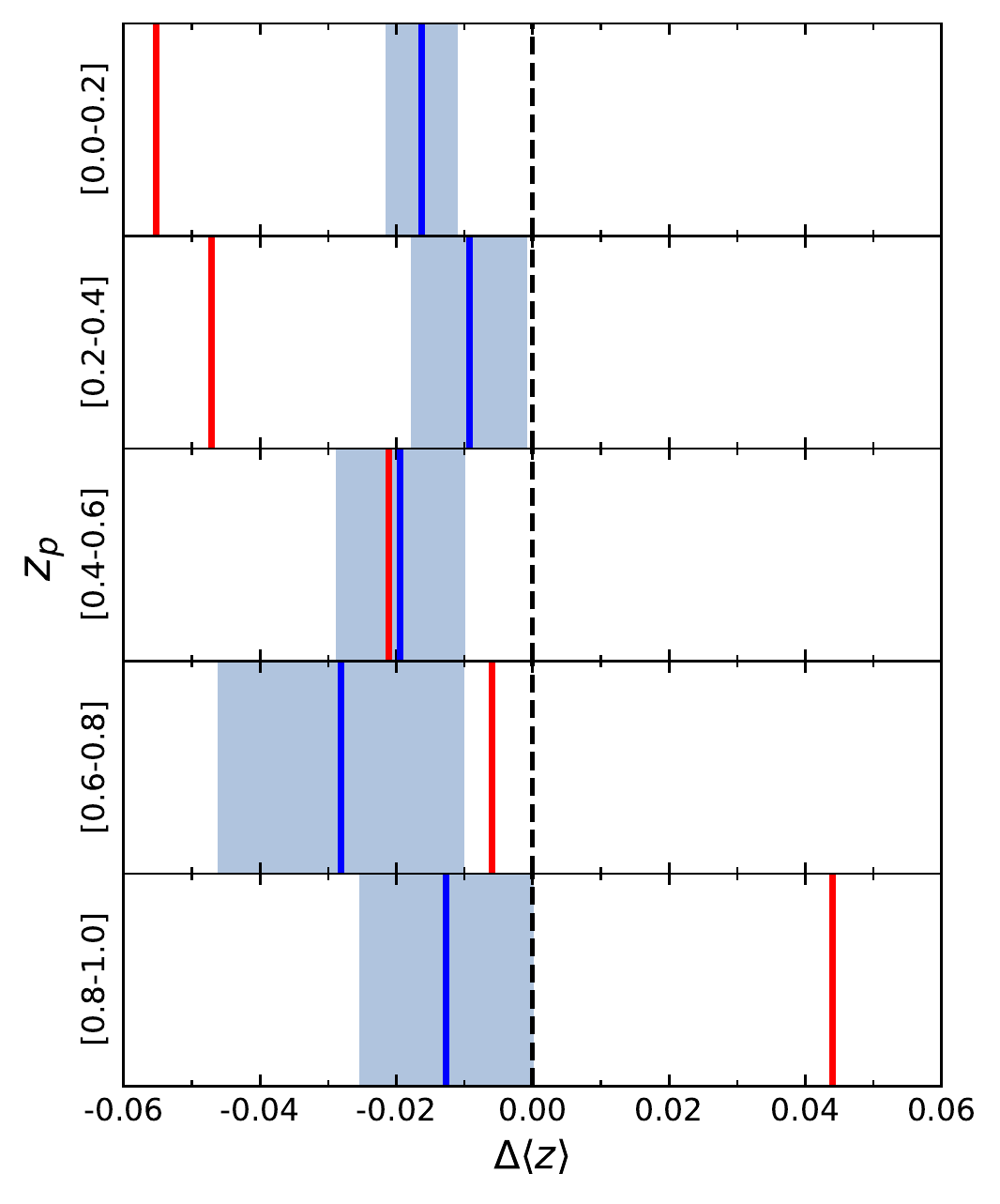}
    \caption{The improvement of the mean redshifts from the self-calibration method.
    The red vertical line represents the bias $\Delta\langle z\rangle$ between the original average of the photo-$z$ and the mean of true-$z$ for the photometric galaxies in each given photo-$z$ bins.
    With the self-calibration technique, the mean redshift for a given photo-$z$ bin can be estimated from equation~(\ref{eqn:estimate}).
    These new mean redshifts plotted in blue vertical lines have smaller biases, except for the 4th photo-$z$ bin.
    The lightsteelblue regions indicate the standard deviations over the the selected solutions.
    As a reference, the vertical black dashed line indicates the position of the mean true-$z$ for each given photometric bin.
    }
\label{fig:sim_normal_delta_z}
\end{figure}

\begin{figure*}
\centering
\subfigure[Results from inputting $C_{\ell}^{gg,P}$]{
    \raisebox{1cm}{\includegraphics[width=\columnwidth]{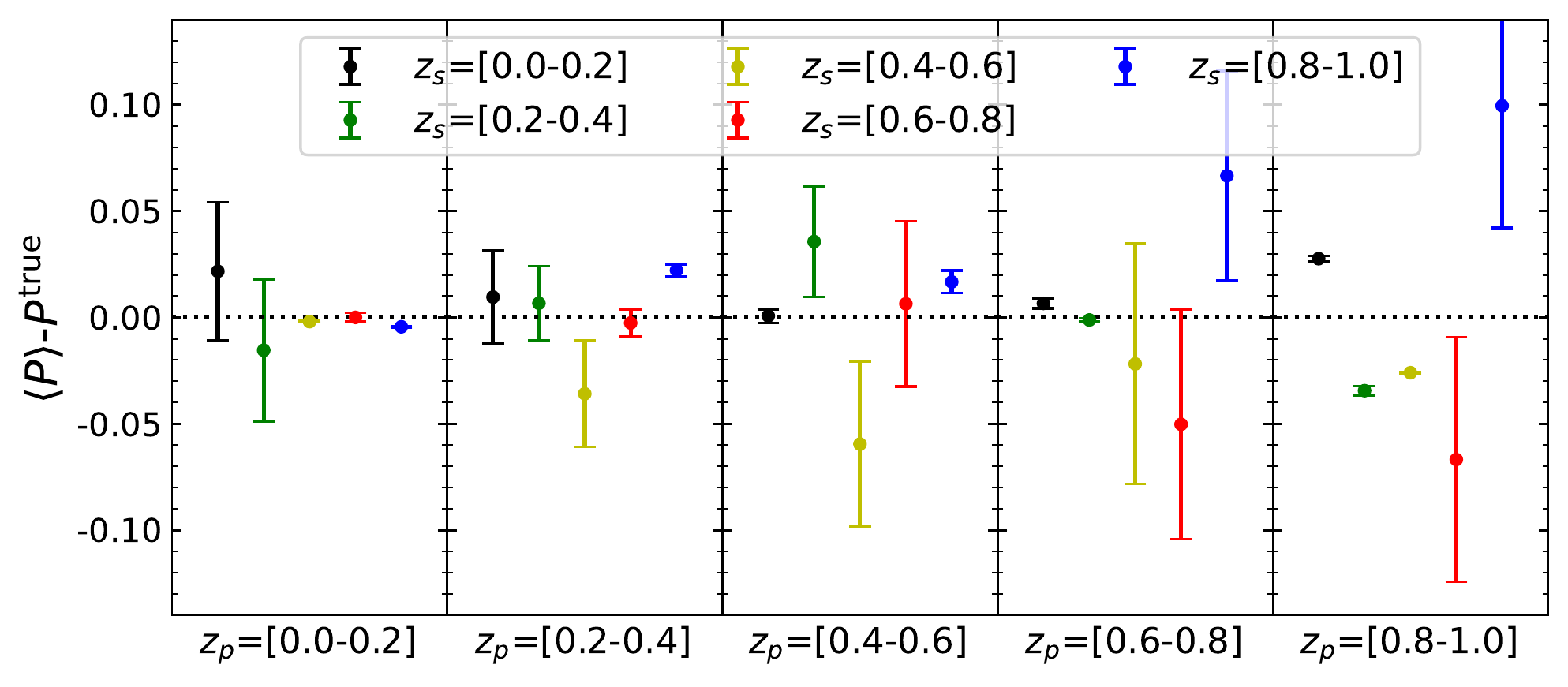}}
    \includegraphics[width=\columnwidth]{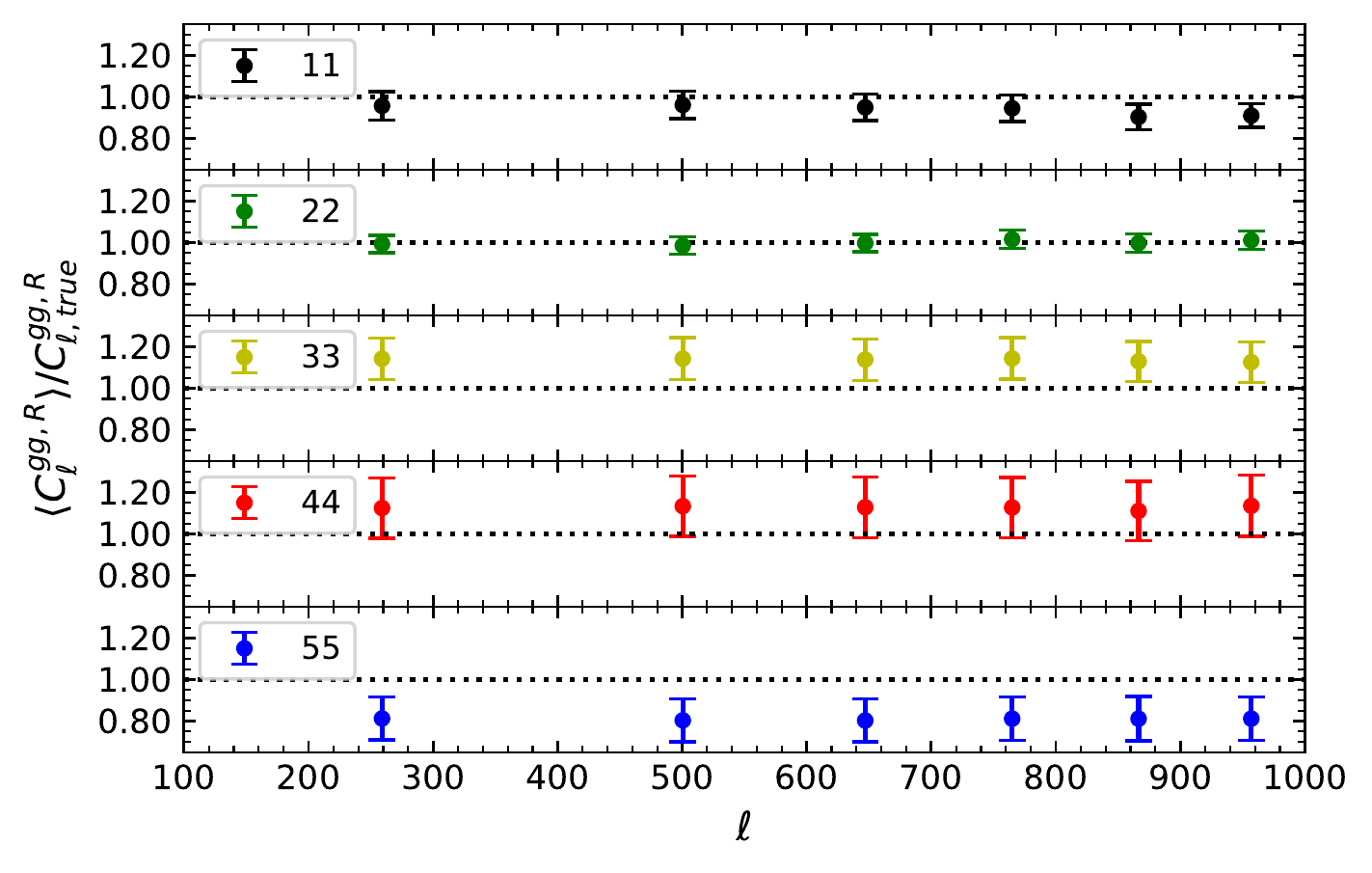}
}
\quad
\centering
\subfigure[Results from inputting $\ell C_{\ell}^{gg,P}$]{
    \raisebox{1cm}{\includegraphics[width=\columnwidth]{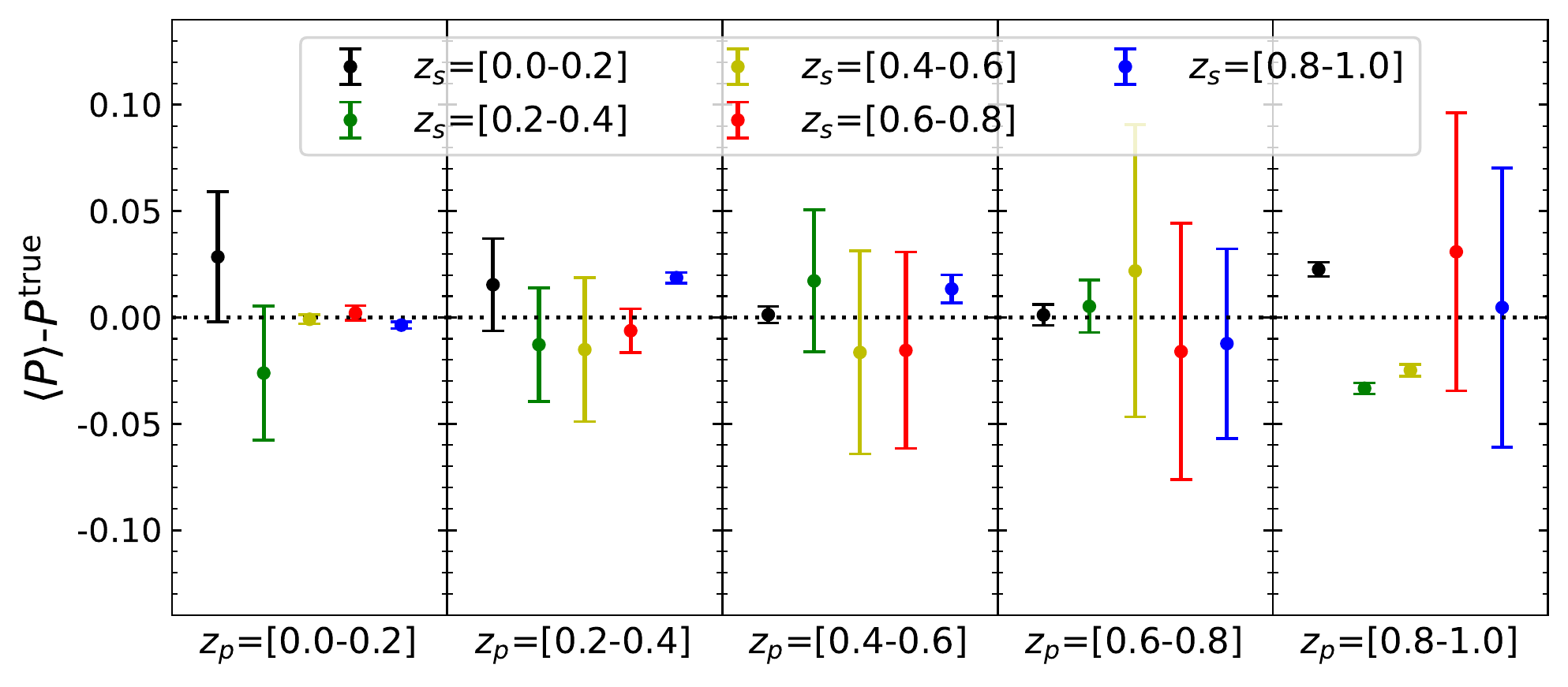}}
    \includegraphics[width=\columnwidth]{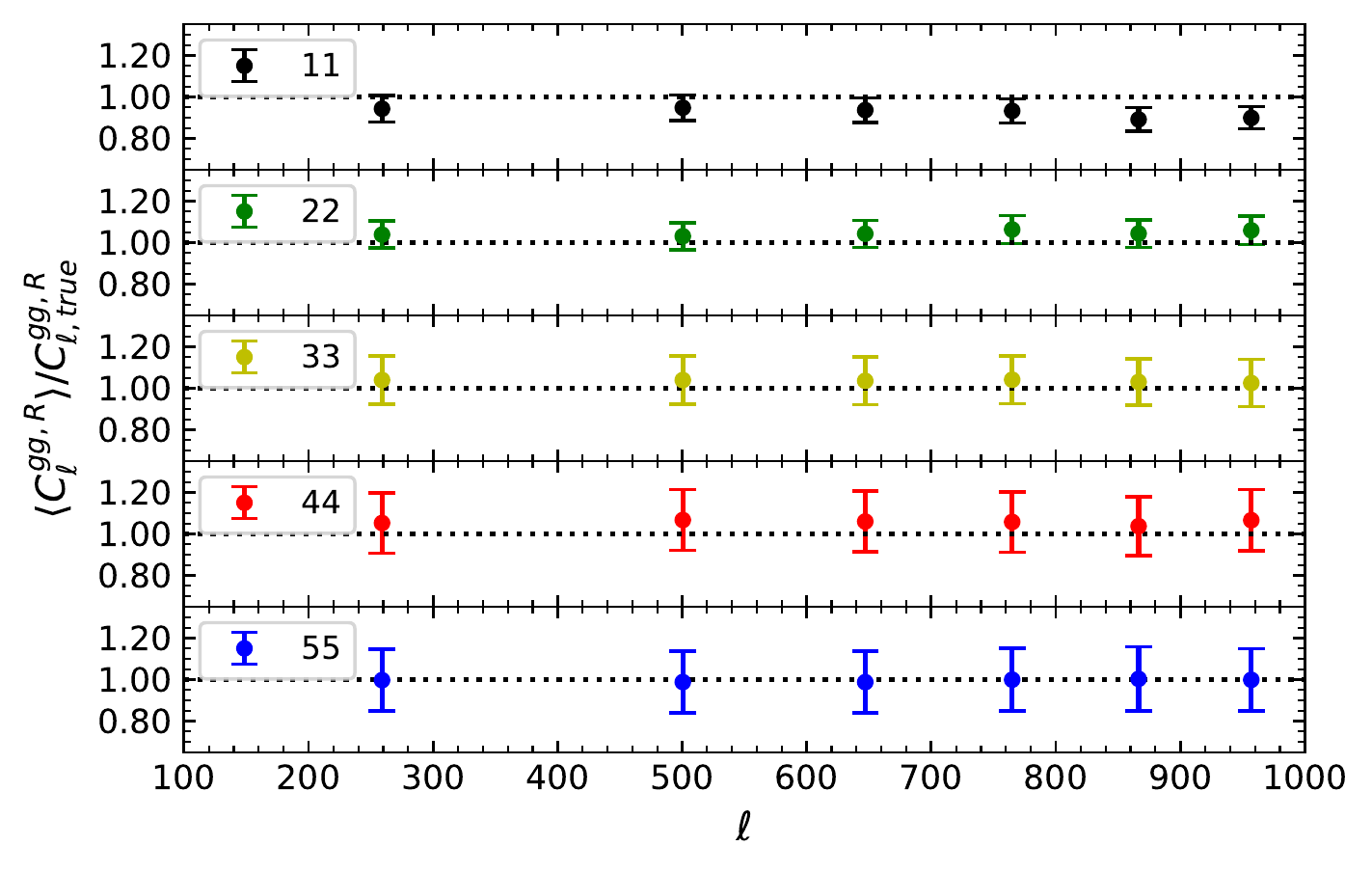}
}
\quad
\centering
\subfigure[Results from inputting $\ell^2 C_{\ell}^{gg,P}$]{
    \raisebox{1cm}{\includegraphics[width=\columnwidth]{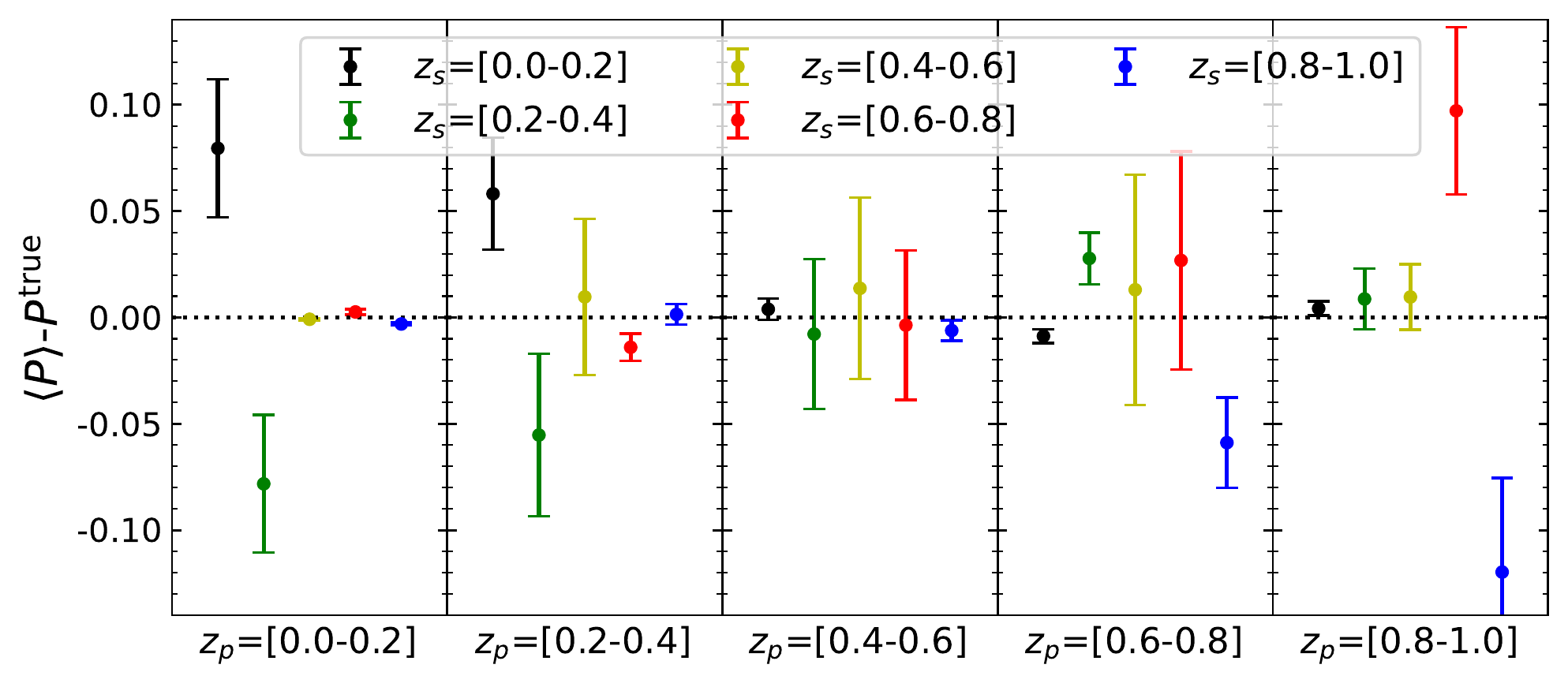}}
    \includegraphics[width=\columnwidth]{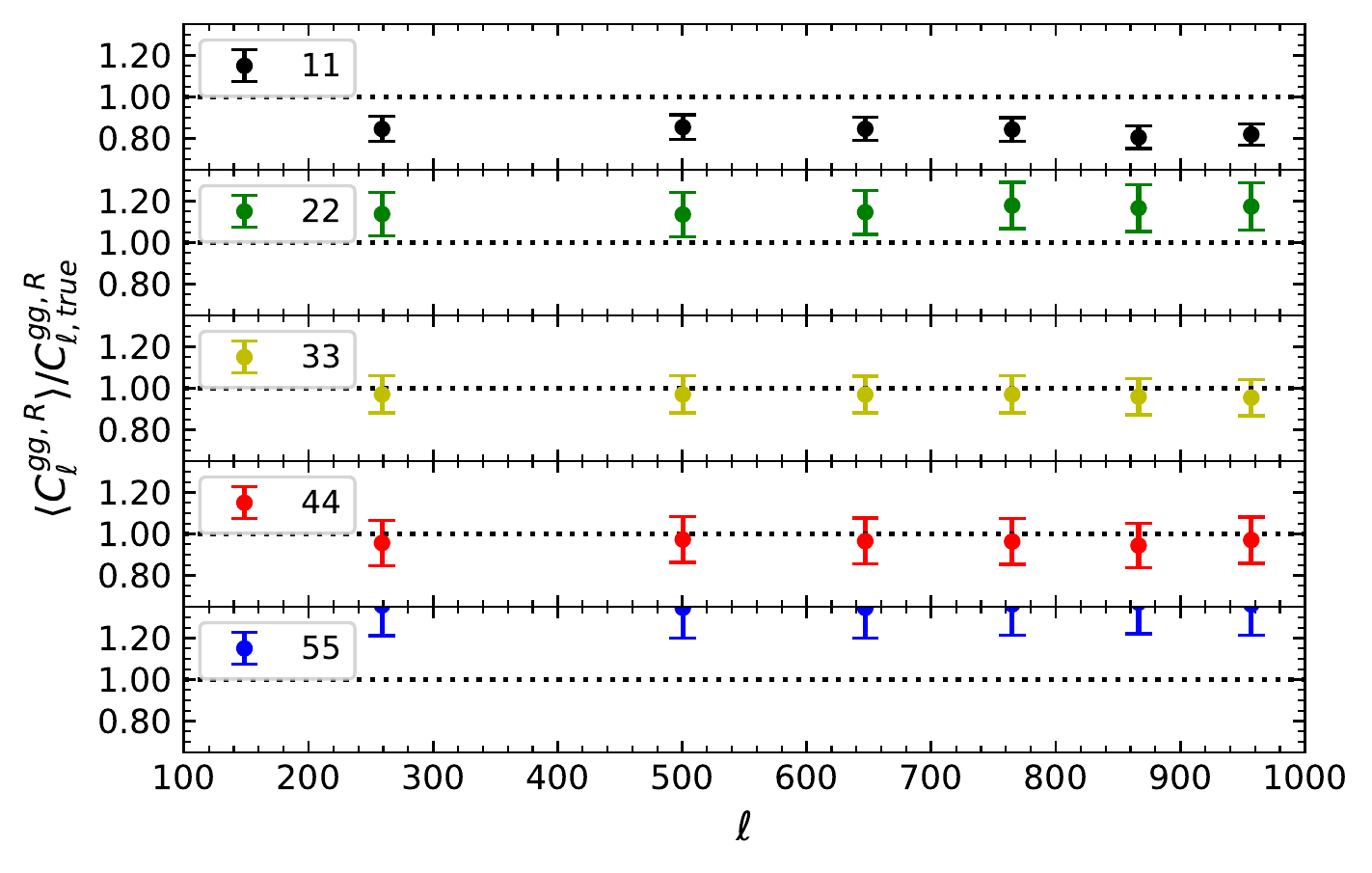}
}
\caption{Comparison of the results when inputting $C_{\ell}^{gg,P}$, $\ell C_{\ell}^{gg,P}$ and $\ell^2 C_{\ell}^{gg,P}$ into the algorithm for the case without catastrophic photo-$z$ errors.
The left panels show the bias and standard deviation of each element in the reconstructed scattering matrix.
The right panels show the performance of the reconstructed $C_{\ell}^{gg,R}$
(divided by $C_{\ell,{\rm true}}^{gg,R}$, the power spectrum between true-$z$ bins measured from simulation).
The legend $ii$ in the right panels indicates the $i$th true-$z$ bin.}
\label{fig:sim_normal_differ_l_result}
\end{figure*}

As shown in Section \ref{sec:obs}, the DECaLS DR8 galaxy sample's photo-$z$ data doesn't show significant outliers or catastrophic error and the scatter is basically concentrated near the diagonal elements in scattering matrix.
Fig.~\ref{fig:sim_normal_pture} shows the matrix $P^{\rm true}$ we preset for the simulation in the case without significant outliers, i.e. $P_{ij} < 0.10$ for $|i-j|\geq 2$.
%, which actually came from a simple test of the DECaLS DR8 data by an earlier imperfect algorithm.

In ideal case, i.e. the observed data can be fully described by equation~(\ref{eqn:Cggp}) with sufficiently low shot noise, the result with minimum $\mathcal{J}$ represents the globally optimal estimate, i.e. the true answer, as long as there is no intrinsic degeneracy in the system\footnote{For example, if the power spectrum of two redshift bins have the exactly same shape over scales, the system is degenerate.}.
However, it is more complicated when applied to mock or survey data with non-negligible noise.
First, in the existence of shot noise, the solution with minimum $\mathcal{J}$ usually does not correspond to the true answer.
Therefore, we think the result of the global minima on the data with large noise has the overfitting problem.
Second, the noise in the measurement flattens the surface of $\mathcal{J}$ function, reducing the efficiency of the iterative algorithm.
Third, the large noise may form multiple local minima around the true answer, reducing the accuracy of the result.
In this situation, initiated from different initial guess, the obviously different solutions of $P$ and $C_{\ell}^{gg,R}$ can have similar small $\mathcal{J}$ values.
It is reasonable to assume that these local minima with sufficiently low $\mathcal{J}$ values distribute around the true answer.

Fig.~\ref{fig:sim_normal_l1_J} shows the distribution of $\mathcal{J}$ when we input the measured $\ell C_{\ell}^{gg,P}$ into the algorithm.
Iterated from different initial guess, the $\mathcal{J}$'s from Algorithm 1 have a quite broad distribution.
It is very clear that Algorithm 2 further reduces $\mathcal{J}$ based on the $P$ estimated by Algorithm 1.
Meanwhile, the $\mathcal{J}$ distribution is more compact from Algorithm 2.
Based on the assumption that the solutions with small $\mathcal{J}$ values are randomly distributed around the true answer, it is straightforward and convenient to average the results over an appropriate range.
we choose the results via
\begin{equation}
    \frac{\mathcal{J}-\mathcal{J}_{\min}}{\mathcal{J}_{\min}}\leq 10\%\ .
    \label{define:J_selection}
\end{equation}
Here, $\mathcal{J}_{\min}$ is smallest $\mathcal{J}$ from Algorithm 2.
In this fiducial analysis, the best 26 solutions are selected and the averaged matrix $\langle P \rangle$ and $\langle \ell C_{\ell}^{gg,R}\rangle$ are regarded as the final results.

In Fig.~\ref{fig:sim_normal_l1_J_pij}, we exam the above assumption about the solution distribution,
and present the output results of $P_{23}$ and $P_{32}$ from Algorithm 2.
According to the distribution of the reconstructed results around the truth values, we find that the above assumption is reasonable and the selection criterion can help us approach to the right answers.
Meanwhile, there is an obvious degeneracy between $P_{23}$ and $P_{32}$, because $C_{i j}^{\mathrm{gg}, \mathrm{P}}(\ell)(i \neq j)$ measures the combination $P_{ji} P_{jj} C_{jj}^{\mathrm{gg}, \mathrm{R}}(\ell)+P_{ii} P_{ij} C_{ii}^{\mathrm{gg}, \mathrm{R}}(\ell)$.
It is not easy to break this degeneracy with galaxy-galaxy correlations alone (see \citet{Zhang:2010wr} for more details).
We notice that there may exist some ways such as the particle swarm optimisation to help us find the global minima.  But the main consideration here is that we think the result of the global minima on the data with large noise has the overfitting problem, so it could be a better and convenient choice to average a set of values via our selection criterion.

Fig.~\ref{fig:sim_normal_result} shows the performance of the self-calibration, including the resulting scattering matrix $\langle P \rangle$ [panel (a)], the bias $\langle P \rangle-P^{\rm true}$ [panel (b)], the uncertainty $\sigma_P$ [i.e. the standard deviation of the selected $P$'s, panel (c)], the reconstructed power spectrum $\langle \ell C_{\ell}^{gg,P}\rangle$ for photo-$z$ bins [panels (d) and (e)], and the reconstructed power spectrum $\langle\ell  C_{\ell}^{gg,R}\rangle$ for true-$z$ bins [panel (f)].
For the power spectrum plots, the reconstruction results are presented in lines along with the error bars showing the r.m.s over the selected solutions.
As the reference, the power spectrum $\ell C_{\ell,{\rm obs}}^{gg,P}$ after the photometric scattering and the power spectrum $\ell C_{\ell,{\rm true}}^{gg,R}$ before the photometric scattering are shown in dots.
The former is the input for the algorithm, and the later is the ground truth for the reconstructed power spectrum for true-$z$ bins.

From panel (b), we find that the bias is small, the mean of the absolute bias over all elements is smaller than $0.015$, and $|\langle P \rangle-P^{\rm true}|<0.03$ for diagonal and sub-diagonal elements.
The diagonal and sub-diagonal elements have relatively large biases for the reason that their true values are much larger than the others.
However, compared to the uncertainties shown in panel (c), this bias is of no significance.
The uncertainties for elements involving the highest two redshift bins are large.
This result is consistent with the observed large uncertainties in the reconstructed $ C_{\ell}^{gg,R}$ in panel (f).
We argue that these large uncertainties come from the relatively large degeneracy for these two high-redshift bins.
The power spectrum shapes of the two bins are quite similar in both panel (d) and (f).

In panels (d) and (e), we compare the reconstructed $C_{\ell}^{gg,P}$ (the lines with error bars) with the input measurement $C_{\ell,{\rm obs}}^{gg,P}$ (the squares).
The reconstructed powers are in good agreement with the inputs.
% This result is consistent with the small $\mathcal{J}$ values observed in Fig.~\ref{fig:sim_normal_l1_J}, implying that the solution is reliable.
Another finding is that the error bars are very small, indicating that all the selected solutions can recover the observation equally well.

Contrary to the small error bars in the reconstructed $\langle C_{\ell}^{gg,P}\rangle$,
the reconstructed $\langle C_{\ell}^{gg,R}\rangle$ shows relatively large bias and uncertainty in panel (f).
The deviation from $C_{\ell,{\rm true}}^{gg,R}$ is 4.4 per cent when averaged over all the 30 data points (6 $\ell$-bins and 5 $z$-bins).
The relatively large uncertainties in the reconstructed matrix $P$ and $C_{\ell}^{gg,R}$ illustrate that the performance is limited by the degeneracy from the noisy measurement.

The power spectrum measurement of a given photo-$z$ bin should be interpreted as the clustering signal at the \denweight mean true-$z$ of these galaxies.
However, in photometric survey this mean true-$z$ of a photo-$z$ bin is unavailable.
Using the mean of the photo-$z$ from the galaxies generally induces bias in the cosmological constraints, due to the biased redshift estimation shown as the red vertical lines in Fig.~\ref{fig:sim_normal_delta_z}.
%As the redshift increases, the bias changes from underestimation to overestimation.
%This is expected from the asymmetry in number of up-scattering and down-scattering between redshift bins when the number of galaxies peaks at $z\sim 0.5$.
%As a result, the fourth redshift bin happens to have the smallest bias.
With the above scattering matrix reconstructed by the self-calibration method,
we can reduce the bias in the mean redshift estimation.
As the mean true-$z$ of the true-$z$ bin is also unknown in observation,
\citet{Zhang:2010wr} proposed to approximate the \denweight mean true-$z$ for a given photo-$z$ bin $i$ by
\begin{equation}
      \langle z_{i}\rangle \simeq \sum_{j} P_{j i}\langle z_{j}^{P}\rangle=\sum_{j} P_{j i} \frac{\int_{j} z^{P} n\left(z^{P}\right) d z^{P}}{\int_{j} n\left(z^{P}\right) d z^{P}}\ .
      \label{eqn:estimate}
\end{equation}
Here, $\langle z_{j}^{P}\rangle$ is the \denweight mean photo-$z$ of the $j$th photo-$z$ bin and it is assumed to be approximately equal to the \denweight mean true-$z$ of the $j$th true-$z$ bin, $\langle z_{j}^{S}\rangle$.
In Fig.~\ref{fig:sim_normal_delta_z}, we also show the bias of the estimated redshift obtained by equation~(\ref{eqn:estimate}).
We find, even under the approximation, the mean redshift deviation from the truth value has been successfully reduced in most bins compared to the original \denweight mean photo-$z$, $\langle z^{P}\rangle$. 
The mean absolute bias of the estimated mean redshift is reduced by more than 50 per cent from 0.035 to 0.017.
% \sout{In Table \ref{tab:delta_z_result}, we show the same exam using the $P^\mathrm{true}$ and $\langle z^{S}\rangle$ which are only available in simulation.
% The comparisons there indicate that the reconstructed scattering matrix from the self-calibration is pretty good in reducing the mean redshift bias, and the residual bias shown in Fig.~\ref{fig:sim_normal_delta_z} is mainly contributed by the above approximation.}

To illustrate the rationality and importance of the weighting,
in Fig.~\ref{fig:sim_normal_differ_l_result} we compare the results from inputting $C_{\ell}^{gg,P}$, $\ell C_{\ell}^{gg,P}$ and $\ell^2 C_{\ell}^{gg,P}$ into the algorithm.
The left panels show the bias and uncertainty of each element in the reconstructed $P$, while the right panels show the results for the reconstructed power spectrum $C_{\ell}^{gg,R}$.
The results are also summarized in Table~\ref{tab:sim_normal_differ_l_result}, including the average absolute bias over all the elements in $P$ and over all the measurements in $C_{\ell}^{gg,R}$.
The weighting $\ell C_{\ell}^{gg,P}$ outperforms the other two.
Therefore, we believe that it is a reasonable and necessary choice to use $\ell C_{\ell}^{gg,P}$ as the input to the algorithm, and this weighting scheme will be adopted in the following analysis both on simulations and DECaLS DR8 observation.

\begin{table}
	\centering
	\caption{The mean absolute bias over all elements in the scattering matrix $P_{ij}$ and the mean absolute deviation for the reconstructed power spectrum $C_{\ell}^{gg,R}$ over all redshift bins and scales.
	The result is for the case without catastrophic photo-$z$ errors.
	Weighting the input power spectrum by $\ell$ outperforms the other two weighting schemes.
	}
	\label{tab:sim_normal_differ_l_result}
	\setlength{\tabcolsep}{5mm}{
	\begin{tabular}{ccc}
		\hline
		 Input & $|\langle P_{ij} \rangle-P_{ij}^{\rm true}|$ & $|\langle C_{\ell}^{gg,R}\rangle/C_{\ell,{\rm true}}^{gg,R}-1|$\\
		\hline
		$C_{\ell}^{gg,P}$ & 0.026 & 10.5\%\\
		$\ell C_{\ell}^{gg,P}$ & 0.015 & 4.4\%\\
		$\ell^2 C_{\ell}^{gg,P}$ & 0.029 & 15.0\%\\
		\hline
	\end{tabular}}
\end{table}

\subsection{Validation with catastrophic photo-$z$ errors}
\label{result:cata}

\begin{figure*}
\centering
\subfigure[$P^{\rm true}$]{
    \includegraphics[width=\columnwidth]{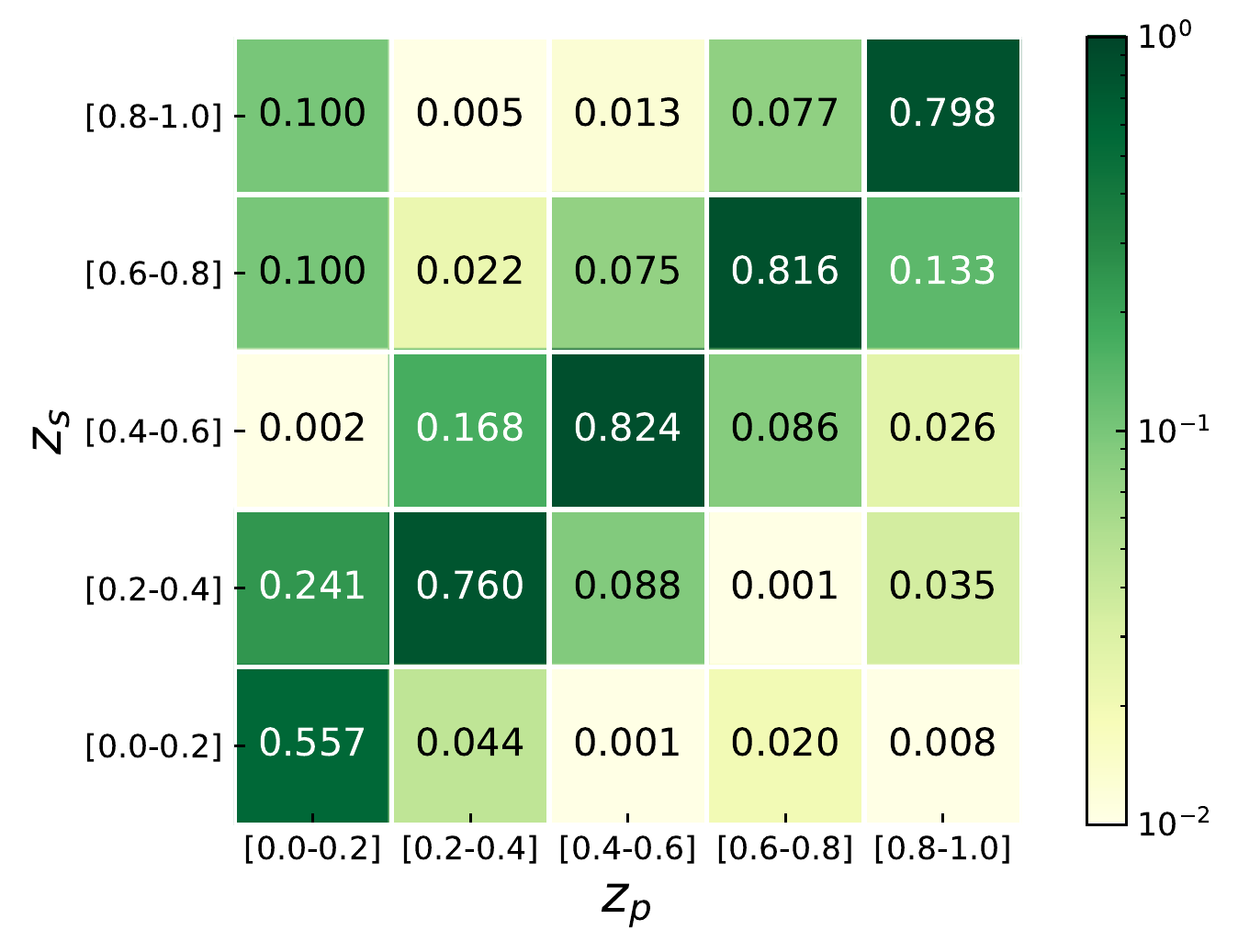}
}
\quad
\centering
\subfigure[$\mathcal{J}$ distribution]{
    \raisebox{0.3cm}{\includegraphics[width=\columnwidth]{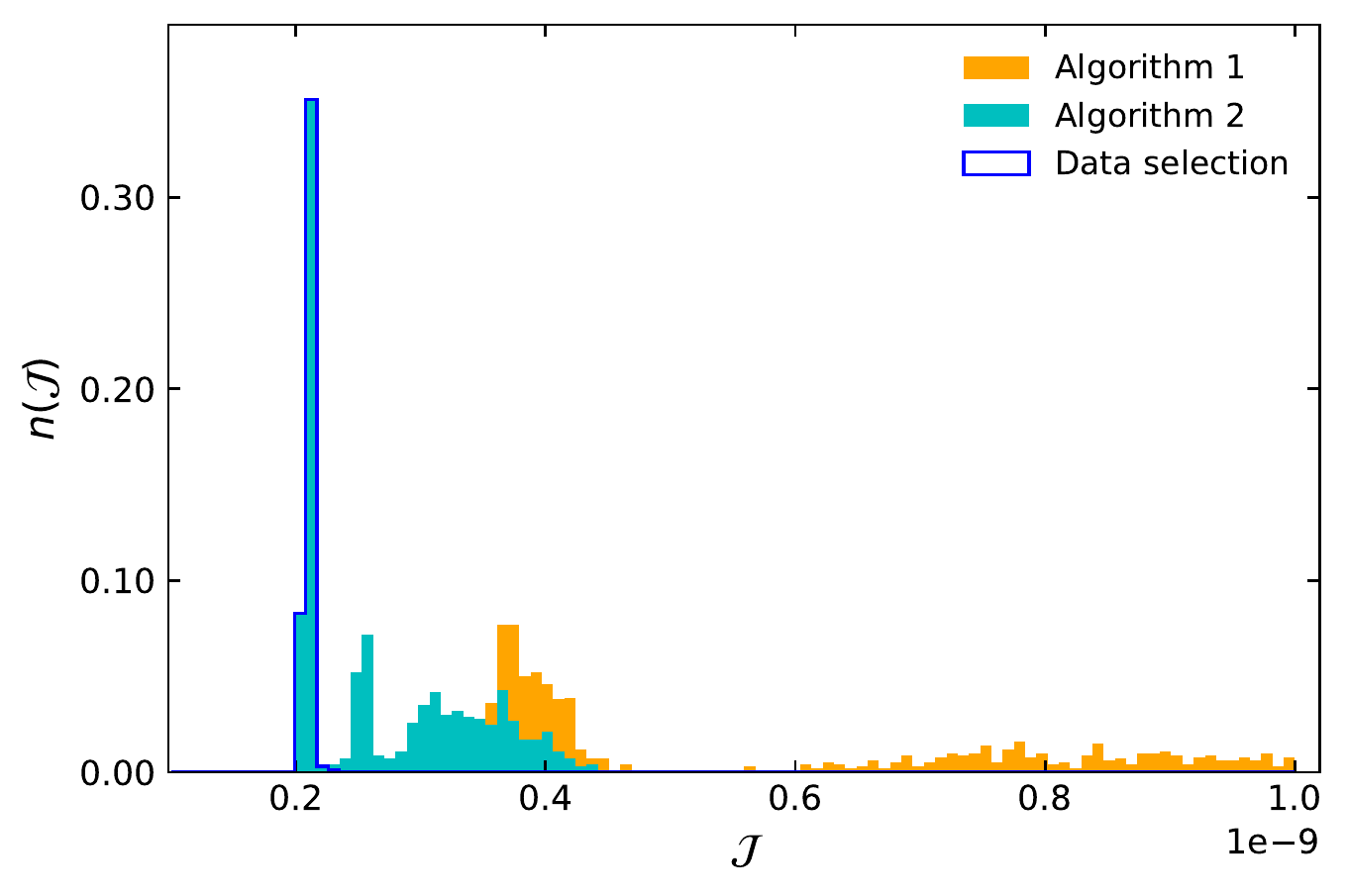}}
}
\quad
\centering
\subfigure[$\langle P \rangle-P^{\rm true}$]{
    \raisebox{1cm}{\includegraphics[width=\columnwidth]{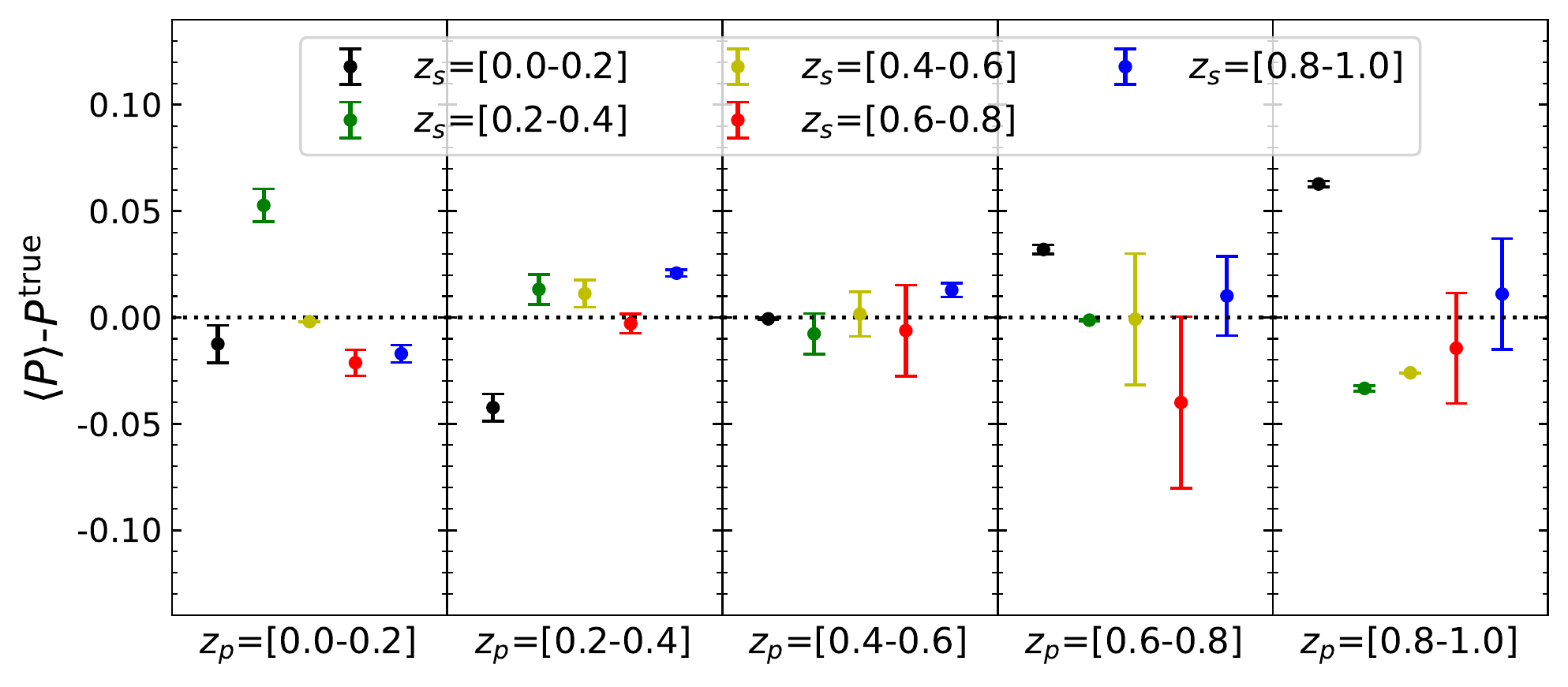}}
}
\quad
\centering
\subfigure[$\langle C_{\ell}^{gg,R} \rangle /C_{\ell,{\rm true}}^{gg,R}$]{
    \includegraphics[width=\columnwidth]{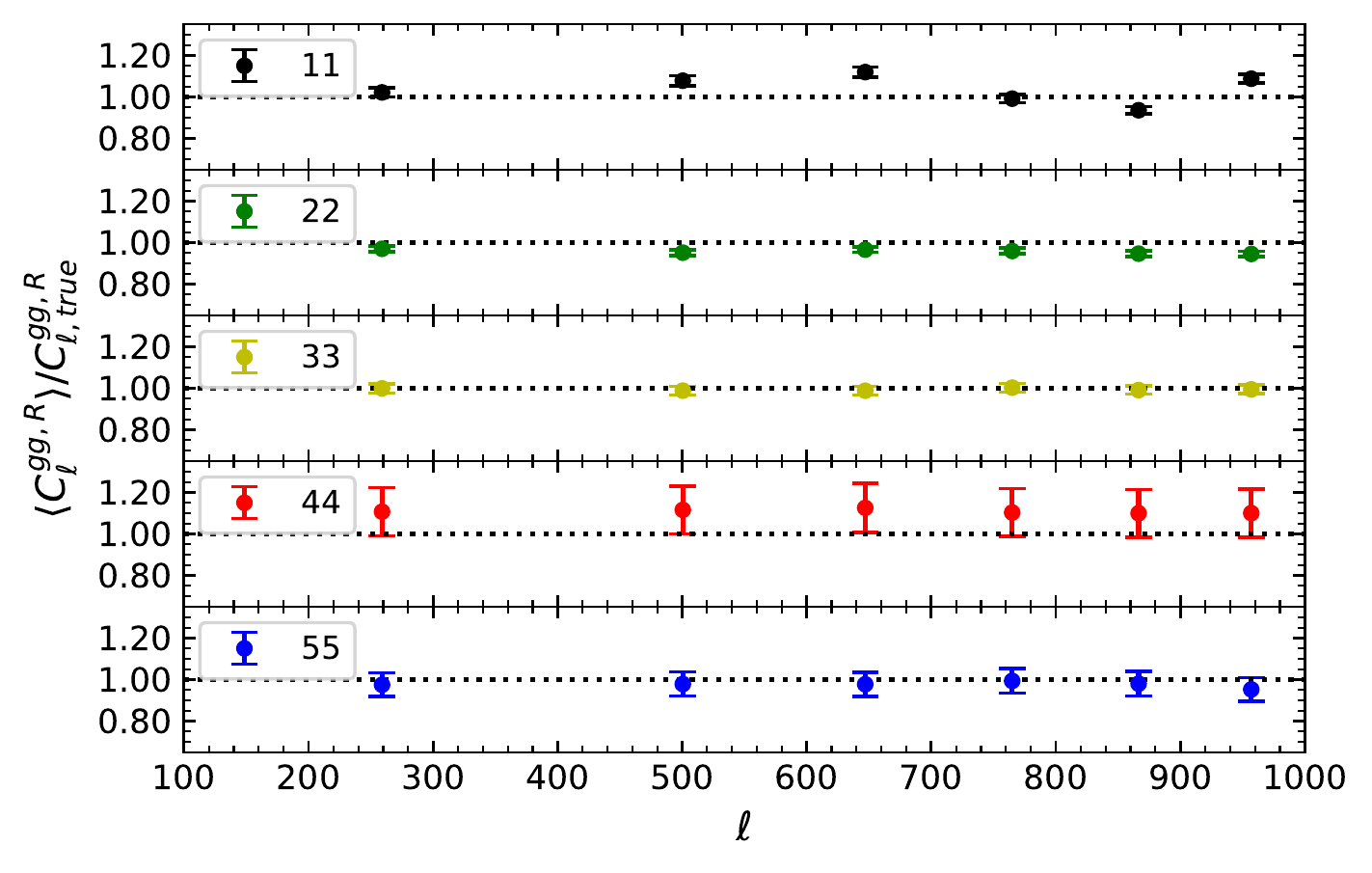}
}
\caption{The $P^{\rm true}$, the $\mathcal{J}$ distribution and the performance of the self-calibration are presented for the case with catastrophic error.
Note that the two elements in the upper left corner are set to $0.1$ in panel (a), indicating the existence of catastrophic photo-$z$ errors.
The self-calibration algorithm passes the test.
}
\label{fig:sim_cata1_result}
\end{figure*}

\begin{figure*}
\centering
\subfigure[$P^{\rm true}$]{
    \includegraphics[width=\columnwidth]{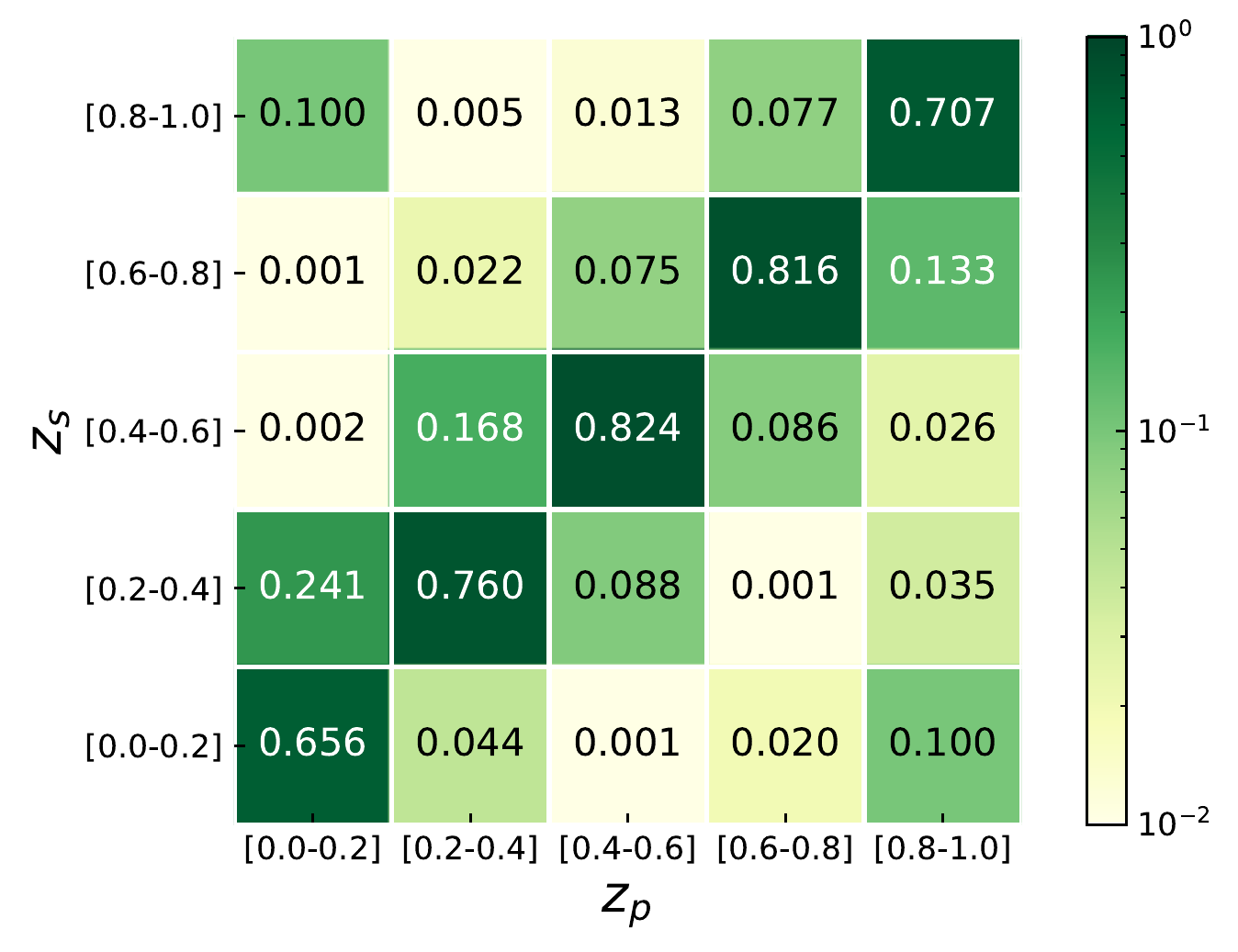}
}
\quad
\centering
\subfigure[$\mathcal{J}$ distribution]{
    \raisebox{0.3cm}{\includegraphics[width=\columnwidth]{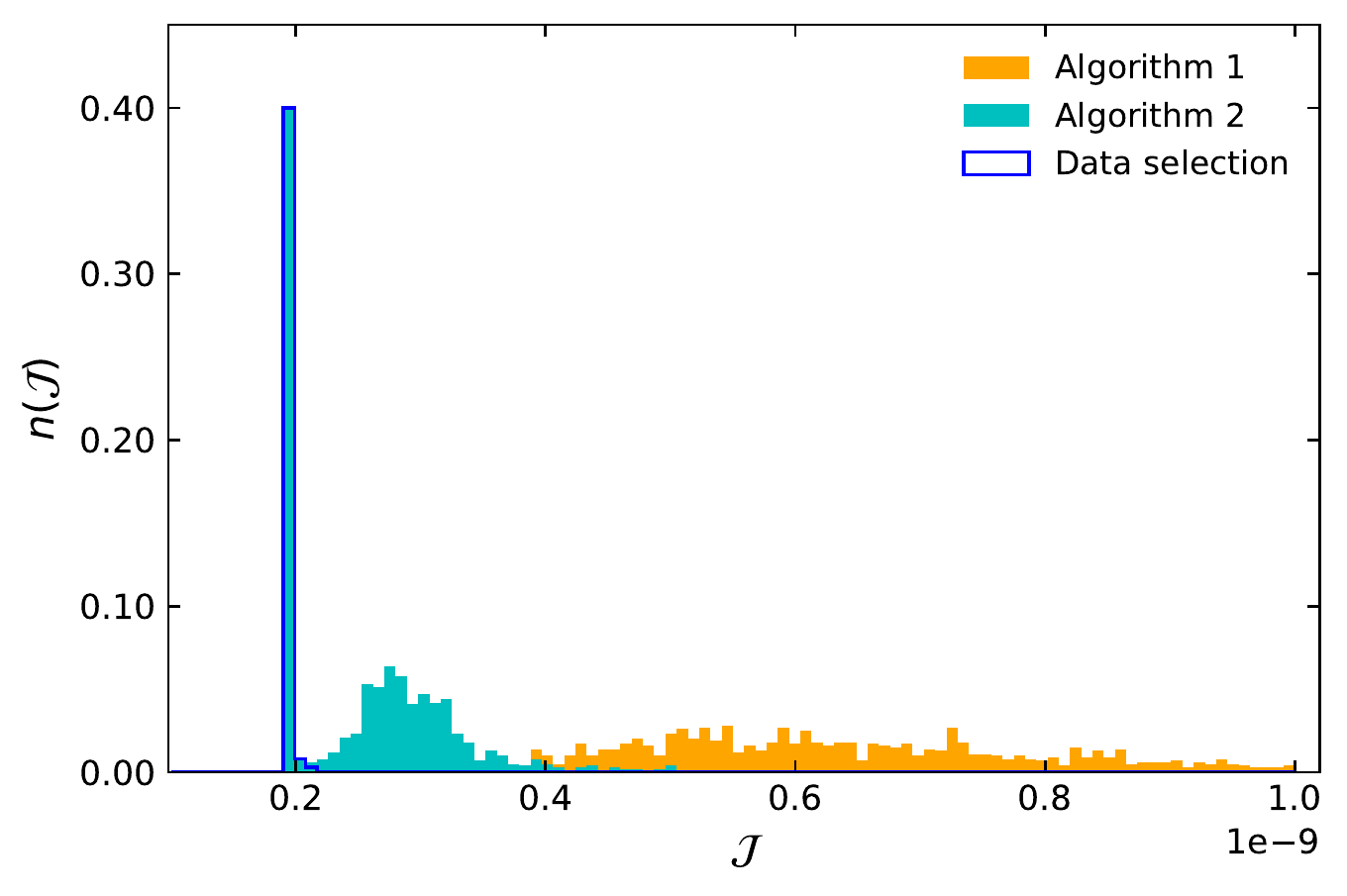}}
}
\quad
\centering
\subfigure[$\langle P \rangle-P^{\rm true}$]{
    \raisebox{1cm}{\includegraphics[width=\columnwidth]{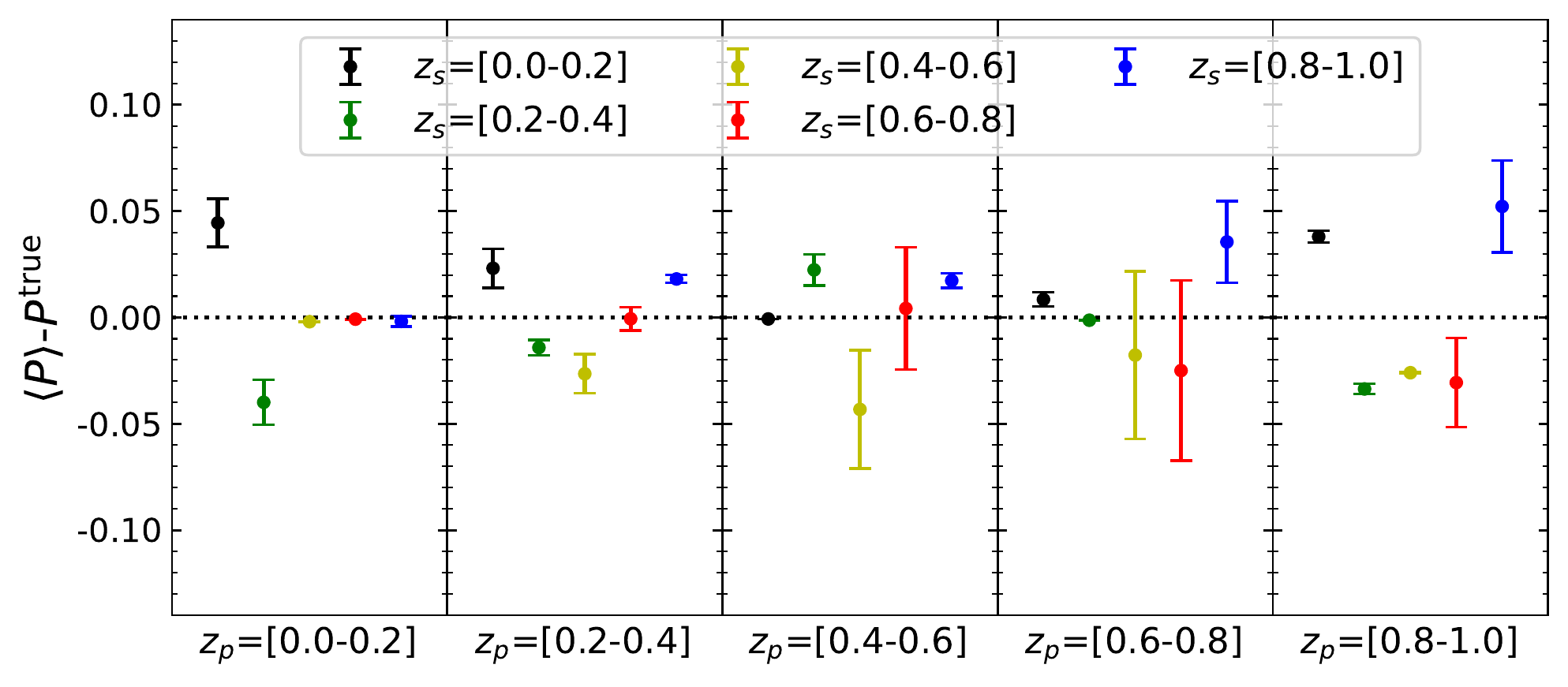}}
}
\quad
\centering
\subfigure[$\langle C_{\ell}^{gg,R} \rangle/C_{\ell,{\rm true}}^{gg,R}$]{
    \includegraphics[width=\columnwidth]{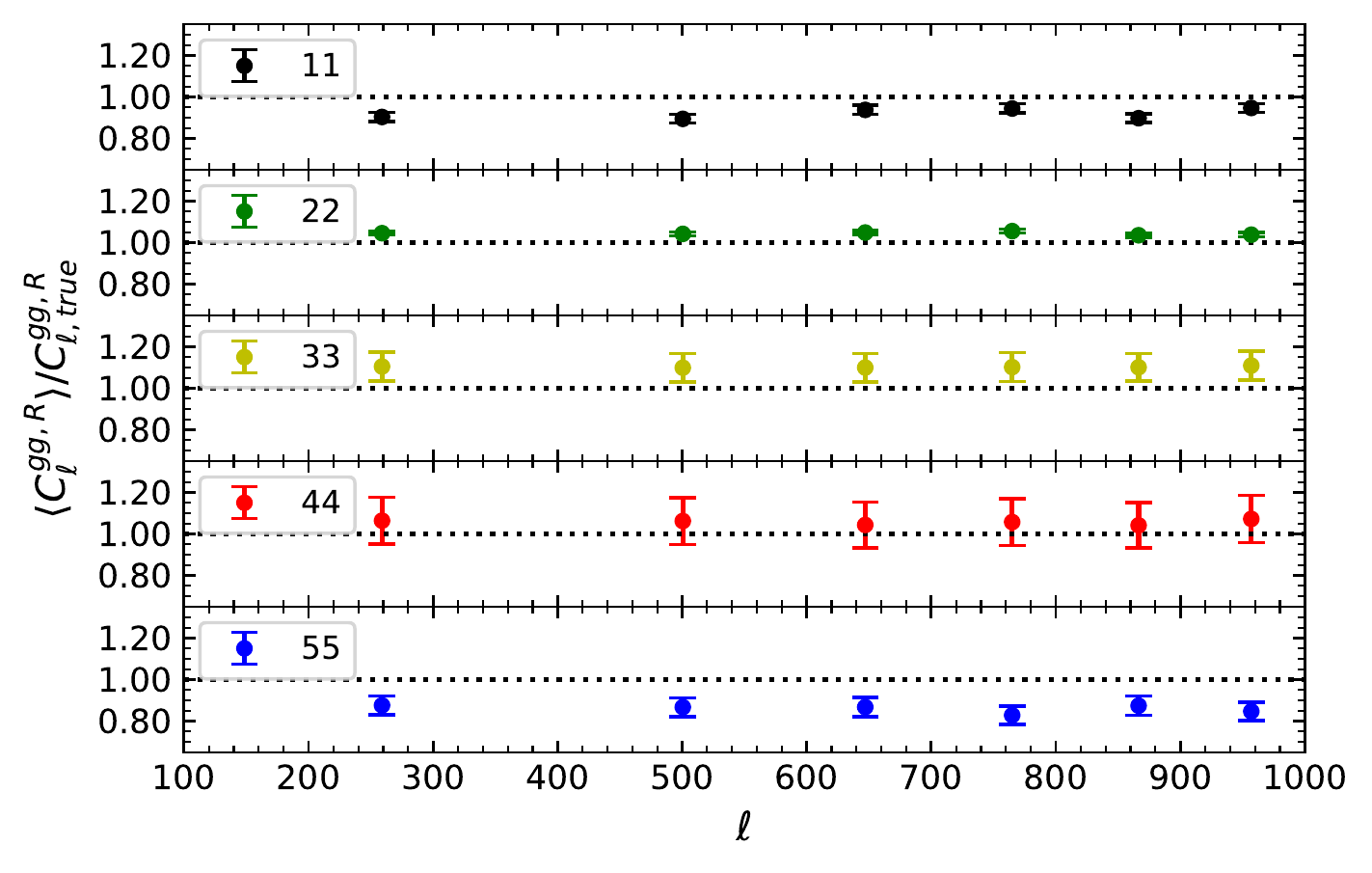}
}
\caption{Same as Fig.~\ref{fig:sim_cata1_result}, 
but the catastrophic photo-$z$ errors happen between the $1$st true-$z$ bin and $5$th photo-$z$ bin, 
and between the $5$th true-$z$ bin and $1$st photo-$z$ bin.}
\label{fig:sim_cata2_result}
\end{figure*}

In previous section, the algorithm shows stable and accurate reconstruction when applied to the case without catastrophic photo-$z$ errors.
Nevertheless, of particular interest is the behavior with non-negligible outliers.
In this section, we present the performance of the algorithm using two sets of photometric catalogues with exaggerated catastrophic photo-$z$ errors.
For a photometric galaxy catalogue, usually the outlier rate is less than $0.1$ for the scattering matrix elements far from the diagonal region.
Therefore, based on the $P^{\rm true}$ shown in Fig.~\ref{fig:sim_normal_pture}, we set $0.1$ for two elements away from the diagonal region and reduce the value(s) in the corresponding diagonal element(s) under the sum-to-one constraint.
The two $P^{\rm true}$'s are presented in panel (a) of Figs~\ref{fig:sim_cata1_result} and \ref{fig:sim_cata2_result}, respectively.

A big difference from the previous result is that the distributions of $\mathcal{J}$ from Algorithm 2 show obvious columnar structures in panel (b) of Figs~\ref{fig:sim_cata1_result} and \ref{fig:sim_cata2_result}.
This implies that the algorithm found a set of solutions that are significantly better than the others.
The selection criterion equation~(\ref{define:J_selection}) is still suitable in this case.
It picks out the solutions within the columnar structure in the distribution.

The results of the reconstructed $P$ and $C^{gg,R}$ are presented in panel (c) and (d), respectively.
We find that the uncertainties in both $P$ and $C^{gg,R}$ are smaller than the previous case without catastrophic error.
This behavior could be understood as follows.
First, the large catastrophic scatters produce significantly non-zero cross power spectra that may not be saturated by the shot noise.
Thus, the information is increased.
Second, the existence of the outliers actually helps to break the degeneracy in some cases.
Assume that the 4th and 5th redshift bin have the same auto power spectrum shape.
Thus, the measurement of $C_{45}^{gg,P}\neq 0$ can not distinguish the contribution of each bin.
If, fortunately, $P_{41}=0$ and $P_{51}\neq 0$, the null measurement of $C_{41}^{gg,P}$ and a non-zero $C_{51}^{gg,P}$ can tell the amplitude of $C_{55}^{gg,R}$, and then the amplitude of $C_{44}^{gg,R}$, as long as $C_{11}^{gg,R}$ does not suffer from other degeneracy.
However, usually there is no exactly zero scatter and the measurement suffers from shot noise in real world.
The degeneracy can not be fully broken.

Nevertheless, the results in both cases show the powerful ability of reconstructing the scattering matrix $P$ and $C_{\ell}^{gg,R}$ even with catastrophic photo-$z$ errors.
The mean absolute bias of the mean redshift estimated by equation~(\ref{eqn:estimate}) is reduced by $\sim$ 60 per cent from 0.060 to 0.023 for the $P^{\mathrm{true}}$ in Fig.~\ref{fig:sim_cata1_result} and $\sim$ 80 per cent from 0.063 to 0.014 for the $P^{\mathrm{true}}$ in Fig.~\ref{fig:sim_cata2_result}, respectively.
% \sout{The details are shown in Table \ref{tab:delta_z_result}.}
Therefore, we believe that the algorithm can be adapted to many complicated cases and finally be applied to the survey data.

\subsection{The reduction of the mean redshift bias}

\label{result:redshift bias}
\begin{table*}
	\centering
	\caption{For different $P^{\mathrm{true}}$ cases, the biases of the mean redshift from 
	$\langle z^{\rm true}\rangle$, the mean of the true-$z$ for the galaxies in a given photo-$z$ bin, are listed.
	For each case, the first row shows the \denweight mean photo-$z$ $\langle z^P \rangle$ without self-calibration.
	The mean redshifts in the second and third rows are the estimates from the preset $P^\mathrm{true}$ and the recovered $P^\mathrm{recover}$ with $\langle z^{S}\rangle$, the mean of the true-$z$ for each true-$z$ bin.
	However, $\langle z^{S}\rangle$ is not an observable.
	Thus, in the 4th and 5th rows, the mean redshifts are estimated using $\langle z^{P}\rangle$.
	In each row, the bias for each photo-$z$ bin is listed in different columns and the mean absolute bias over all the bins is listed in the last column.
	}
	\label{tab:delta_z_result}
	\setlength{\tabcolsep}{2.4mm}{
	\begin{tabular}{cccccccc}
		\hline		  \multicolumn{2}{c|}{\multirow{2}{*}{Case}}&\multicolumn{5}{c}{$z_p$ range}&\multirow{2}{*}{|$\Delta\langle z\rangle|$}\\
		\cline{3-7}
		  &&$[0.0-0.2]$& $[0.2-0.4]$ & $[0.4-0.6]$ & $[0.6-0.8]$ & $[0.8-1.0]$&\\
		\hline
		\multirow{5}{*}{Normal case (Fig.~\ref{fig:sim_normal_pture})} &$\langle z^{P}\rangle-\langle z^{\rm true}\rangle$&$-$0.055 &$-$0.047 &$-$0.021 &$-$0.006 &0.044 &0.035\\
		&$\sum_{j} P_{j i}^{\rm true}\langle z_{j}^{S}\rangle-\langle z^{\rm true}\rangle$&$3.5\times10^{-5}$ &$-2.6\times10^{-6}$ &$-1.2\times10^{-5}$  &$2.7\times10^{-5}$ &$-4.3\times10^{-5}$ &$2.4\times10^{-5}$\\
		&$\sum_{j} P_{j i}^{\rm recover}\langle z_{j}^{S}\rangle-\langle z^{\rm true}\rangle$&$-$0.006 &0.003 &$-$0.002 &$-$0.009 &0.006 &0.005\\
		&$\sum_{j} P_{j i}^{\rm true}\langle z_{j}^{P}\rangle-\langle z^{\rm true}\rangle$&$-$0.010 &0.012 &$-$0.018 &$-$0.019 &$-$0.019 &0.016\\
		&$\sum_{j} P_{j i}^{\rm recover}\langle z_{j}^{P}\rangle-\langle z^{\rm true}\rangle$&$-$0.016 &$-$0.009 &$-$0.019 &$-$0.028 &$-$0.013 &0.017\\
		\hline
		\multirow{5}{*}{Catastrophic case (Fig.~\ref{fig:sim_cata1_result})}&$\langle z^{P}\rangle-\langle z^{\rm true}\rangle$& $-$0.183 &$-$0.047 &$-$0.021 &$-$0.006 &0.044 &0.060\\
		&$\sum_{j} P_{j i}^{\rm true}\langle z_{j}^{S}\rangle-\langle z^{\rm true}\rangle$&$-1.9\times10^{-5}$ &$1.2\times10^{-5}$ &$-1.9\times10^{-6}$  &$-1.0\times10^{-5}$ &$8.5\times10^{-6}$ &$1.0\times10^{-5}$\\
		&$\sum_{j} P_{j i}^{\rm recover}\langle z_{j}^{S}\rangle-\langle z^{\rm true}\rangle$&$-$0.016 &0.020 &0.006 &$-$0.015 &$-$0.015 &0.014\\
		&$\sum_{j} P_{j i}^{\rm true}\langle z_{j}^{P}\rangle-\langle z^{\rm true}\rangle$&$-$0.012 &$-$0.012 &$-$0.018 &$-$0.019 &$-$0.019 &0.016\\
		&$\sum_{j} P_{j i}^{\rm recover}\langle z_{j}^{P}\rangle-\langle z^{\rm true}\rangle$&$-$0.028 &0.008 &$-$0.013 &$-$0.034 &$-$0.033 &0.023\\
		\hline
		\multirow{5}{*}{Catastrophic case (Fig.~\ref{fig:sim_cata2_result})}&$\langle z^{P}\rangle-\langle z^{\rm true}\rangle$& $-$0.127 &$-$0.047 &$-$0.021 &$-$0.006 &0.113 &0.063\\
		&$\sum_{j} P_{j i}^{\rm true}\langle z_{j}^{S}\rangle-\langle z^{\rm true}\rangle$&$3.2\times10^{-5}$ &$-4.2\times10^{-6}$ &$-1.9\times10^{-7}$ &$-5.0\times10^{-6}$ &$-1.6\times10^{-5}$ &$1.1\times10^{-5}$\\
		&$\sum_{j} P_{j i}^{\rm recover}\langle z_{j}^{S}\rangle-\langle z^{\rm true}\rangle$&$-$0.009 &0.001 &0.004 &0.006 &0.007 &0.006\\
		&$\sum_{j} P_{j i}^{\rm true}\langle z_{j}^{P}\rangle-\langle z^{\rm true}\rangle$&$-$0.011 &$-$0.012 &$-$0.018 &$-$0.019 &$-$0.018 &0.016\\
		&$\sum_{j} P_{j i}^{\rm recover}\langle z_{j}^{P}\rangle-\langle z^{\rm true}\rangle$&$-$0.020 &$-$0.011 &$-$0.014 &$-$0.013 &$-$0.011 &0.014\\
		\hline
	\end{tabular}}
\end{table*}

Usually for the galaxies in a given photo-$z$ bin, the mean of the photo-$z$ is different from the mean of the true-$z$, causing the bias in the cosmological constraints. 
Thus, a result of particular interest is the reduction of the bias in the mean redshift by the self-calibration method.

In Table~\ref{tab:delta_z_result}, we list the biases of various mean redshift estimates.
From top to bottom, the results for different $P^\mathrm{true}$ are listed.
$\langle z^{\rm true}\rangle$ is the \denweight mean true-$z$ of each photo-$z$ bin.
This is the true answer measured from the simulated catalogue.
$\langle z^P\rangle$ is the mean of photo-$z$, which roughly differs from $\langle z^{\rm true}\rangle$ about $0.053$ (averaged over all the cases).
To test the performance of the self-calibration, we can use the recovered scattering matrix $P^{\mathrm{recover}}$ along with the simulated mean true-$z$ $\langle z^{S}\rangle$ to check whether the mean redshift is correctly de-biased.
The result is presented in the third row for each case.
On average over all the cases, the mean bias is about $0.008$.
As a sanity check, the mean redshift from $P^{\mathrm{true}}$ and $\langle z^{S}\rangle$ is shown in the second row for each case.
The result implies that using the true scattering matrix, the bias is of order $10^{-5}$ and is indeed negligible.
However, in the real observations we do not have $\langle z^{S}\rangle$.
Instead, we have to use $\langle z^{P}\rangle$ to approximate $\langle z^{S}\rangle$ [equation~(\ref{eqn:estimate})].
The results using $\langle z^{P}\rangle$ and $P^{\mathrm{true}}$ are listed in the fourth row, which tell us the accuracy of the above approximation.
The bias from this approximation is roughly 0.016 on average.
The new mean redshifts from the self-calibration method we can obtain from the real observation, i.e. the mean redshifts estimated from $\langle z^{P}\rangle$ and $P^{\mathrm{recover}}$, are listed in the fifth row.
The averaged number over all cases is roughly 0.018.

For each case, both the small mean redshift bias listed in the third row, and the small difference in the mean redshift bias between the fourth and fifth rows indicate the success of the self-calibration.
The residual bias in the mean redshift estimated from equation~(\ref{eqn:estimate}) is mainly contributed by using $\langle z^{P}\rangle$ to approximate $\langle z^{S}\rangle$.
Comparing the new mean redshift estimated from $\langle z^{P}\rangle$ and $P^{\mathrm{recover}}$ with the original mean redshift without self-calibration, i.e. $\langle z^{P}\rangle$, the bias is reduced by more than 50 per cent in all three simulation cases.

\subsection{Photo-$z$ scatter and power spectrum for DECaLS}
\label{result:DECaLS}
%outline

\begin{figure*}
\centering
\setcounter {subfigure} {0}
\subfigure[$\mathcal{J}$ distribution]{\includegraphics[width=\columnwidth]{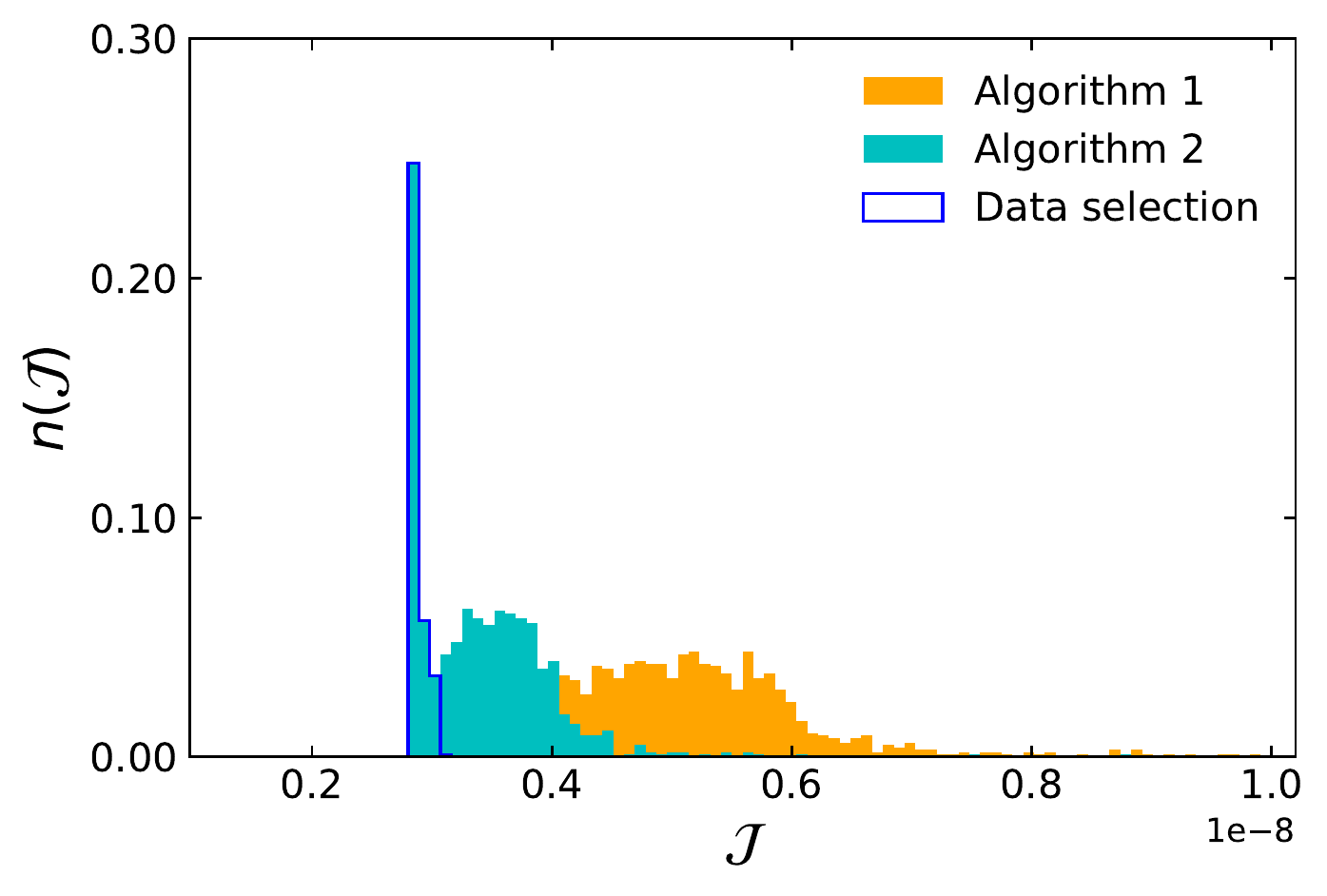}
}
%\quad
\centering
\setcounter {subfigure} {3}
\subfigure[reconstructed diagonal $C_{\ell}^{gg,P}$ and $C_{\ell,{\rm obs}}^{gg,P}$]{
    {\includegraphics[width=\columnwidth]{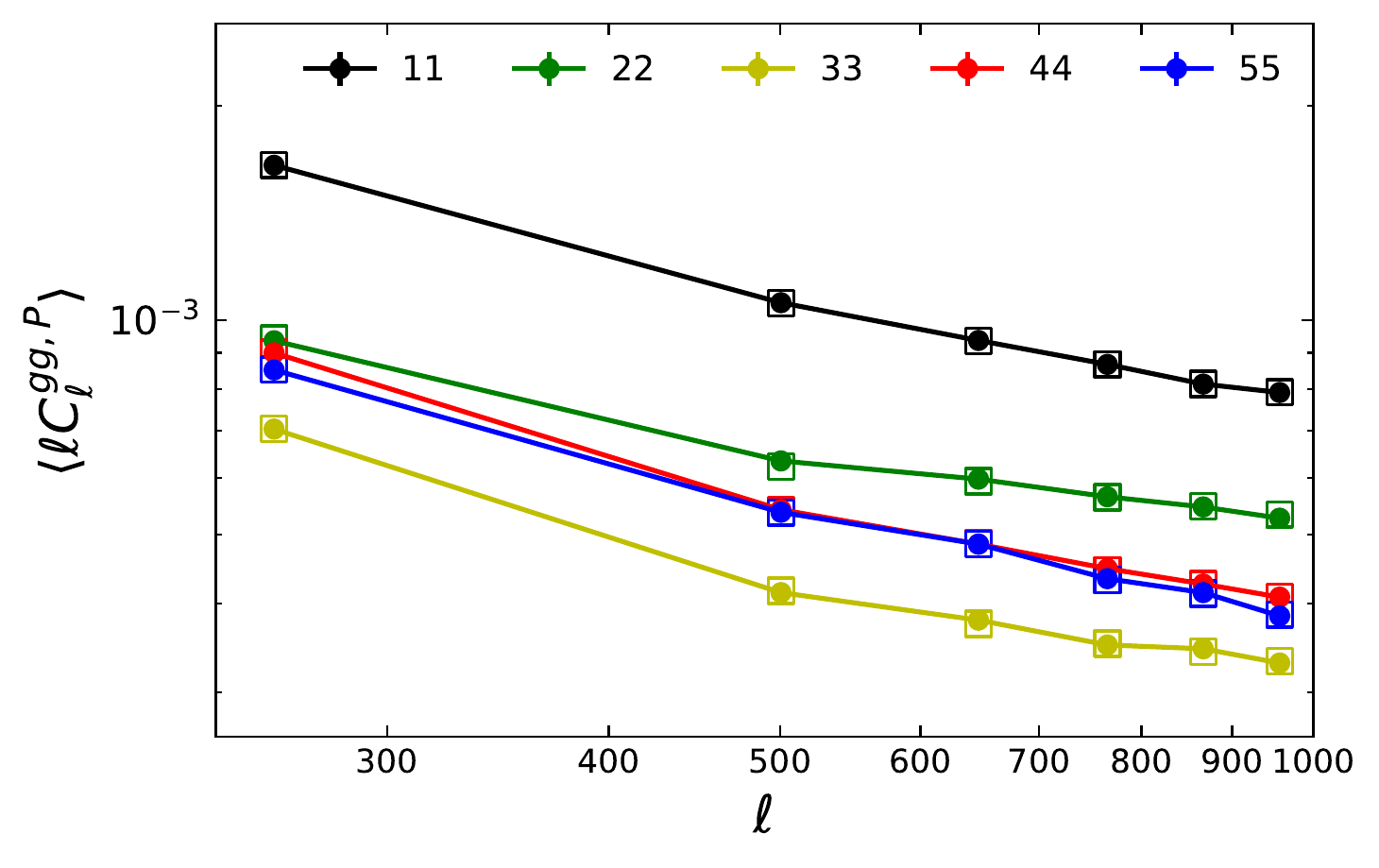}}
}
%\quad
\centering
\setcounter {subfigure} {1}
\subfigure[average value: $\langle P \rangle$]{
    \includegraphics[width=\columnwidth]{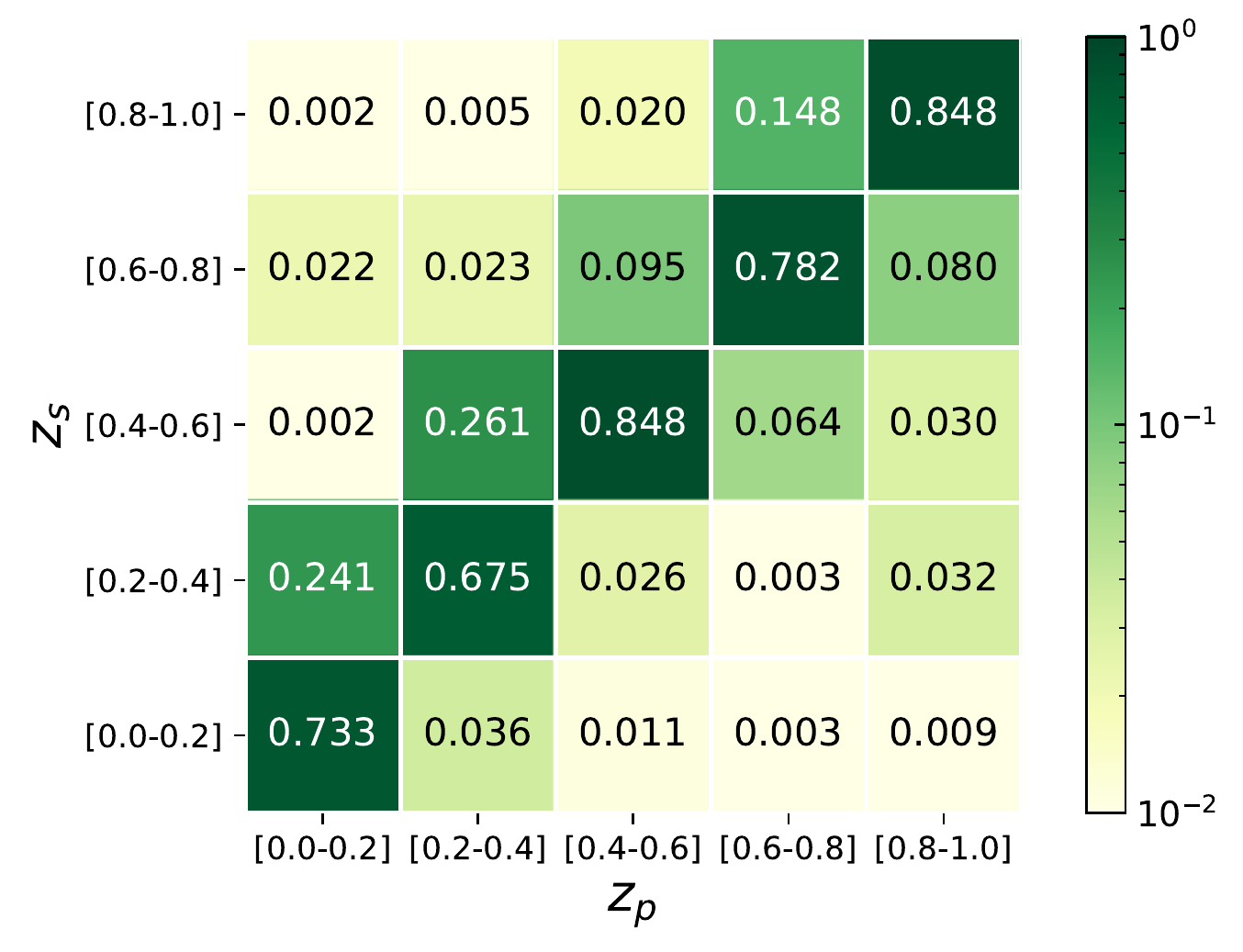}
}
%\quad
\centering
\setcounter {subfigure} {4}
\subfigure[reconstructed sub-diagonal $C_{\ell}^{gg,P}$ and $C_{\ell,{\rm obs}}^{gg,P}$]{\raisebox{0.3cm}
    {\includegraphics[width=\columnwidth]{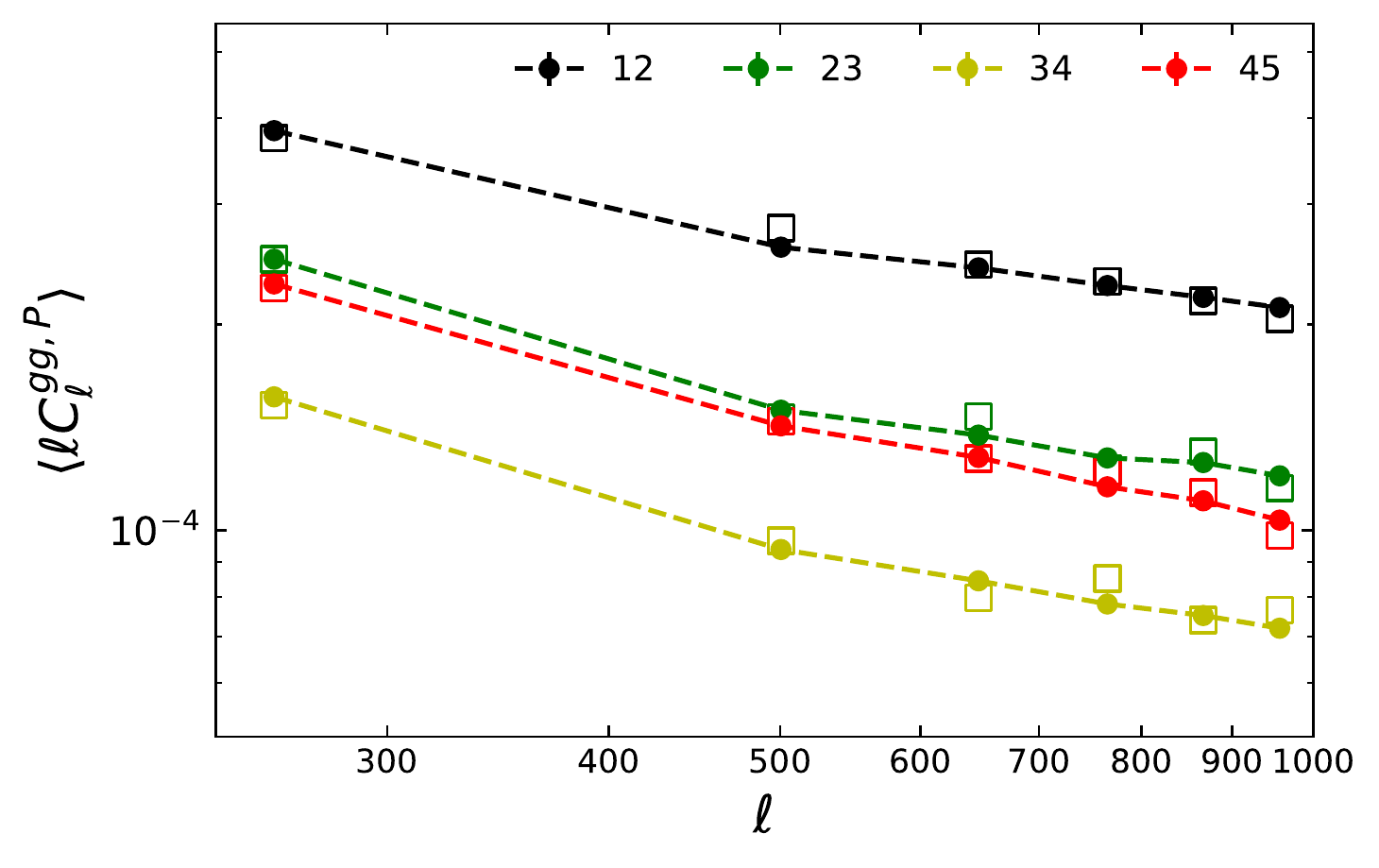}}
}
%\quad
\centering
\setcounter {subfigure} {2}
\subfigure[standard deviation: $\sigma_P$]{
    \includegraphics[width=\columnwidth]{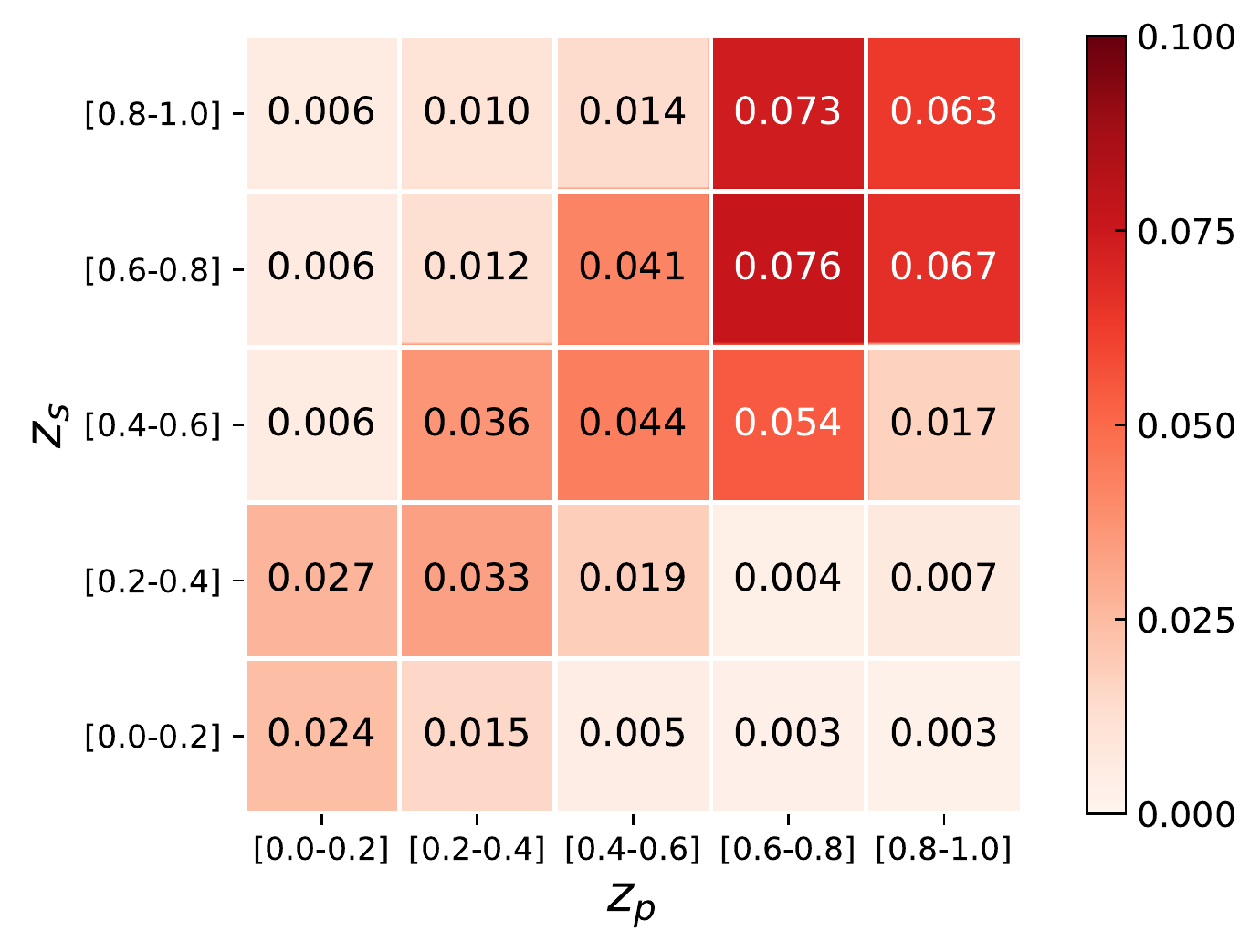}
}
%\quad
\centering
\setcounter {subfigure} {5}
\subfigure[reconstructed $C_{\ell}^{gg,R}$]{\raisebox{0.3cm}
    {\includegraphics[width=\columnwidth]{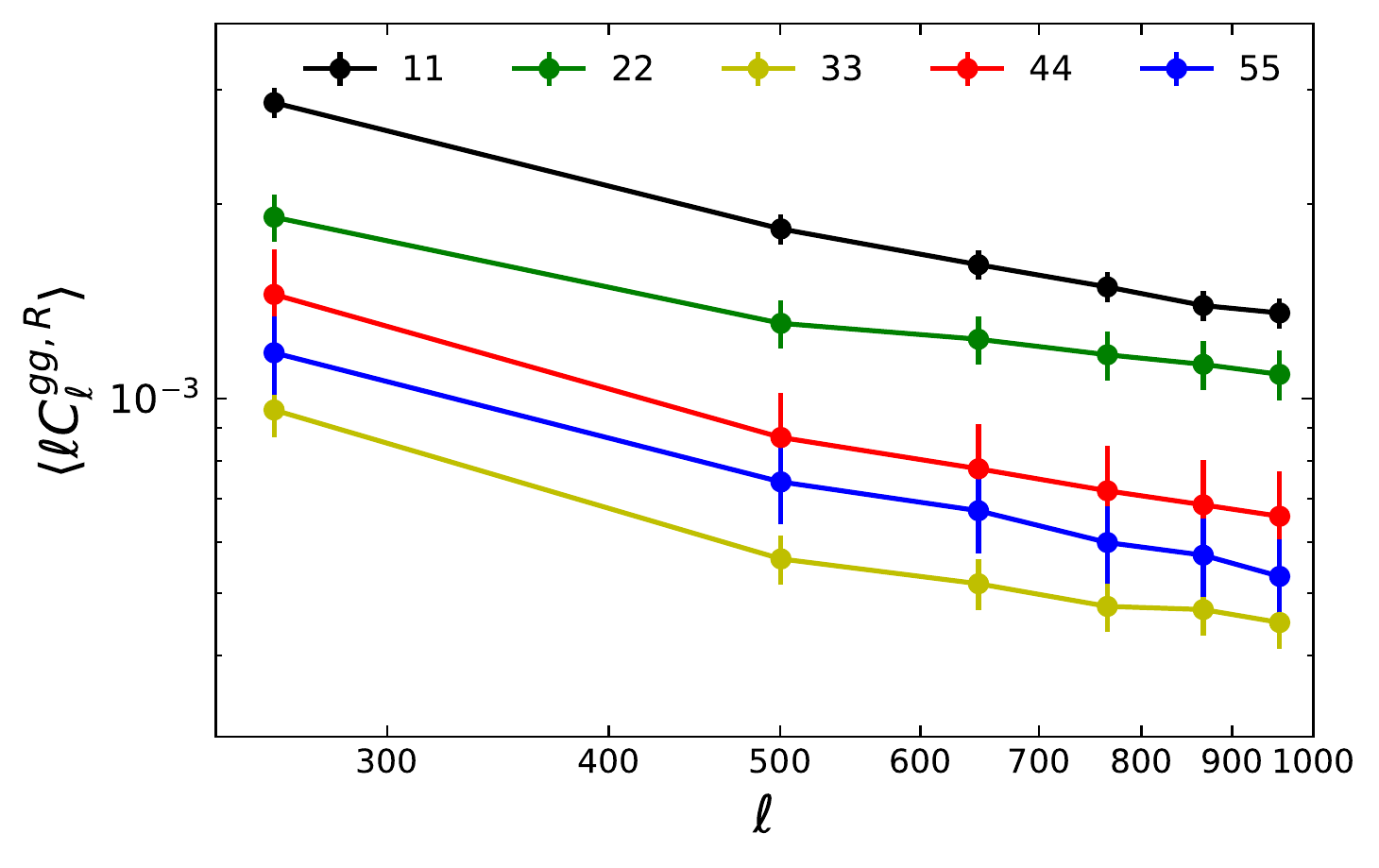}}
}
\caption{The $\mathcal{J}$ distribution [panel (a)], the reconstructed scattering matrix $P$ [panel (b)], the uncertainty in the recovered $P$ [panel (c)], the reconstructed power spectrum for photo-$z$ bins [panels (d) and (e)] and for true-$z$ bins [panel (f)].
Note that for the observations we do not know the values of $P^{\mathrm{true}}$ and $C_{\ell,\mathrm{true}}^{gg,R}$.
}
\label{fig:DR8_result}
\end{figure*}

\begin{figure*}
\centering
\setcounter {subfigure} {0}
\subfigure[$\mathcal{J}$ distribution]{\includegraphics[width=\columnwidth]{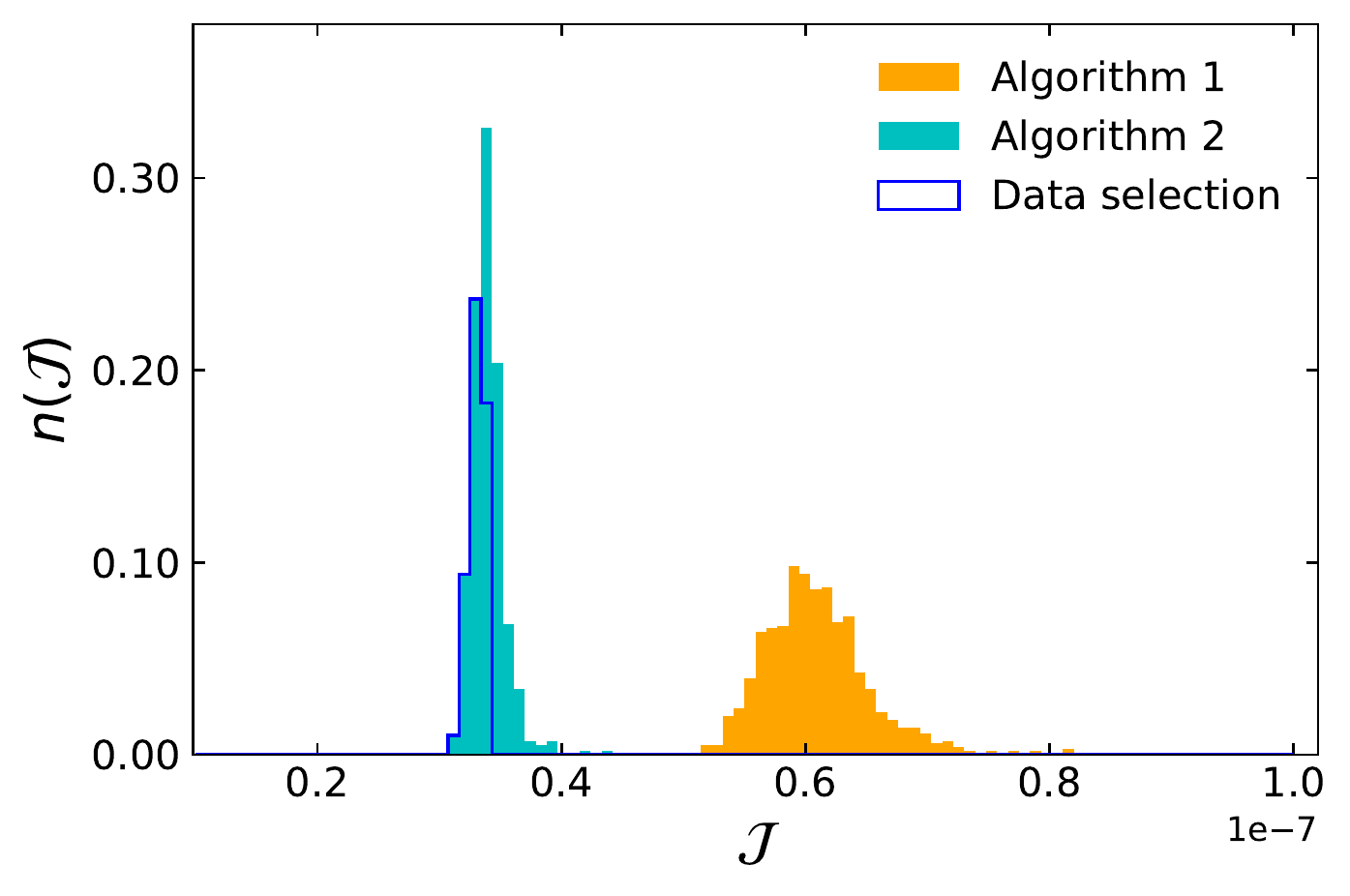}}
%\quad
\centering
\setcounter {subfigure} {3}
\subfigure[reconstructed diagonal $C_{\ell}^{gg,P}$ and $C_{\ell,{\rm obs}}^{gg,P}$]{
    {\includegraphics[width=\columnwidth]{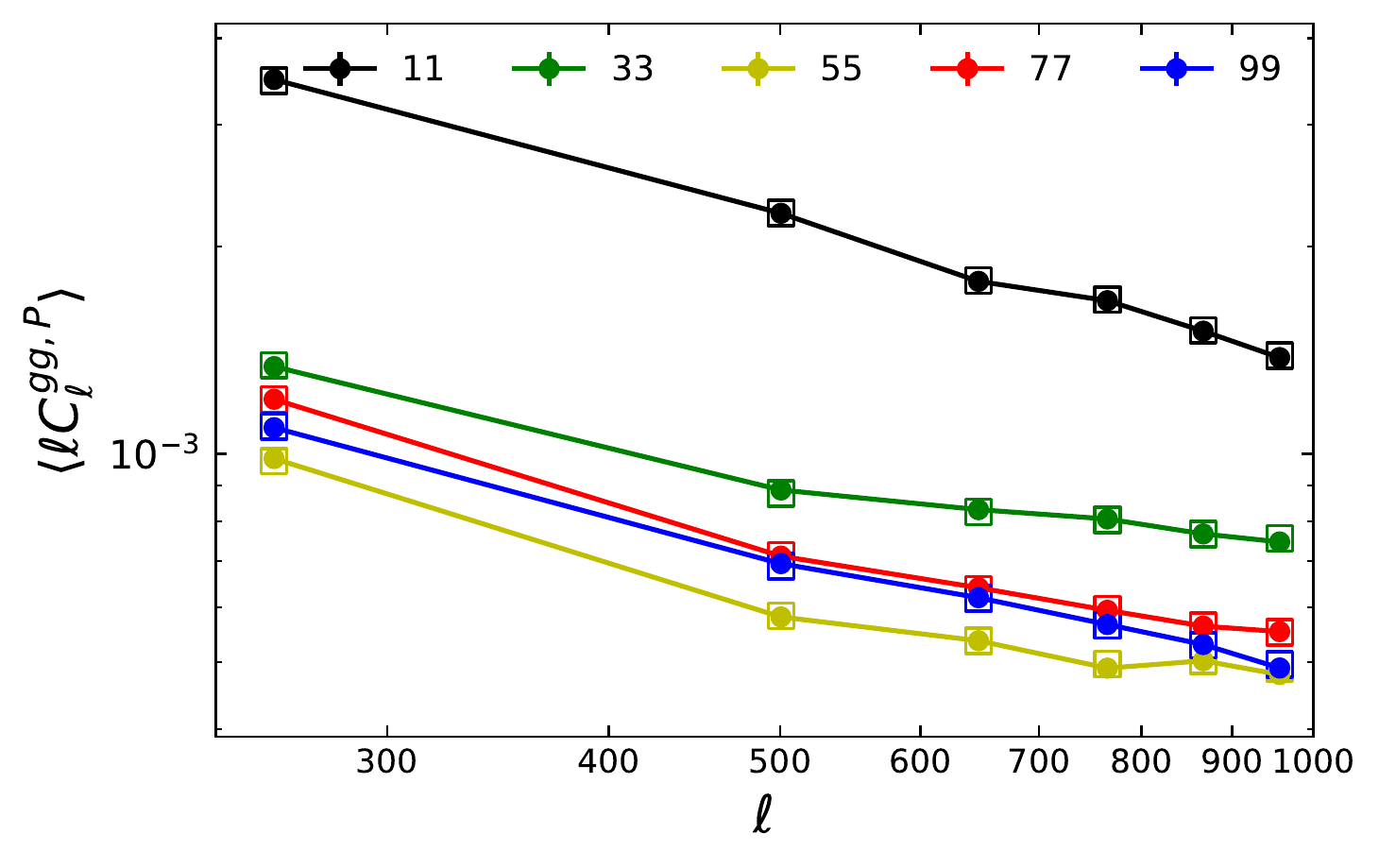}}
}
\centering
\setcounter {subfigure} {1}
\subfigure[average value: $\langle P \rangle$]{
    \includegraphics[width=\columnwidth]{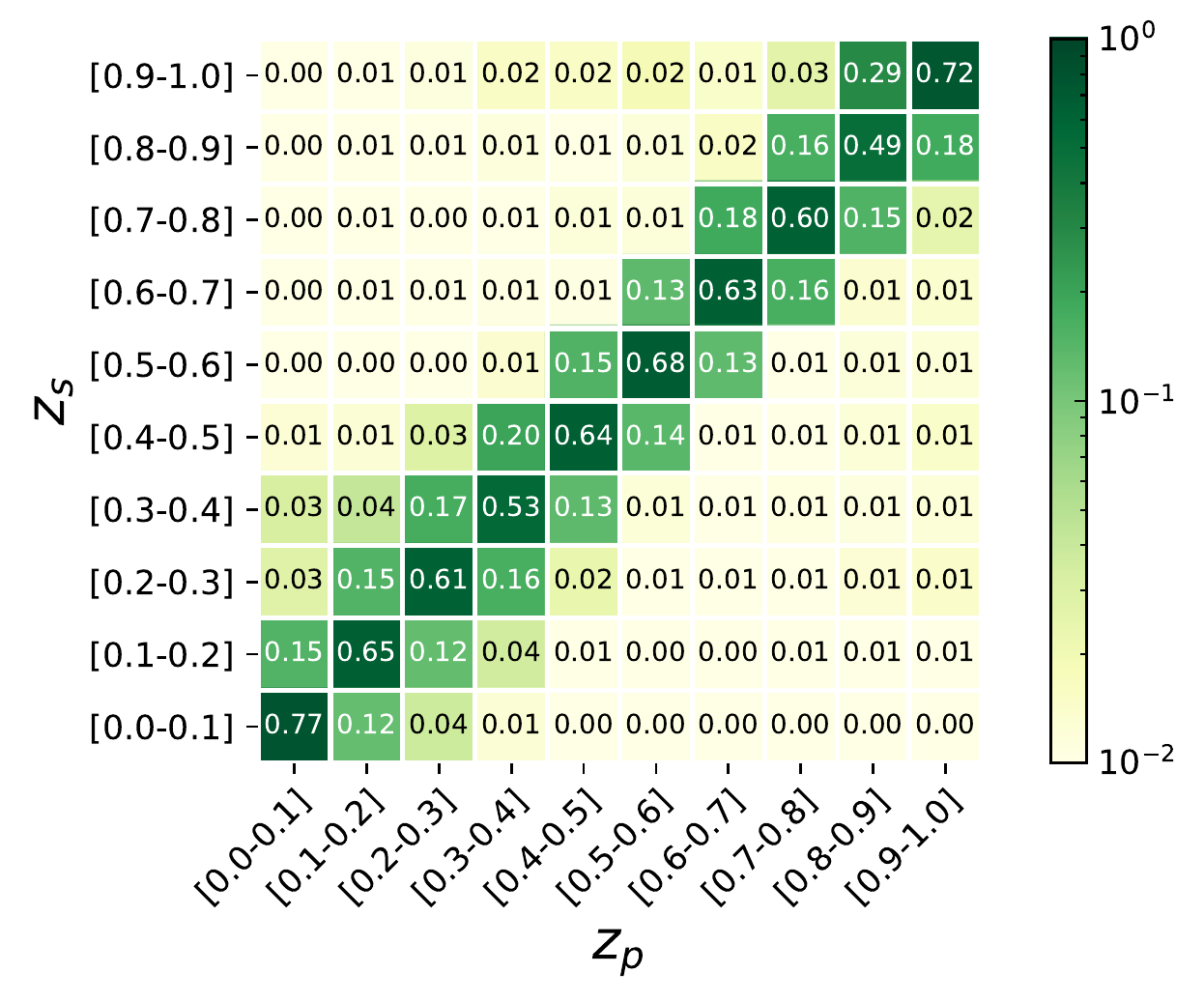}
}
%\quad
\centering
\setcounter {subfigure} {4}
\subfigure[reconstructed sub-diagonal $C_{\ell}^{gg,P}$ and $C_{\ell,{\rm obs}}^{gg,P}$]{
    \raisebox{0.8cm}{\includegraphics[width=\columnwidth]{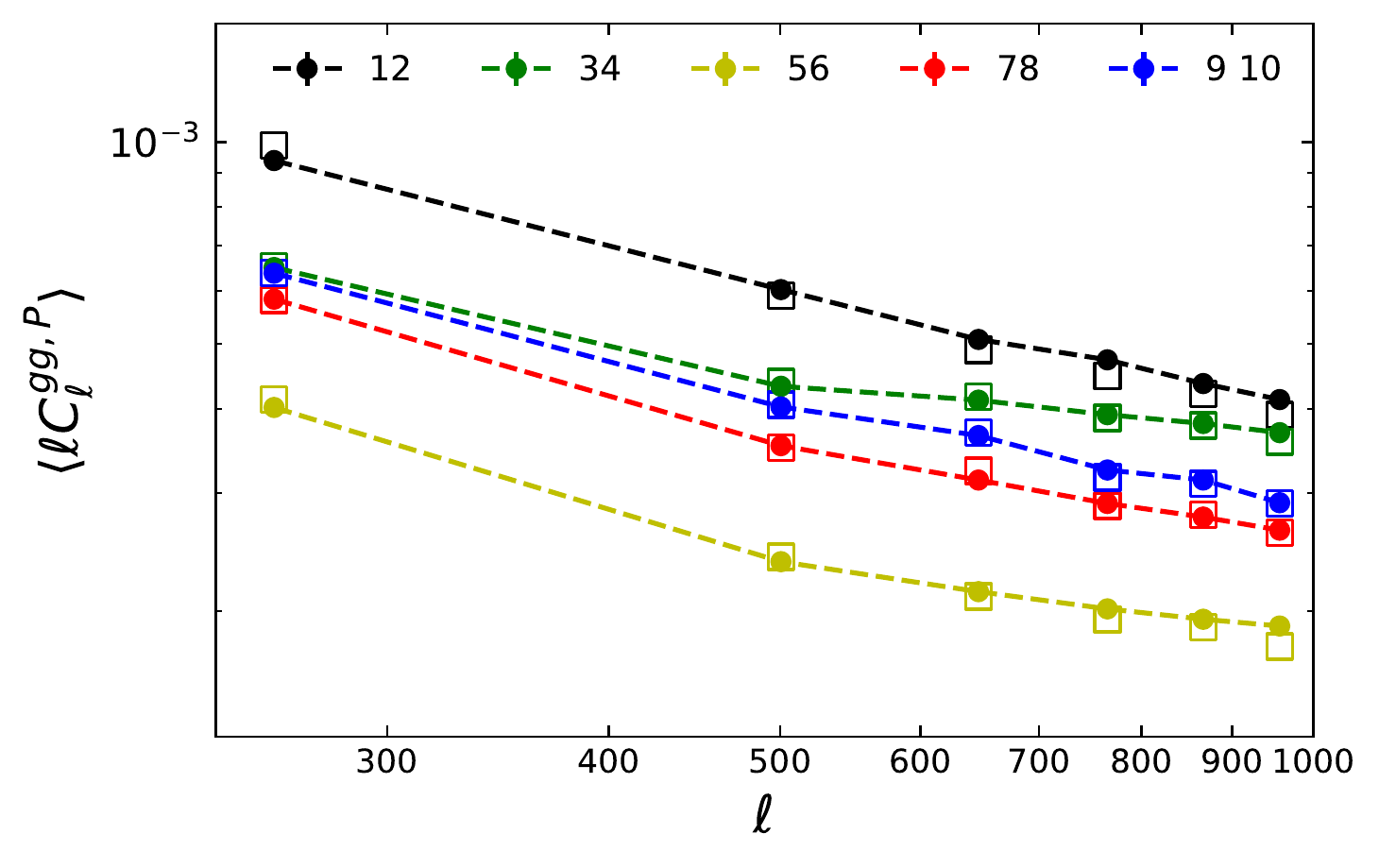}}
}
%\quad
\centering
\setcounter {subfigure} {2}
\subfigure[standard deviation: $\sigma_P$]{
    \includegraphics[width=\columnwidth]{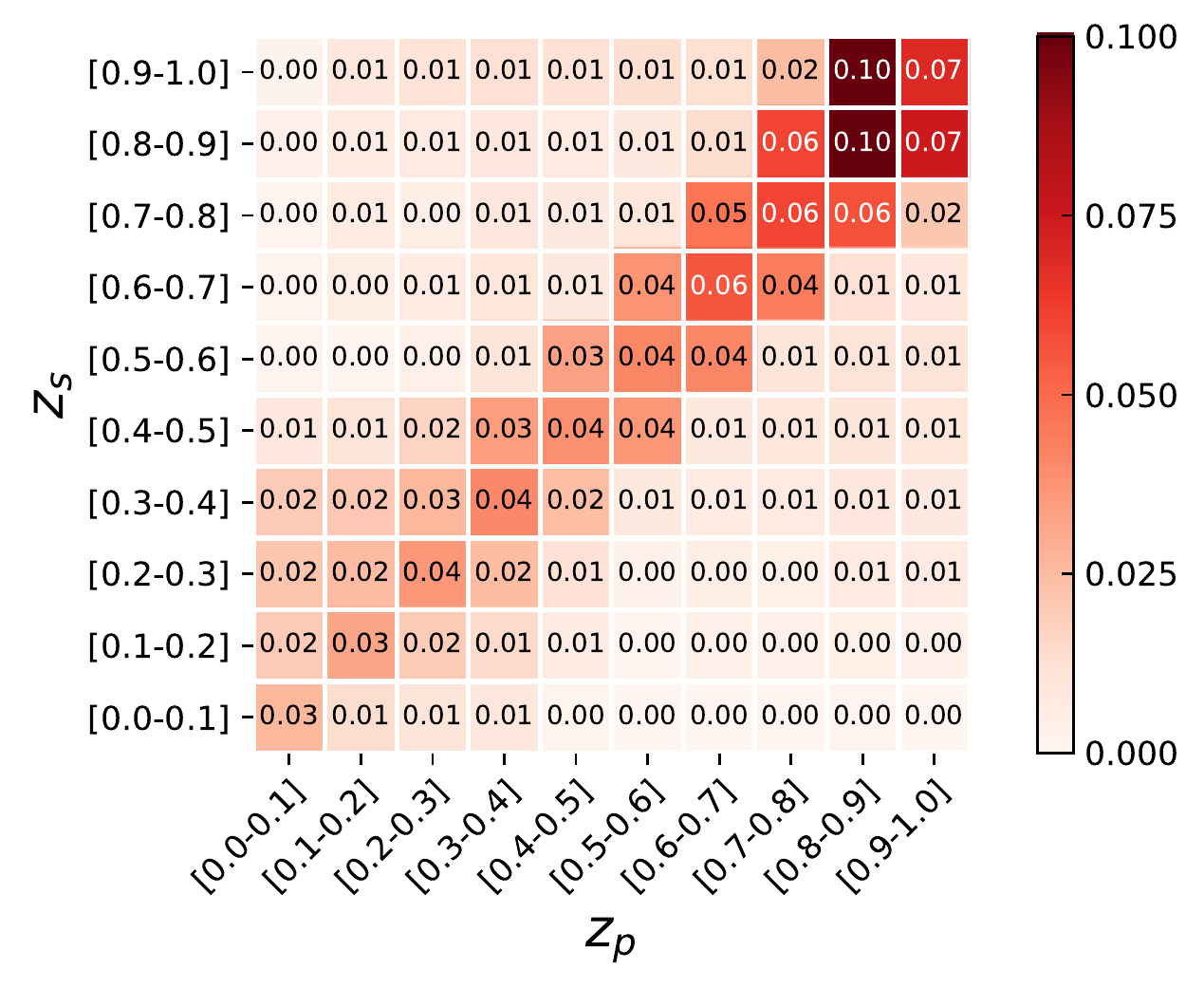}
}
%\quad
\centering
\setcounter {subfigure} {5}
\subfigure[reconstructed $C_{\ell}^{gg,R}$]{\raisebox{0.8cm}
    {\includegraphics[width=\columnwidth]{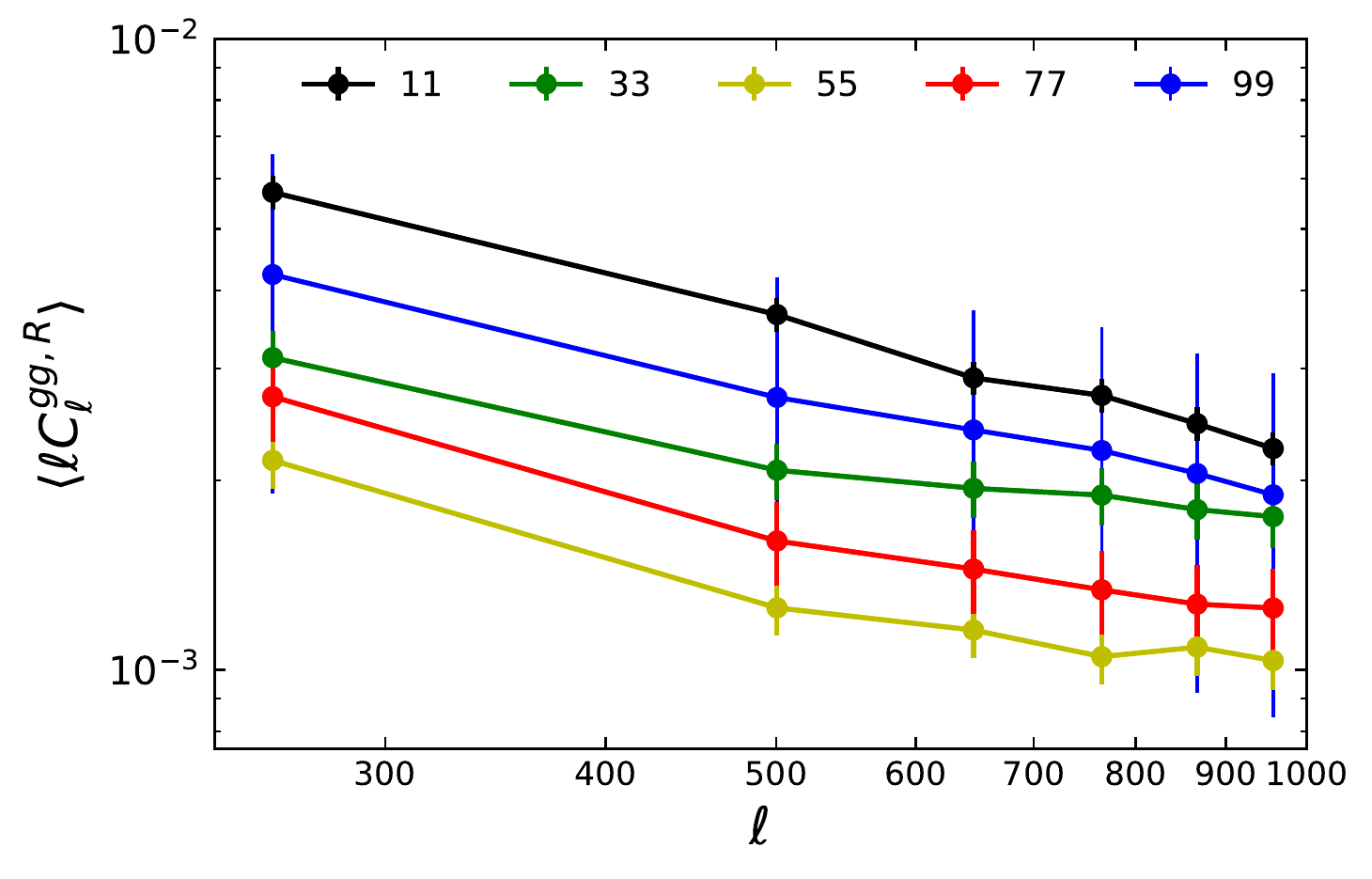}}
}
\caption{Same as Fig~\ref{fig:DR8_result}, but we divide the galaxies into 10 redshift bins.
For ease of view, we only show some results from five bins.
Compared to the results with 5 redshift bins above, the uncertainties in the reconstructed $P$ and the reconstructed power spectrum $C_{\ell}^{gg,R}$ are increased,
due to the increase of the noise level in the power spectrum measurement and the increase of the number of unknown variables.
}
\label{fig:DR8_bin10_result}
\end{figure*}

% \begin{figure}
% \centering
%     \includegraphics[width=\columnwidth]{figures/DR8_compare.pdf}
%     \caption{Comparisons between the reconstruction results $P^{\rm CF}$ using two-point correlation function in Xu et al. (in preparation) and ours ($P^{\rm PS}$). 
%     We set our result (the circles) as the reference.
%     The squares show the difference between the two results.
%     The error bars are the uncertainties from both approaches.
%     }
% \label{fig:DR8_compare}
% \end{figure}

After the validation on several different simulated catalogues, we finally apply the algorithm to the DECaLS DR8 photometric galaxy catalogue introduced in Section~\ref{sec:obs}.
In Fig.~\ref{fig:DR8_result}, we show the distribution of $\mathcal{J}$, the reconstructed scattering matrix and its uncertainty, the recovered $C_\ell^{gg,P}$ and the derived $C_\ell^{gg,R}$.
Here we do not compare our results with $P^{\rm true}$ or $C_{\ell,{\rm true}}^{gg,R}$ because in observation both are unknown.
We find the error bars of the reconstructed $C_{\ell}^{gg,P}$ are still very small, which validates the rationality of our method for solution selection.
Meanwhile, the reconstructed scattering matrix $\langle P \rangle$ does not show significant outliers, and the scattering is only significant between adjacent bins.
This result is consistent with the relationship between the photo-$z$ and spec-$z$ for $z_{\text{mag}} < 21$ samples in DR8 South \citep{Zhou:2021uw}.

%It should be noted that although the distribution of $\mathcal{J}$ has similar columnar structures as the case of simulation with catastrophic photo-$z$ errors in Fig.~\ref{fig:sim_cata1_result} and Fig.~\ref{fig:sim_cata2_result}, there is no catastrophic error in the reconstructed scattering matrix $\langle P \rangle$.
%It implies that the columnar structure in the $\mathcal{J}$ distribution is potentially caused by \orange{many factors} in survey, not just the existence of the catastrophic photo-$z$ errors.

\citet{Ma:2006wj} shows that there is only very mild gain on the cosmological parameter constraints by increasing the number of tomographic bins over 5. 
Here we redo the analysis using 10 redshift bins,
and the results are presented in Fig.~\ref{fig:DR8_bin10_result}.
This investigation helps us to explore the ability, stability, robustness and limitation of the algorithm.
The precise reconstruction of  $C_{\ell}^{gg,P}$ still shows that the algorithm successfully solves equation~(\ref{eqn:Cggp}).
Although the result is consistent with the fiducial case with 5 redshift bins,
we observe the increase in the uncertainty of the reconstructed $P$ and $C_{\ell}^{gg,R}$.
This is expected, for two reasons.
First, the number of galaxies in each photo-$z$ bin is decreased, leading to very noisy power spectrum measurement.
Second, the number of unknown variables is increased dramatically, causing the decrease of the reconstruction precision.
There is another reason to stop at 10 redshift bins.
The approximation $C_{k{\neq}m}^{gg,R}(\ell)=0$ is suspicious to hold with bin size less than $0.1$. 

Careful readers might notice that in some cases the $\mathcal{J}$ distribution shows a columnar structure very close to the minimum $\mathcal{J}$.
For example, the result from the DECaLS DR8 observation using 5 redshift bins and the result from simulated catalogue with catastrophic errors has this obvious feature in the $\mathcal{J}$ distribution.
However, there is no columnar feature for the DECaLS DR8 result using 10 redshift bins and the simulation result without catastrophic error.
Note that the presence/absence of this feature is related to the degeneracy of the system, which determines the degree of difficulty to find the good solution.
Furthermore, the degeneracy of the system is at least related to the shape of the power spectra on tomographic bins.
For the same fiducial analysis configuration, the existence of this feature in the observation and the absence of this feature in the simulation validation is not conflicting.
We argue that this difference is a result from the different degree of degeneracy in the two cases.
We constructed the simulation catalogue to mimic the photo-$z$ distribution of the observation, but the clustering is not exactly same.
By considering detailed halo-galaxy connection, such as using halo occupation distribution to plant galaxies into haloes, we can obtain the same clustering signal as the observed one.
In this case, the constructed system is expected to have similar degree of degeneracy as the observation.
We emphasize that our solution selection criteria is valid in both cases, and we leave the detailed investigation of the system degeneracy dependence on the halo-galaxy connection to future investigation.

\section{Conclusions and Discussion}
\label{sec:conclusion}

We improved the photo-$z$ self-calibration algorithm originally proposed by \citet{Zhang:2017um} and applied it to the simulation and survey data for the first time.
The modifications of the algorithm, including the use of new random initial guess, the optimization by weighting different scales, the adjustments in convergence criterion and the solution selection, enormously enhance the stability and robustness of the algorithm on the noisy measurements.
We applied the algorithm in three representative scenarios, one for normal photo-$z$ scatter simulation without catastrophic error, another for simulations with catastrophic photo-$z$ errors, and the last one for the DECaLS DR8 photometric galaxy sample.

For the case without catastrophic photo-$z$ errors, the diagonal and sub-diagonal elements of the reconstructed scattering matrix deviate less than 0.03 from the true ones.
The mean absolute bias over all elements in $P$ is smaller than 0.015.
In terms of the mean redshift estimation, the self-calibration method reduce the bias more than 50 per cent.
For the reconstructed power spectrum, the mean deviation is about 4.4 per cent.
We observe that the uncertainties for the highest two redshift bins are larger than the other's.
We suspect the main reason for this behavior is due to the degeneracy in the power spectrum shape.
For the case with catastrophic error and for the observation, the behavior and the results are similar.
The scattering matrix from the DECaLS data shows no conflict with the result in \citet{Zhou:2021uw}.
Also, we add the analysis using 10 redshift bins, and we found consistent results with the fiducial analysis, but with larger uncertainties.

In different cases, we found that the $\mathcal{J}$ distribution from Algorithm 2 appears strong columnar structures or not.
The proposed solution selection criterion equation~(\ref{define:J_selection}) is suitable for both situations.
The appearance of the columnar structure in the low-end of the $\mathcal{J}$ distribution implies that the algorithm successfully find reliable answers, leading to small uncertainties in the reconstructed scattering matrix and auto power spectrum in true-$z$ bins.
We also compared different weighting scheme and found that inputting $\ell C_\ell^{gg,P}$ to the algorithm has the best performance.
Finally, the performance is not sensitive to the $\ell$ binning as long as the binning scheme is reasonable.

So far our work is carried out based on the sample of DECaLS DR8 photometric galaxy catalogue.
The galaxy number density and the fraction sky coverage are not big enough and thus cause a large error in the power spectrum measurements especially in the high-$\ell$ region.
This worse situation inevitably affects the accuracy of the reconstruction.
However, the algorithm succeeded, although so far the results do not satisfy the requirement of precision cosmology.
We expect that the algorithm presented in this paper will be extremely useful for future photometric surveys, for which the measurement error in power spectrum diminishes as the significant increase in the sample size.
Furthermore, precise small-scale information in power spectrum can effectively breaks the degeneracy caused by the shape of galaxy power spectrum.
We would include the information of power spectrum at $\ell\geq1000$ to further improve photo-$z$ self-calibration accuracy with better data in the future.

The definition of $\mathcal{J}$ in our algorithm differs from the usual form in the $\chi^2$ analysis, in which we need to consider the covariance of the power spectrum.
Also, the update rule in the NMF iteration is different from the usual Markov Chain Monte Carlo analysis.
In future work we will try to use more comprehensive definition of $\mathcal{J}$ to better handle the difference in the measurement errors and the correlation between them.
The solution selection method we used in this paper is not meant to be unchanged.
There may be some other reasonable ways.
People can decide on their own which method is more meaningful according to the distribution of $\mathcal{J}$ when the algorithm is applied to other survey data.

It is worth noting that so far we do not use the lensing-galaxy correlations which are considered in the original Fisher analysis work in \citet{Zhang:2010wr}.
In fact, if available, the lensing-galaxy correlations are very useful to break some degeneracy, for example, caused by the similar shape of galaxy power spectrum, and ultimately improve the accuracy of reconstruction.
Future work will consider to use the full sample of DESI imaging survey data and to really incorporate the lensing-galaxy correlations as part of algorithm inputs.
Also, the three-point correlation function or high-order two-point correlation function may be helpful to alleviate the degeneracy trouble.
We delay these studies to future work.

\section*{Acknowledgements}

We thank Pengjie Zhang for useful suggestions.
This work made use of the Gravity Supercomputer at the Department of Astronomy, Shanghai Jiao Tong University.
This work is supported by the National Key Basic Research and Development Program of China (No. 2018YFA0404504, 2018YFA0404601, 2020YFC2201600), 
the National Science Foundation of China (grant Nos. 11621303, 11890691, 11773048, 11653003),
the science research grants from the China Manned Space Project with NO. CMS-CSST-2021-B01, the “111” Project of the Ministry of Education under grant No. B20019, the CAS Interdisciplinary Innovation Team (JCTD-2019-05), and the Ministry of Science and Technology of China (2020SKA0110100).

\section*{Data availability}
All data included in this study are available upon request by contacting with the corresponding author.



\bibliographystyle{mnras}
\bibliography{example} % if your bibtex file is called example.bib

%%%%%%%%%%%%%%%%% APPENDICES %%%%%%%%%%%%%%%%%%%%%

\appendix

\section{The effect of the binning scheme and magnification bias}
\label{sec:appendix}

\begin{figure}
\centering
\subfigure[$\mathcal{J}$ distribution]{
    \includegraphics[width=\columnwidth]{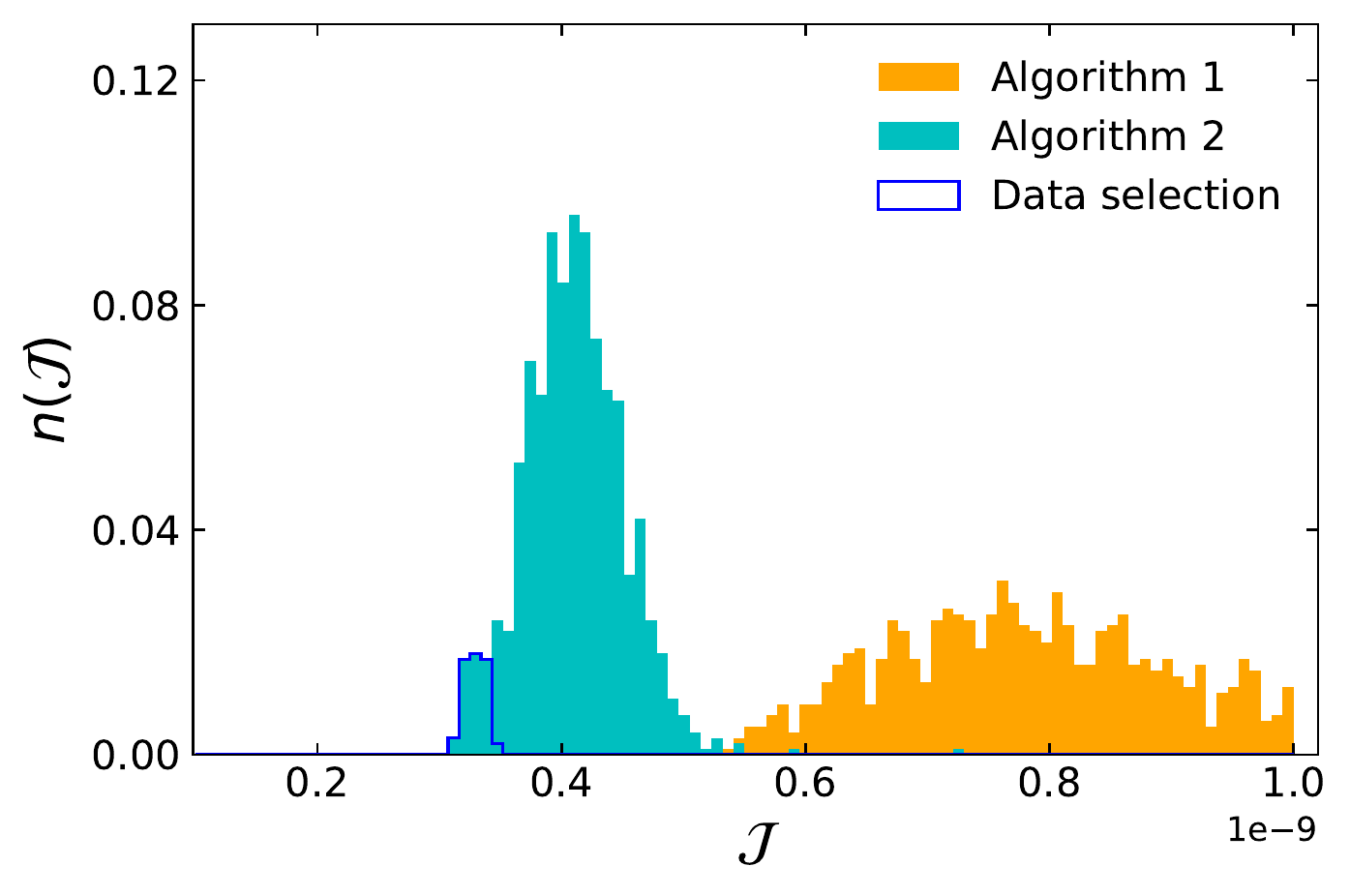}
}
%\quad
\centering
\subfigure[$\langle P \rangle-P^{\rm true}$]{
    \raisebox{0.6cm}{\includegraphics[width=\columnwidth]{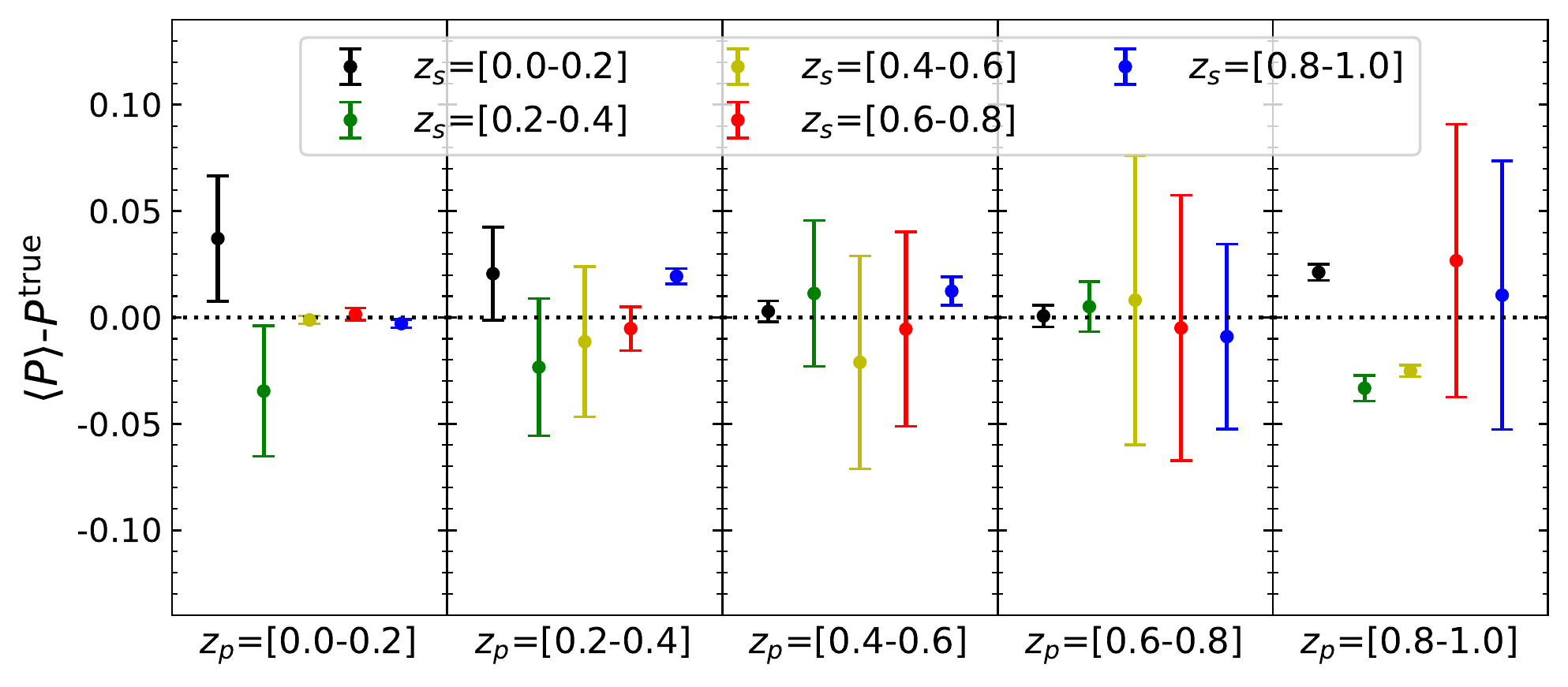}}
}
%\quad
\centering
\subfigure[$\langle C_{\ell}^{gg,R} \rangle /C_{\ell,{\rm true}}^{gg,R}$]{
    \includegraphics[width=\columnwidth]{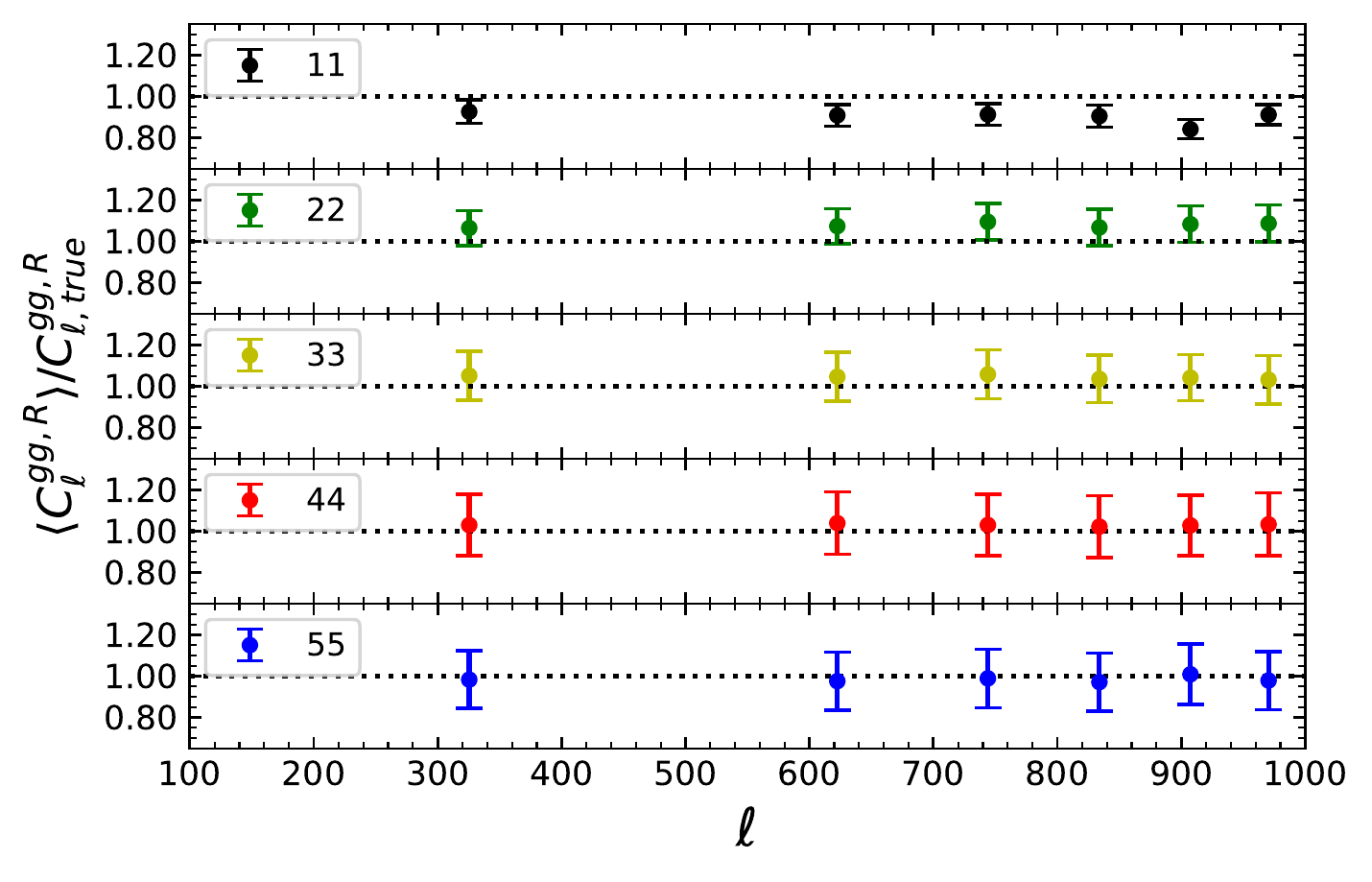}
}
\caption{The $\mathcal{J}$ distribution [panel (a)], the performance of the reconstructed scattering matrix [panel (b)] and the reconstructed power spectrum for true-$z$ bins [panel (c)] are presented for another $\ell$ binning scheme in the power spectrum measurement.
The performance is very similar to the results presented in Fig.~\ref{fig:sim_normal_l1_J} and the panel (b) of Fig.~\ref{fig:sim_normal_differ_l_result}.
}
\label{fig:appendix_law3}
\end{figure}

\begin{figure}
    \centering
    \includegraphics[width=0.95\columnwidth]{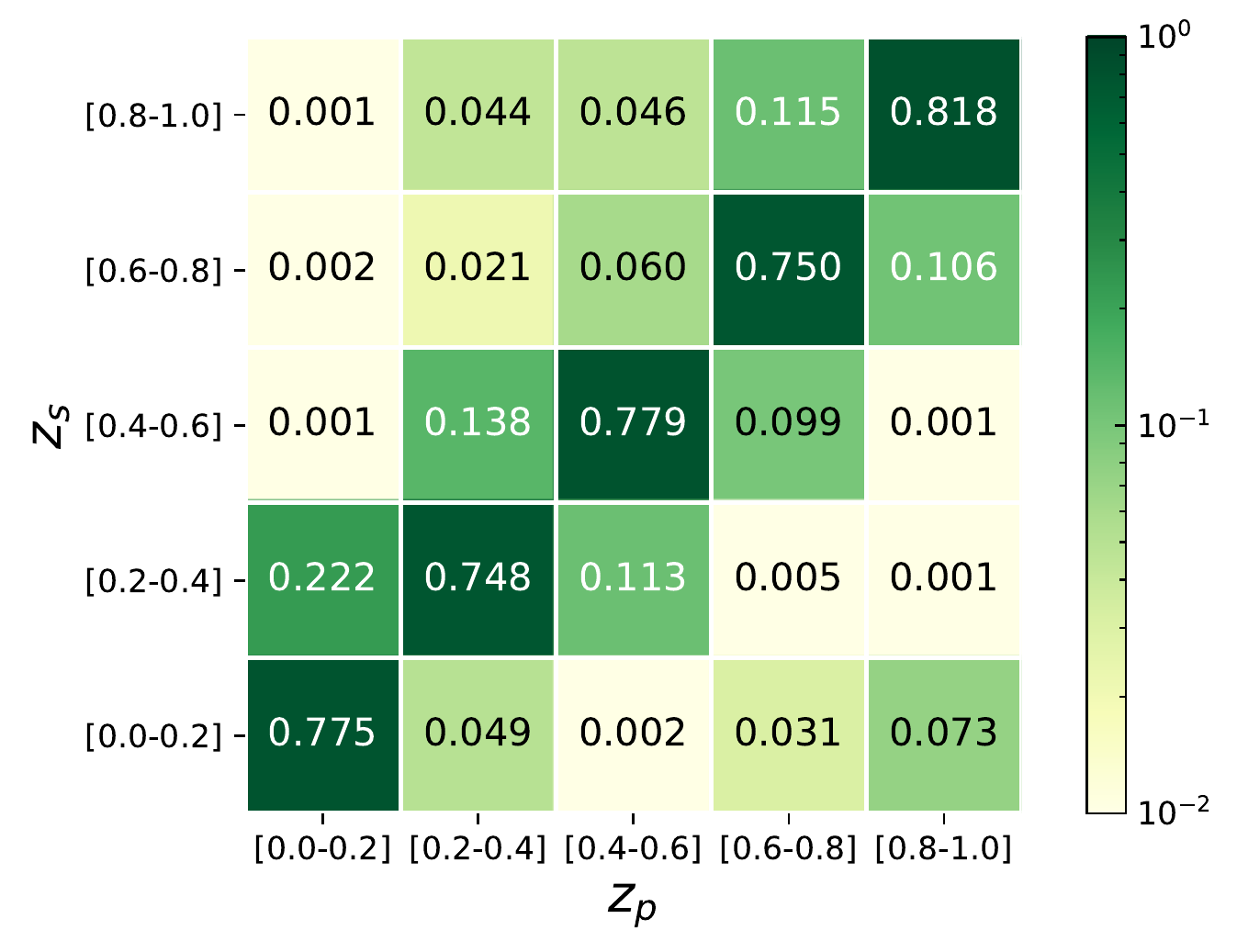}
    
	\includegraphics[width=\columnwidth]{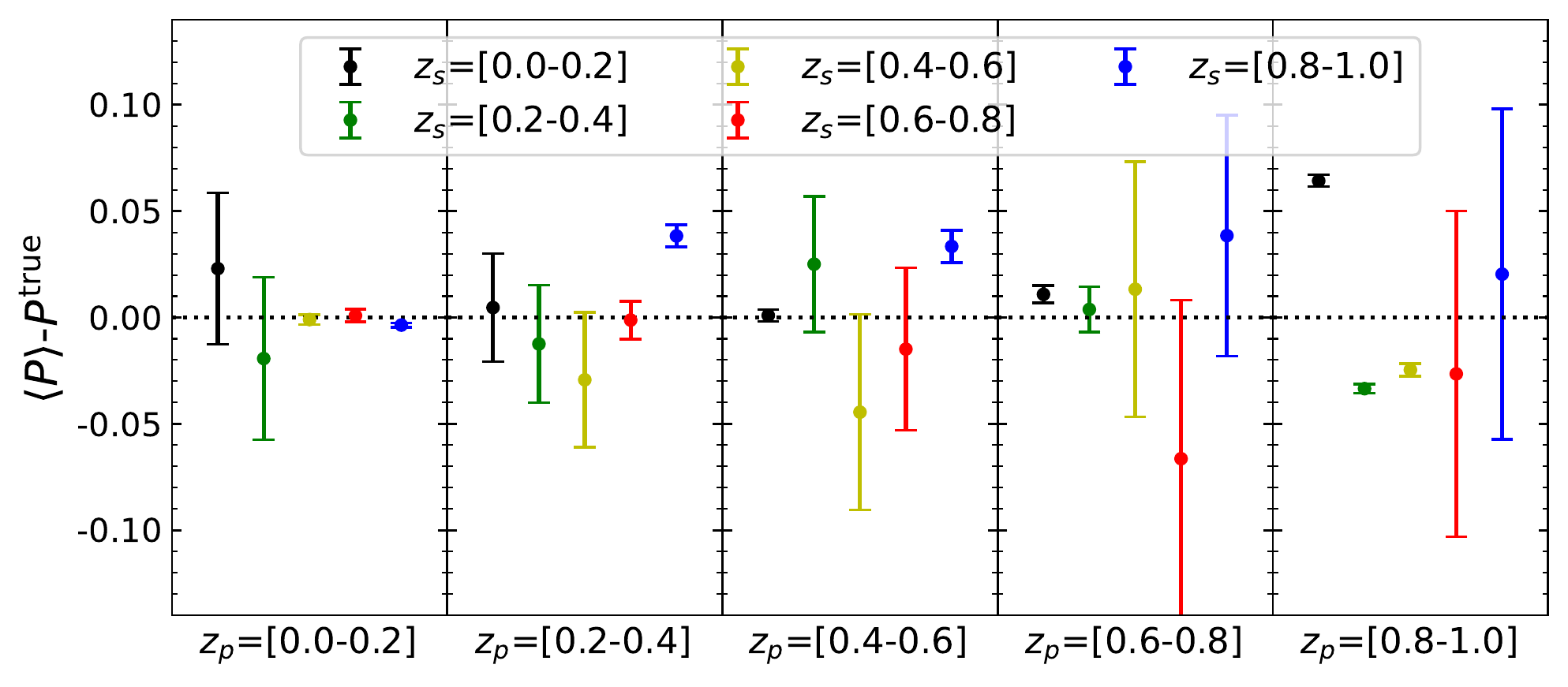}
    \caption{Top panel: the scattering matrix reconstructed by the algorithm after adding the magnification bias.
    Lower panel: the bias and standard deviation of each element in the reconstructed scattering matrix.}
    \label{fig:mag_bias}
\end{figure}

The dependence of the algorithm on power spectrum data is a problem that needs to be considered. Different binning scheme will lead to different data forms, which may affect the reconstruction ability of the self-calibration algorithm.
We arbitrarily choose a different binning scheme on the scale to explore the impact of this aspect.
Here the power spectrum is measured over another 6 broad bands, $[100,551)$, $[551,694)$, $[694,794)$, $[794,874)$, $[874,941)$, and $[941,1000)$.
The results are presented in Fig.~\ref{fig:appendix_law3}.
However, we find almost no difference from the results in the fiducial analysis shown in panel (b) of Fig.~\ref{fig:sim_normal_differ_l_result}.
So our algorithm is robust against the choice in the binning scheme as long as it is reasonable.

An important source of systematic error is the magnification bias. Gravitational lensing alters the galaxy flux and positions, and thus changes the observed galaxy clustering. The galaxy number overdensity after lensing takes the form
\begin{equation}
    \delta^{g}_{L}=\delta^{g}+g \kappa\ .
    \label{eqn:delta_g}
\end{equation}
Here $\kappa$ is the lensing convergence, with prefactor $g \equiv 2(\alpha-1)$, and $\alpha \equiv-d \ln n(F) / d \ln F -1$.
$n(F)$ is the number of galaxies per ﬂux interval.
% And the power spectrum of the lensed galaxy distribution can be approximated by
% \begin{equation}
%     C_{L}^{g g} \approx C^{g g}+2 g C^{g \kappa}\ .
%     \label{eqn:lensed_cgg}
% \end{equation}
Then the cross power spectrum between two photo-$z$ bins with the lensed galaxy distribution can be approximated by
\begin{equation}
    C_{ij,L}^{g g, P} \approx C_{ij}^{g g, P}+ g_j C_{ij}^{g \kappa, P}+g_i C_{ji}^{g \kappa, P}\ ,
    \label{eqn:lensed_cggp}
\end{equation}
where $C_{ij}^{g \kappa, P}$ is the galaxy-galaxy lensing power spectrum between photo-$z$ bins.
To roughly estimate the lensing magnification terms, we use the measured galaxy bias $b=[1.20, 1.25, 1.40, 1.70, 1.90]$ for DECaLS DR8 sample in \citet{Wang:2021wg}. 
$\alpha$ measured from the five photo-$z$ bins are [0.23, 0.41, 1.09, 1.52, 2.41].

Here we take the simulation case without catastrophic photo-$z$ errors (Fig.~\ref{fig:sim_normal_pture}) for example.
We use the \texttt{CCL}\footnote{\href{https://github.com/LSSTDESC/CCL}{https://github.com/LSSTDESC/CCL}}\citep{Chisari:2019wv} to estimate the $C^{g g, P}$ and $C^{g \kappa, P}$ with the same cosmological parameters and redshift distribution in the simulation, and the galaxy bias and $\alpha$ above.
%%The error on $C^{g g, P}$ is given by equation~(\ref{sigma_gg}).
%%We find that, the value of magnification bias is close to the $C_{ij}^{gg,P}$ only when the scatter rate is about 1\%.
%In order to further explore its influence, 
We theoretically calculate the ratio $C_{ij,L}^{g g, P}/C_{ij}^{g g, P}$, 
and then we multiply the measured power spectrum of the simulation case by the ratio, as a rough approximation of magnification bias.
In Fig.~\ref{fig:mag_bias}, we show the reconstructed results after adding the magnification bias.
The overall result is still good, though the values of some off-diagonal elements in the scattering matrix become slightly larger as expected, especially at high redshift bin due to large logarithmic luminosity slope $\alpha$.
Therefore, we believe that the magnification bias will not have a significant impact on the current reconstruction accuracy.
Furthermore, as mentioned in \cite{Zhang:2010wr}, the same weak-lensing surveys contain the right information to correct for the magnification bias.
The primary goal of this paper is to provide a proof of the self-calibration algorithm under the current accuracy.
We leave the delicate investigation on the magnification removal in the future work.

%%%%%%%%%%%%%%%%%%%%%%%%%%%%%%%%%%%%%%%%%%%%%%%%%%

% Don't change these lines
\bsp	% typesetting comment
\label{lastpage}
\end{document}